\documentclass[11pt,preprint]{aastex}
\def\lta{\lower2pt\hbox{$\buildrel {\scriptstyle <} 
   \over {\scriptstyle\sim}$}}
\def\gta{\lower2pt\hbox{$\buildrel {\scriptstyle >} 
   \over {\scriptstyle\sim}$}}
\def\propa{\hbox{$\buildrel {\scriptstyle\propto} 
   \over {\scriptstyle\sim}$}}
\def\Gr{\Gamma_{rel}}
\def\grba{{\em GRB}-$\natural$}
\def\fiveD{$(\gamma_i, \Gamma, B, N, \tau)$ }

\begin{document}

\title{A general scheme for modeling gamma-ray burst prompt emission}

\author{Pawan Kumar \& Erin McMahon \\
     Astronomy Department, University of Texas, Austin, TX 78712}

\begin{abstract}

We describe a general method for modeling gamma-ray burst prompt
emission, and determine the range of magnetic field strength, electron
energy, Lorentz factor of the source, and the distance of the source
from the central explosion that is needed to account for the prompt
$\gamma$-ray emission of a typical long duration burst. We
find that for the burst to be produced via the synchrotron process
unphysical conditions are required -- the distance of the source from
the center of the explosion ($R_\gamma$) must be larger than $\sim
10^{17}$cm and the source Lorentz factor $\gta 10^3$; for such a high
Lorentz factor the deceleration radius ($R_d$) is less than $R_\gamma$
even if the number density of particles in the surrounding medium is as
small as $\sim 0.1$ cm$^{-3}$. The result, $R_\gamma > R_d$, is in
contradiction with the early x-ray and optical afterglow data that show
that $\gamma$-rays precede the afterglow flux that is produced
by a decelerating forward shock. This problem for the synchrotron
process applies to all long-GRBs other than those that have the low
energy spectrum precisely $\nu^{-1/2}$.  In order for the synchrotron
process to be a viable mechanism for long-bursts, the energy of
electrons radiating in the $\gamma$-ray band needs to be continuously
replenished by some acceleration mechanism during much of the observed
spike in GRB lightcurve -- this is not possible if GRB prompt radiation
is produced in shocks (at least the kind that has been usually
considered for GRBs) where particles are accelerated at the shock front
and not as they travel down-stream and emit $\gamma$-rays, but {\it
might} work in some different scenarios such as magnetic outflows.

The synchrotron-self-Compton (SSC) process fares much better. There is a
large solution space for a typical GRB prompt emission to be
produced via the SSC process. The prompt optical emission accompanying the 
burst is found to be very bright ($\lta$ 14 mag; for $z\sim2$) in the SSC 
model, which exceeds the observed flux (or upper limit) for most GRBs. 
The prompt optical is predicted to be even brighter for the sub-class 
of bursts that have the spectrum $f_\nu\propto \nu^\alpha$ with $\alpha\sim 
1$ below the peak of $\nu f_\nu$. Surprisingly, there are no SSC solutions 
for bursts that have $\alpha\sim1/3$; these bursts might require continuous
or repeated acceleration of electrons or some physics beyond the
simplified, although generic, SSC model considered in this work.
Continuous acceleration of electrons can also significantly reduce the
optical flux that would otherwise accompany $\gamma$-rays in the SSC
model.

\end{abstract}


\section{Introduction}

The last 10 years have seen a rapid advance in our understanding of
gamma-ray bursts, due mainly to the study of GRB afterglows.  We now
know that at least some of the long duration GRBs (that last for more
than about 5s) are produced in the collapse of massive (young) stars
\citep{galama98,stanek03,hjorth03,kawabata03,malesani04,dellavalle03,
pian06,dellavalle06} as proposed by \citet{woosley93} and
\citet{paczynski98}, and short duration bursts are associated with old
stellar populations and are a likely product of merging neutron-star
binaries \citep{paczynski91,narayan92,gehrels05,fox05,berger05,gorosabel06,
nakar07}; for recent reviews please see
\citet{piran05,meszaros02,woosley06,zhang07}. We also have good
estimates of the total energy and beaming for these explosions as well
as the property of the medium within about 1 pc of the explosion
(Rhoads 1999; Sari, Piran \& Halpern, 1999; Frail et al 2001;
Panaitescu \& Kumar, 2001). However, 
our understanding of how the prompt $\gamma$-ray emission is generated, 
and the mechanism for energy transport from the central engine (via 
magnetic field or kinetic energy of protons-neutrons and/or 
electron-positron pairs) remains highly uncertain.

The goal of this paper is to provide a nearly model independent way of
modeling the prompt $\gamma$-ray emission with synchrotron or
synchrotron-self-Compton (SSC) processes. We determine the basic
properties of the $\gamma$-ray source from the data, and then determine
how these can be interpreted in currently popular models such as the
internal/external shock model.

In the next section we provide the basic idea and details of the technique we
use to model $\gamma$-ray emission (the idea in its early form can be found 
in Kumar et al. 2006), and in \S3 \& \S4 we describe the synchrotron and
SSC results, respectively.

\section{Modeling $\gamma$-ray emission: basic idea and technical formalism}

\begin{figure}[h!]
\begin{center}
\includegraphics[height=3.9in,width=5.4in]{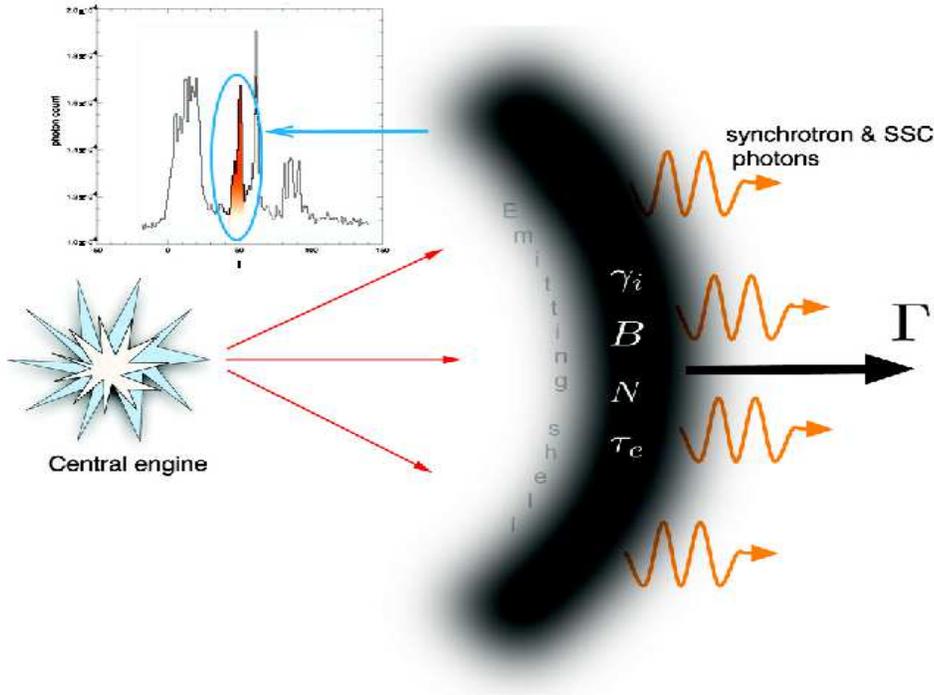}
\caption{\small A schematic representation of our model. Assuming that
radiation is synchrotron and inverse Compton, the $\gamma$-ray source
properties can be described by five parameters \fiveD~  
that determine the observed flux at
one instance in time.  We take this time to be the peak of a pulse in a GRB 
lightcurve. All of the calculations presented in this work apply to 
one single pulse in a typical GRB prompt lightcurve, as shown in the top 
left corner.}
\label{fig:shellpars}
\end{center}
\end{figure}

The starting point for our modeling of the prompt $\gamma$-ray emission
in GRBs is the assumption that the radiation is produced via the
synchrotron or synchrotron-self-Compton processes\footnote{Mechanisms
such as the inverse-Compton scattering of ``photospheric" emission from
a
hot fireball (cf. Lazzati et al. 2000; Broderick, 2005) are not modeled
by
the approach we have
adopted. And if it were to turn out that the GRB prompt emission is
produced by such a mechanism then the work presented here is of little
relevance.} in a source moving relativistically outward from the inner
engine.  Figure~\ref{fig:shellpars} provides a cartoon description of
our
model.  For a simple GRB light-curve (LC) consisting of a single
peak we determine the average source properties corresponding to the
time
when the observed light curve peaks, and for a multi-peak GRB LC our
calculation applies to individual pulses or spikes in the lightcurve.

The source property can be uniquely described by the following set of 5
parameters: the magnetic field strength ($B$) in Gauss, the number of
radiating particles ($N$) i.e., electrons and positrons, the optical
depth of the source to Thomson scattering ($\tau$), the Lorentz factor
of the source with respect to the rest frame of the GRB host galaxy
($\Gamma$), and the minimum electron energy\footnote{The electron energy
is $\gamma_i m_e c^2$, however, for convenience we suppress the factor
$m_e c^2$.} $\gamma_i$ at the location where particles are accelerated
(all the variables we use in this paper are defined in table 1 for easy
reference).  In addition, the particle distribution above $\gamma_i$ is
taken to be a power-law function: $dn/d\gamma\propto \gamma^{-p}$.
Particles cool as a result of radiative losses and with time, or as they
travel away from the acceleration site, and the distribution function
becomes steeper than the index $p$ at some energy where radiative losses
become important. We calculate the modified distribution
self-consistently as discussed below.  We constrain this 5D parameter
space with at least 4 observed quantities -- the $\nu f_\nu$ peak
frequency $\nu_\gamma$, the spectral index below $\nu_\gamma$, the flux
$f_\gamma$ at $\nu_\gamma$, the decay time of a single pulse in a GRB LC
$t_\gamma$; $p$, the power law index, is constrained by the high energy
spectral index, for $\nu > \nu_\gamma$.

\begin{deluxetable}{ll}
\tabletypesize{\small}
\tablecolumns{2}
\tablewidth{0pc}
\tablecaption{Definition of variables}
%\tablehead{
%\colhead{Variable} & \colhead{Definition} }
\startdata
\hline\hline
$\gamma_i$ & minimum electron LF in source comoving frame\\
$\Gamma$ & bulk LF of the source \\
$B$      & magnetic field strength, in Gauss, in source comoving frame\\
$\tau$ & optical depth to Thomson scattering \\
$N$      & number of radiating electrons(isotropic equivalent)\\
$\alpha$ & The spectral index below the peak of $\nu f_\nu$ i.e.
$f_\nu\propto
      \nu^\alpha$ \\
$\gamma_a$ & LF of electrons emitting synchrotron at $\nu_a$\\
$\gamma_c$ & LF of electrons emitting synchrotron at $\nu_c$\\
$\Gamma_{sh}$ & LF of the shocked gas wrt the unshocked gas \\
$\nu_\gamma$ & observed peak frequency of GRB $\nu f_\nu$ spectrum
 ($\nu_{\gamma_5} \equiv \nu_\gamma/10^5$eV) \\
$\nu_i$ & synchrotron injection frequency in observer frame ($\nu_{i_5}
\equiv \nu_i / 10^5$ eV) \\
$\nu_c$ & synchrotron cooling frequency in observer frame ($\nu_{c_5}
\equiv \nu_c / 10^5$ eV) \\
$\nu_a$ & synchrotron self absorption frequency in observer frame\\
$\nu_a^{ic}$ & SSC self absorption frequency, below which the
$f^{ic}_\nu$
   spectral index is +1\\
$A_*$ & external medium wind parameter, $n = (A/m_p) r^{-2}$; $A_*\equiv
  A/5 \times 10^{11}$g cm$^{-1}$\\
$d_{L_{28}} $ & luminosity distance in units of $10^{28}$ cm \\
$E_\pm$ & kinetic energy of electrons and positrons (lab frame;
isotropic equivalent)\\
$E_B$ & energy in magnetic field (lab frame; isotropic equivalent) \\
$E_{53}$ & isotropic equivalent of outflow energy in units of $10^{53}$
ergs\\
$f_{B/ke}$ & $E_B/E_\pm$ -- ratio of magnetic to $e^\pm$ energy \\
$f_R$ & synchrotron prompt optical flux (in $R$ band, at 2 eV) \\
$f_x$ & synchrotron prompt x-ray flux, at 1 keV \\
$f_\gamma$ & observed flux (in mJy) at $\nu_\gamma$\\
$f_{\nu_p}$ & synchrotron flux at peak -- $\min(\nu_i,\nu_c)$ \\
$n_0$ & density of circum-burst medium \\
$n_e$ & comoving electron density in unshocked shell \\
$p$   & power law index of electron energy distribution\\
$R_\gamma$ & distance from center-of-explosion at which the radiation is
produced\\
$R_d$ & deceleration radius\\
$t_\gamma$ & duration of one pulse in GRB light-curve (observer frame)
\\
% $\Gamma_1$ \& $\Gamma_2$ & bulk LF of shell `1' \& `2' before
% collision \\
$t_a$ &  the time available for electrons to cool before being
    re-accelerated  \\
% $n_1$ and $n_2$ & unshocked densities of shells before collision\\
$Y$ & Compton parameter \\
$z$  & redshift\\
$\eta_i$ & $\gamma_i / \gamma_c$ \\
$\eta_a$ & $\gamma_a / \gamma_i$ \\
\enddata
%\tablecomments{These are comments.}
\end{deluxetable}

A relativistic moving source of finite angular size $\theta_j$ (as seen
by an observer at the center of explosion) can be treated as spherically
symmetric as long as $\Gamma^{-1} < \theta_j$. The angular size
determined from afterglow modeling suggests that $\theta_j$ is larger
than about two degrees for all bursts for which we have good data (Frail
et al.  2001; Panaitescu \& Kumar, 2001) and a number of lines of
argument and evidence suggests that $\Gamma$ is greater than about 100
(cf. Piran 1992; Lithwick \& Sari, 2001). Therefore, we can treat the
source for prompt $\gamma$-ray emission as spherically symmetric, and
the numerical values we quote in this paper are all isotropic equivalent
quantities; for instance $N$ is the total number of radiating particles
in the source assuming the source to be spherically symmetric.

\subsection{Synchrotron and inverse-Compton radiations: basic equations}

The synchrotron injection frequency, $\nu_i$, corresponding to electron 
minimum energy $\gamma_i$, is 
\begin{equation}
\nu_i={qB\gamma_i^2\Gamma \over 2\pi m_e c(1+z)},
\label{nui}
\end{equation}
(eg. Rybicki \& Lightman, 1979; Wijers \& Galama, 1999),
where $q$ is electron charge, $m_e$ the electron mass, $c$ the speed of
light, and $z$ is the burst redshift. The synchrotron cooling
frequency, $\nu_c$, the characteristic frequency at which electrons
cooling on a time scale $t_a$ (observer frame) radiate, is
\begin{equation}
\nu_c=\frac{18 \pi q m_e c (1+z)}{\sigma_T^2 B^3 \Gamma t_a^2 (1+Y)^2}.
\label{nuc}
\end{equation}
where $\sigma_T$ is the Thomson scattering cross-section, and $Y$ the
Compton parameter.

For most of the calculations in this work we assume that electrons are
accelerated only once, and the time scale for acceleration is taken to be
much less than the duration of a pulse in the GRB light-curve
($t_\gamma$). One time acceleration is, for instance, believed to apply
to shocks where electrons are accelerated at the shock front (by crossing
the front back and forth multiple times) and not
while they travel downstream; the picture is likely very different in
magnetic reconnection/dissipation.  To capture some of the effects of
multiple-times particle acceleration in time period of a pulse duration
in GRB LC we introduce a time scale, $t_a$, which is the average time in
between two successive episodes of particle acceleration or the time
available for electrons to cool in between acceleration; for one shot
acceleration $t_a=t_\gamma$, and in the opposite limit of continuous
acceleration $t_a=0$ when the rate of energy gain is balanced by
radiative loss rate.  The electron distribution function resulting from
the acceleration process is taken to be $dn/d\gamma_e\propto
\gamma_e^{-p}$.  The distribution function in the source as a whole is
different due to the radiative cooling of electrons with time. The
electron distribution function averaged over the source is described by
two characteristic energies viz.  $\gamma_i$ and $\gamma_c$; $\gamma_c
m_e c^2$ is the energy of electrons that cool on time scale $t_a$.
The electron distribution for $\gamma_e> \max(\gamma_i,
\gamma_c)$ is proportional to $\gamma_e^{-p-1}$, and the distribution
between $\gamma_c$ and $\gamma_i$ (for $\gamma_c<\gamma_i$) is
proportional to $\gamma_e^{-2}$.  Electrons cool via synchrotron and
inverse-Compton losses. The rate of loss of energy is affected by the
synchrotron self-absorption frequency $\nu_a$ --- electrons with
characteristic synchrotron frequency below $\nu_a$ lose energy only via
the inverse-Compton scattering process. We calculate $\gamma_c$ and
$\nu_a$ by solving a coupled set of equations as described in
\citet{mcmahon06}.

The synchrotron flux at the peak of the $f_\nu$ spectrum, at
$\min\left[\nu_i,\nu_c\right]$, is given by
\begin{equation}
\label{f_syn_nup}
f_{\nu_p} = { \sqrt{3} q^3 B N\Gamma\over 4\pi d_L^2 m_e c^2},
\end{equation}
where $d_L$ is the luminosity distance to the source.
The effect of synchrotron self absorption is not
included in the above expression for $f_{\nu_p}$, and therefore the
observed flux, in general, would be different from $f_{\nu_p}$. The flux
at other frequencies are calculated as described in Sari, Piran and
Narayan (1998).

The inverse-Compton flux (in observer frame) at frequency $\nu$,
$f^{ic}(\nu)$, is calculated using the following equation
(cf. Rybicki \& Lightman, 1979)
\begin{equation}
f^{ic}(\nu) = {3\over 4} \sigma_T\delta r\int{d\nu_s\over\nu_s}
{\nu\over\nu_s}
   f(\nu_s) \int_{\gamma_1}^{\infty} d\gamma_e{d n\over\gamma_e^2d
    \gamma_e} F\left( {\nu\over 4 \gamma_e^2\nu_s}\right),
\end{equation}
where $f(\nu_s)$ is the synchrotron flux per unit frequency in the
observer
frame, $\delta r$ is the radial extent of the source (comoving frame)
which
is related to the optical depth $\tau$
(one of the five parameters we use to characterize the source),
the function $F(x)$ is
\begin{equation}
F(x) = 2x\ln x + x + 1 - 2x^2, \quad\quad{\rm for}\,\, 0<x<1,
\end{equation}
and $\gamma_1$ is the minimum LF for electron distribution.
We include the Klein-Nishina correction to the above expression when
$\nu_s\gamma_e/\Gamma > m_e c^2$.

The expression for the Compton Y parameter is
\begin{equation}
Y = \sigma_T\int dr'\int d\gamma_e \gamma_e^2 {d n_e\over d\gamma_e} = 
   {4 \over 3} \tau \left({p-1 \over p-2}\right) \gamma_i^2 \times
\left\{
\begin{array}{lll}
\hskip -5pt\left(\nu_{c}/\nu_{i}\right)^{\frac{1}{2}} & \gamma_c \ll
\gamma_i & p>2 \\
\hskip -5pt\left(\nu_{c}/\nu_{i}\right)^{3-p} \left(3-p\right)^{-1} &
\gamma_i \ll
\gamma_c & 2<p<3 \\
\hskip -5pt\left(p-3\right)^{-1} & \gamma_i \ll \gamma_c & p>3 \\
\hskip -5pt p \left(p-1\right)^{-1} & \gamma_i \sim \gamma_c & p>2
\end{array} \right. ,\label{s5}
\end{equation}
where the $r'$-integral is over the comoving radial width of the source.
For ease of future use we rewrite the above expression for $Y$ as $Y =
4\tau\gamma_i^2 \xi/3$ where
\begin{equation}
\xi \equiv \left({p-1 \over p-2}\right) \times \left\{
\begin{array}{lll}
\hskip -5pt \left(\nu_{c}/\nu_{i}\right)^{\frac{1}{2}} & \gamma_c
\ll \gamma_i & p>2 \\
\hskip -5pt\left(\nu_{c}/\nu_{i}\right)^{3-p} \left(3-p\right)^{-1} &
   \gamma_i \ll \gamma_c & 2<p<3 \\
\hskip -5pt\left(p-3\right)^{-1} & \gamma_i \ll \gamma_c & p>3 \\
\hskip -5pt p \left(p-1\right)^{-1} & \gamma_i \sim \gamma_c & p>2 .
\end{array} \right.  \label{xieq}
\end{equation}
In our 5D parameter space search, we limit
the Compton $Y$-parameter for a synchrotron solution to be less than 10
and for an inverse-Compton solution $Y\lta100$. The rationale for the
constraint on the $Y$-parameter is that we want an efficiency of
$\gta10$\% in the $\gamma$-ray energy band of $\sim$ 10--400 keV;
observations suggest this efficiency for a typical long duration burst
from a comparison of energy in the $\gamma$-ray radiation and the
kinetic energy of the ejecta determined from the afterglow observations
(Panaitescu \& Kumar, 2002).

We can calculate the distance of the source from the center of explosion
with 2 of the 5 parameters,
$N$ and $\tau$:
\begin{equation}
\label{r_source}
 R_\gamma= \left( {N \sigma_T \over 4\pi \tau} \right)^{1/2}.
\end{equation}

\subsection{Relation between $R_\gamma$ and pulse-width}

We consider a $\gamma$-ray source of a finite lifetime, at a distance
$R_\gamma$ from the central explosion, that is
responsible for generating one pulse in the observed GRB prompt
lightcurve. Electrons in the source are heated during some time interval, set
by the central engine variability/activity time, and subsequently the source
undergoes adiabatic expansion.
The width of an observed GRB pulse is determined by a number of different
factors -- the central engine variability time, the adiabatic expansion
and cooling time, and the curvature time -- which are described below.
\begin{enumerate}
\item{\bf Central engine variability time:
\it sets the observed GRB pulse width if the variability time is
larger than $\sim R_\gamma/2c\Gamma^2$ and if the distance where
$\gamma$-ray photons are generated does not increase with time; in this
case GRB pulse duration is independent of $R_\gamma$. However, when
the source is turned off the $\gamma$-ray flux would decline
on a timescale of $R_\gamma/2c\Gamma^2$, the adiabatic expansion
time scale; for all the calculations in this work we use the
light-curve decline time although for simplicity we continue to
refer to it as pulse width. }

\item{\bf Curvature timescale: \it is the time interval between arrival
of photons with angular separation of $\Gamma^{-1}$ as seen by an
observer at the center of explosion. It is
the minimum time scale for a $\gamma$-ray pulse width, as long as the
outflow from the GRB has an angular size larger than $\Gamma^{_1}$, and
is equal to $R_\gamma/(c\Gamma^2)$.}

\item {\bf Adiabatic expansion time scale -- \it this is the timescale for
  electrons/protons to cool because of expansion of the source. As the
  distance of the source from the center doubles its volume increases
  by a factor $\gta4$, and electron/proton energy drops by a factor ~2.
  This timescale in the observer frame is $\sim R_\gamma/(c\Gamma^2)$}.

  We note that if electrons are heated by coupling with protons,
  then the time scale for electrons to cool down can be larger than
  $R_\gamma/(c\Gamma^2)$; $e^\pm$ cooling time in this case can be as large as
  $R_\gamma/(c\Gamma^2)\times$(proton energy/electron energy) provided
  that protons transfer energy to electrons for this time duration,
  and the energy transfer rate
  balances the loss of energy for electrons to adiabatic and
  radiative coolings. However, the coupling between $p^+$s and $e^\pm$s is
  unlikely to increase the pulse width by a large factor ($\gta5$) unless
  the energy in electrons is much smaller than in $p^+$s, but in that case
  the efficiency of $\gamma$-ray generation would be small which is not
  supported by observations.

\item {\bf Radiative cooling timescale.} For $\nu_c>\nu_i$ the radiative
  cooling time scale is larger than the adiabatic time scale and in that
  case GRB pulse width is equal to the adiabatic or curvature time. For
  $\nu_c<\nu_i$ electrons cool on a smaller time scale, and once electron
  heating stops, the lightcurve falls off on the curvature timescale.
  We note that the observed pulse duration cannot be larger than
  $R_\gamma/(2c\Gamma^2)$ which corresponds to an elapsed time of
  $R_\gamma/c$ in the center of explosion frame, and during this time
  the source has moved to $R_\gamma$ from the center.
\end{enumerate}
Thus we see that the observed decay time for a pulse in GRB LC, produced by a
relativistic source, is
\begin{equation}
\label{t_grb}
 t_\gamma \approx {R_\gamma (1+z) \over 2 c \Gamma^2},
\end{equation}
when the GRB redshift is $z$.

\subsection{Energy etc.}

The total energy in the source consists of
the kinetic energy of electrons and positrons ($E_\pm$) and the magnetic
field ($E_B$):
\begin{equation}
E_\pm = N (p-1)\gamma_i m_e c^2 \Gamma/(p-2), \quad\quad
E_B = R_\gamma^3 B^2/6. \label{energies}
\end{equation}
Note that $(p-1)\gamma_i m_e c^2/(p-2)$ is the average energy per
electron/positron in the source comoving frame at the acceleration site,
and in the calculation of $E_B$ we took the comoving radial thickness
of the source to be $R_\gamma/\Gamma$ which is roughly what one expects
for a causally connected source where the signal speed is close to
the speed of light.

We do not make any assumptions regarding the energy in protons
since protons do not contribute to the observed $\gamma$-ray radiation.
This has the effect that the parameter space we determine is larger than
it would be if protons carried a substantial amount of energy since the
energy available to $e^\pm$ would be smaller than the upper limit of
10$^{55}$ erg (isotropic equivalent) we impose in our search for
solutions
in the 5D parameter space.

For a $\gamma$-ray source that arises from shock heated
gas, the minimum electron energy behind the shock front, $\gamma_i$
(one of the five parameters we use), can be related to
the Lorentz factor of the shocked gas wrt the unshocked gas,
$\Gamma_{sh}$. The minimum $\Gamma_{sh}$ needed to produce $\gamma_i$
is
\begin{equation}
\Gamma_{sh} = \left[ {m_e(p-1)\over m_p(p-2)}\right]2\gamma_i,
\label{gamsh}
\end{equation}
where $m_p$ ($m_e$) is proton (electron) mass. The factor 2 in the
above expression is for the case where there is an energy
equipartition between electrons and
protons and there are no $e^\pm$ pairs in the plasma; $\Gamma_{sh}$
will be larger if there are pairs or if electrons have less energy than
protons.\footnote{If there are ``cold" protons and electrons in the
shocked
gas, i.e. only a fraction of particles in the shocked gas are
accelerated,
and electrons have more than $m_p c^2\Gamma_{sh}$ energy, we would in
that
case overestimate $\Gamma_{sh}$.  However, in this work we do not
consider
that there is a cold component to the $\gamma$-ray source since such a
component would not radiate and affect observations and the solutions
in the 5-D parameter space. Therefore, such a cold component,
if present, would have little effect on all of the major results in this
work; the only quantity affected by the cold component is the value for
$\Gamma_{sh}$ which is a peripheral quantity and not part of the central
flow of the logic in this paper.}

\subsection{The basic technique for finding source properties}

 We determine the properties of the $\gamma$-ray source for a
GRB by finding the region in the 5-D parameter space ~\fiveD~ that satisfies the
following set of observational constraints: 1) the frequency at the peak
of the $\nu f_\nu$ spectrum ($\nu_\gamma$); 2) the peak flux at
$\nu_\gamma$; 3) the spectral index above $\nu_\gamma$ -- which
constrains electron index $p$ -- and the index below $\nu_\gamma$; 4)
the burst duration -- for a GRB with a single pulse in the LC -- or the
duration of an individual pulse ($t_\gamma$) for GRBs with complicated
LC; 5) optical and x-ray prompt flux or limit if available.  The flux at
a given observer time reflects the property of the source averaged over
equal-arrival-time volume, therefore, the observed peak flux depends on
the evolution of the source and this introduces uncertainty in the flux
calculation by a factor of about two.  For this reason we only require
the theoretical flux to match the observed value to within a factor of
$\sim2$.

We now use this technique to find the 5D solution space and source
property for GRBs produced via synchrotron (\S3) and SSC (\S4).

\section{Synchrotron solutions}

We consider in this section the parameter space of solutions when the
observed $\gamma$-rays are produced via the synchrotron process\footnote{
It has been suggested that another radiation process, called jitter, might 
be responsible for $\gamma$-ray generation for those bursts that have
low energy spectrum $f_\nu\propto \nu$ (Medvedev, 2000). 
We show in appendix {\bf B} that whenever jitter radiation dominates
the observed flux to produce a $f_\nu\propto \nu$ spectrum the Compton-Y
parameter is extremely large -- $Y\gta10^6$ -- and most of the energy 
of the explosion comes out in $\sim 100$ GeV SSC photons.}.  First,
we determine approximate solutions by analytically solving a system of 
equations for our 5 parameters \fiveD~for the generic synchrotron case.  The
solutions for each parameter are expressed in terms of the Compton $Y$
parameter and three observed quantities: the frequency $\nu_\gamma$ where
$\nu f_\nu$ peaks, the $\gamma$-ray flux at this frequency ($f_\gamma$; in 
mJy), and the duration of a pulse in GRB LC ($t_\gamma$);  $Y$ is a
convenient and useful parameter because its value is expected to lie
in a limited range, e.g. $Y\lta1$ for the synchrotron solutions \& 
$1\lta Y \lta 10$ for the SSC process.  Having the
general synchrotron solutions in hand, we then apply the analytical
results to the low energy spectral index cases of $\alpha=1/3$,
$\alpha=-1/2$ and $\alpha=-(p-1)/2$, compare the analytical and
numerical results, and draw conclusions as to the process by which 
$\gamma$-rays are generated in GRBs;  the spectral index $\alpha$ is 
defined by $f_\nu\propto \nu^{\alpha}$ for $\nu<\nu_\gamma$.

The 5 equations that we solve are those for the observer frame
synchrotron injection frequency $\nu_i$ (\ref{nui}), the cooling
frequency $\nu_c$ (\ref{nuc}), the pulse duration ($t_\gamma$)
(\ref{r_source} \& \ref{t_grb}), synchrotron flux $f_{\nu_p}$ in mJy at
$\nu_p\equiv \min(\nu_i, \nu_c)$ (\ref{f_syn_nup}), and the Compton $Y$
parameter (\ref{s5}):
\begin{eqnarray}
&& \nu_{i_5} = 1.1\times 10^{-13} B \gamma_i^2 \Gamma (1+z)^{-1} \label{s1}\\\nonumber\\[-.5cm] 
&& \nu_{c_5} = 6.6 \times 10^4 (1+z) B^{-3} \Gamma^{-1} t_a^{-2}
(1+Y)^{-2} \label{s2} \\\nonumber\\[-.5cm] 
&& \tau \approx 1.5 \times10^8 N_{55} (1+z)^2 \Gamma^{-4} t_\gamma^{-2}
\label{s3} \\\nonumber\\[-.5cm] 
&& f_{\nu_p} = 110 B \Gamma N_{55} d_{L_{28}}^{-2} (1+z) \, {\rm mJy}\label{s4}
\end{eqnarray}
where $\nu_{i_5}\equiv \nu_i/10^5$eV,
$\nu_{c_5}\equiv\nu_c/10^5$eV, and $N_{55}\equiv N/10^{55}$.  

To solve Equations \ref{s1}-\ref{s4} \& \ref{s5}, we first eliminate $N_{55}$
from equation (\ref{s4}) using equation (\ref{s3}), then eliminate $\tau$
using equation (\ref{s5}) to find
\begin{equation}
B \Gamma^5 \gamma_i^{-2} \approx 1.9 \times 10^6 f_{\nu_p} d_{L_{28}}^2 (1+z)
t_\gamma^{-2} Y^{-1} \xi. 
\label{s6}
\end{equation}  
Next, combining (\ref{s6}) \& (\ref{s1}) we get
\begin{equation}
\gamma_i^{-4} \Gamma^4 \approx 2.1 \times 10^{-7} \nu_{i_5}^{-1} f_{\nu_p}
d_{L_{28}}^2 Y^{-1} \xi t_\gamma^{-2}
\label{s7}.
\end{equation} 
Multiplying the square root of equations (\ref{s1}) and (\ref{s2}) together, we have
\begin{equation}
B \gamma_i^{-1} \approx 8.5 \times 10^{-5} t_a^{-1} \nu_{c_5}^{-\frac{1}{2}}
\nu_{i_5}^{-\frac{1}{2}} (1+Y)^{-1}.
\label{s8}
\end{equation}
We can eliminate $\gamma_i$ from equations (\ref{s7}) and (\ref{s8}) by
dividing equation (\ref{s7}) by (\ref{s8}) to the fourth power:
\begin{equation}
\Gamma^4 B^{-4} \approx 4.0 \times 10^9 \nu_{i_5} \nu_{c_5}^2 f_{\nu_p}
d_{L_{28}}^2 Y^{-1} (1+Y)^4 \xi t_a^4 t_\gamma^{-2} 
\label{s9}
\end{equation}
And finally, if we multiply equation (\ref{s9}) by the fourth power of 
(\ref{s6}) and divide by the square of (\ref{s7}) we find the solution 
for $\Gamma$ to be
\begin{equation} 
\Gamma \approx 10^3 \nu_{i_5}^{\frac{3}{16}} \nu_{c_5}^{\frac{1}{8}}
f_{\nu_p}^{\frac{3}{16}} t_\gamma^{-\frac{3}{8}} t_a^{\frac{1}{4}}
Y^{-\frac{3}{16}}
(1+Y)^{\frac{1}{4}} \xi^{\frac{3}{16}} 
d_{L_{28}}^{\frac{3}{8}} (1+z)^{\frac{1}{4}} 
\label{s10}.
\end{equation}
Using $\Gamma$, we can solve for $\gamma_i$, $B$, and $\tau$:
\begin{eqnarray}
& & \gamma_i \approx 4.7 \times 10^4 \nu_{i_5}^{\frac{7}{16}}
\nu_{c_5}^{\frac{1}{8}} f_{\nu_p}^{-\frac{1}{16}}
t_\gamma^{\frac{1}{8}} t_a^{\frac{1}{4}} Y^{\frac{1}{16}}
(1+Y)^{\frac{1}{4}} \xi^{-\frac{1}{16}}
d_{L_{28}}^{-\frac{1}{8}} (1+z)^{\frac{1}{4}} 
\label{s11}  \\  \nonumber\\[-.4cm] 
& & B \approx 4.0  \nu_{i_5}^{-\frac{1}{16}} \nu_{c_5}^{-\frac{3}{8}}
f_{\nu_p}^{-\frac{1}{16}} t_\gamma^{\frac{1}{8}} t_a^{-\frac{3}{4}}
Y^{\frac{1}{16}}
(1+Y)^{-\frac{3}{4}} \xi^{-\frac{1}{16}}
d_{L_{28}}^{-\frac{1}{8}} (1+z)^{\frac{1}{4}} 
\, {\rm Gauss}\label{s12} \\ \nonumber\\[-.4cm] 
& & \tau \approx 3.3 \times 10^{-10} \nu_{i_5}^{-\frac{7}{8}}
\nu_{c_5}^{-\frac{1}{4}} f_{\nu_p}^{\frac{1}{8}}
 t_\gamma^{-\frac{1}{4}} t_a^{-\frac{1}{2}} Y^{\frac{7}{8}}
(1+Y)^{-\frac{1}{2}} \xi^{-\frac{7}{8}}
d_{L_{28}}^{\frac{1}{4}} (1+z)^{-\frac{1}{2}} 
\label{s13}.
\end{eqnarray}
Equations (\ref{s10})--(\ref{s13}) provide approximate solutions for \fiveD~
when the synchrotron process produces the observed $\gamma$-ray radiation; more
accurate solutions for these parameters are obtained by numerical calculations
and the results are shown in Figures 
\ref{fig:syncp12}--\ref{fig:synchalfxo}. These general solutions can be 
used to investigate different cases of low energy spectral indices ($\alpha$)
by adopting appropriate values for $\nu_{i_5}$ and $\nu_{c_5}$.  
 The full dependences on these two frequencies are not completely shown here --
each case of $\alpha$ has a different functional dependence on $\xi$, 
and $\xi$ is a function of $\nu_i$ and $\nu_c$.  

Note that $f_{\nu_p}$ is not the observed flux at $\nu_\gamma$, the peak of 
$\gamma$-ray spectrum, but is the flux at $\min(\nu_i,\nu_c)\equiv\nu_p$, 
and the effect of synchrotron-self-absorption, if any, at $\nu_p$ has
been ignored. Since the dependence of the parameters $\Gamma$, $\gamma_i$ etc. 
on $f_{\nu_p}$ is very weak (eqs. \ref{s10}--\ref{s13}), we do not worry 
about the difference between $f_{\nu_p}$ \& $f_\gamma$ at this point, even 
though $f_{\nu_p}$ can be much greater than $f_\gamma$ (the flux at 
$\nu_\gamma$); $\nu_\gamma$ is the peak of $\nu f_\nu$ -- for $p<3$, 
$\nu_\gamma=\max[\nu_i, \nu_c]$ \& for $p>3$ $\nu_\gamma=\min[\nu_i, \nu_c]$).  

Using the parameter solutions, we can derive the distance of the $\gamma$-ray 
source from the center of explosion ($R_\gamma$), and the energy in 
the magnetic field and electrons. The radius $R_\gamma=2 c 
\Gamma^2 t_\gamma (1+z)^{-1}$ is found to be 
\begin{equation}
R_\gamma \approx 6.0 \times 10^{16} \nu_{i_5}^{\frac{3}{8}} \nu_{c_5}^{\frac{1}{4}}
f_{\nu_p}^{\frac{3}{8}} 
t_\gamma^{\frac{1}{4}} t_a^{\frac{1}{2}} Y^{-\frac{3}{8}}
(1+Y)^{\frac{1}{2}} \xi^{\frac{3}{8}} d_{L_{28}}^{\frac{3}{4}}
(1+z)^{-\frac{1}{2}}\,  \mathrm{cm}
\label{s14}
\end{equation}
and should be compared to the deceleration radius ($R_d$) of the GRB outflow
in both a homogeneous external medium with particle number density $n_0$, 
and a wind external medium where the particle number density is given by 
$(A/m_p)r^{-2}$
(these are two special cases of a power law density stratification -- 
the density varying as $r^{-s}$ -- corresponding to $s=0$ \& $s=2$)
\begin{equation}
R_d = \left\{\begin{array}{lll}
\hskip -7pt 1.2 \times 10^{17} E_{53}^{\frac{1}{3}} n_0^{-\frac{1}{3}}
\Gamma_2^{-\frac{2}{3}} &
\mathrm{cm}& s=0 \\
  & \\
\hskip -7pt 1.8 \times 10^{15} E_{53} A_*^{-1} \Gamma_2^{-2} &
  \mathrm{cm}& s=2
\end{array} \right.
\label{rd1}
\end{equation}
where $E_{53}$ is the isotropic equivalent energy in GRB-ejecta in units of
$10^{53}$ ergs, $\Gamma_2 = \Gamma/100$, and $A_* = A/(5\times10^{11}
\mathrm{g}$ cm$^{-1}$). Substituting in the solution for $\Gamma$, we find
$R_d$ to be 
\begin{equation}
R_d \approx \left\{\begin{array}{lll}
\hskip -7pt 2.6 \times 10^{16} E_{53}^{1 \over 3} n_0^{-{1\over 3}}
 \nu_{i_5}^{-{1\over 8}} \nu_{c_5}^{-{1 \over 12}} f_{\nu_p}^{-{1 \over 8}}
t_\gamma^{1 \over 4}
t_a^{-{1\over 6}} Y^{1 \over8} (1+Y)^{-{1\over6}} \xi^{-{1\over 8}}
d_{L_{28}}^{-{1 \over 4}} (1+z)^{-{1\over 6}} &
       \mathrm{cm}& s=0 \\
  &  \\
\hskip -7pt 1.8\times 10^{13}  E_{53} A_*^{-1} \nu_{i_5}^{-{3 \over8}}
\nu_{c_5}^{-{1 \over 4}} f_{\nu_p}^{-{3 \over 8}} t_\gamma^{{3 \over 4}}
t_a^{-{1\over 2}} Y^{3\over8} (1+Y)^{-{1\over 2}} \xi^{-{3 \over8}}
d_{L_{28}}^{-{3 \over 4}} (1+z)^{-{1\over 2}} &
   \mathrm{cm}& s=2
\end{array} \right.
\label{rd2}
\end{equation} 

The magnetic and $e^\pm$ energies, given by equation (\ref{energies}),
are found to be
\begin{equation}
E_B \approx5.8 \times 10^{50} \, \nu_{i_5} f_{\nu_p}
t_\gamma Y^{-1} \xi \label{E_Bs} d_{L_{28}}^2
(1+z)^{-1}\,\,\, {\rm ergs}
\label{EB1}
\end{equation}
and 
\begin{equation}
E_\pm \approx 8.5 \times 10^{50} \,\left({p-1 \over
p-2}\right) \nu_{i_5}^{\frac{1}{2}} \nu_{c_5}^{\frac{1}{2}} f_{\nu_p} 
 t_a (1+Y)  d_{L_{28}}^2
(1+z)^{-1}\,\,\, {\rm ergs} \label{E_ps}.
\label{EP1}
\end{equation}
Since the dependence of the above two quantities on $f_{\nu_p}$ is
linear, we should replace $f_{\nu_p}$ with $f_\gamma$, the flux observed
at $\nu_\gamma$.  This will be done in the following sections, 
since the expression for $f_\gamma$ depends on $\alpha$.

We can relate the solution-subspace we find to parameters for the shock model
for GRBs; if electrons are accelerated in a relativistic shock then the LF of
shock front ($\Gamma_{sh}$) wrt to the unshocked material is related
to $\gamma_i$ (one of the 5 parameters) and is given by equation (\ref{gamsh}) 
\begin{equation}
\Gamma_{sh} \approx 50 \,\epsilon_e^{-1} \left({p-1 \over p-2}\right)
\nu_{i_5}^{\frac{7}{16}} \nu_{c_5}^{\frac{1}{8}} f_{\nu_p}^{-\frac{1}{16}}
t_\gamma^{\frac{1}{8}} t_a^{\frac{1}{4}} Y^{\frac{1}{16}}
(1+Y)^{\frac{1}{4}} \xi^{-\frac{1}{16}} d_{L_{28}}^{-\frac{1}{8}}
(1+z)^{\frac{1}{4}},
\end{equation}
where $\epsilon_e$ is the ratio of energy in electrons and the total thermal
energy in the $\gamma$-ray source.

We now apply the results obtained in this section to each possible 
synchrotron low energy spectral index $\alpha$.

%************************************************************
\subsection{Synchrotron solutions when the low energy spectrum is
$\nu^{-\frac{(p-1)}{2}}$} 

We use equation (\ref{xieq}) to eliminate $\xi$ from the analytical
solutions given by equations \ref{s10}--\ref{s13} for the $\gamma_i \ll
\gamma_c$ \& $2<p<3$ case, and substitute $\nu_{\gamma_5}=\nu_{c_5}$ \&
$f_{\nu_p} = f_\gamma (\nu_{\gamma_5}/\nu_{i_5})^{\frac{(p-1)}{2}}$, to
find that synchrotron solutions for $\alpha=-\frac{(p-1)}{2}$ are:
\begin{eqnarray}
&& \Gamma \approx 10^3
\nu_{\gamma_5}^{19-3p \over 32} f_\gamma^{\frac{3}{16}}
t_\gamma^{-\frac{3}{8}}  \nu_{i_5}^{3p-9 \over 32}
t_a^{\frac{1}{4}} Y^{-\frac{3}{16}} (1+Y)^{\frac{1}{4}}
(1+z)^{\frac{1}{4}} d_{L_{28}}^{\frac{3}{8}} 
A_{1p}^{\frac{3}{16}} \label{p12-1}\\ \nonumber\\[-.4cm] 
&& \gamma_i \approx 4.7 \times 10^4 
\nu_{\gamma_5}^{p-1 \over 32} f_\gamma^{-\frac{1}{16}}
t_\gamma^{\frac{1}{8}} \nu_{i_5}^{19-p \over 32}
t_a^{\frac{1}{4}} Y^{\frac{1}{16}} (1+Y)^{\frac{1}{4}}
(1+z)^{\frac{1}{4}} d_{L_{28}}^{-\frac{1}{8}}
A_{1p}^{-\frac{1}{16}}  \label{p12-2} \\ \nonumber\\[-.4cm] 
&& B \approx 4.0 
\nu_{\gamma_5}^{p-17 \over 32} f_\gamma^{-\frac{1}{16}}
t_\gamma^{\frac{1}{8}} \nu_{i_5}^{3-p \over 32}
t_a^{-\frac{3}{4}} Y^{\frac{1}{16}} (1+Y)^{-\frac{3}{4}}
(1+z)^{\frac{1}{4}} d_{L_{28}}^{-\frac{1}{8}} 
A_{1p}^{-\frac{1}{16}} \, {\rm Gauss} \label{p12-3} \\ \nonumber\\[-.4cm] 
&& \tau \approx 3.3 \times 10^{-10} 
\nu_{\gamma_5}^{15p-47 \over 16} f_\gamma^{\frac{1}{8}}
 t_\gamma^{-\frac{1}{4}} \nu_{i_5}^{29-15p \over 16}
t_a^{-\frac{1}{2}} Y^{\frac{7}{8}} (1+Y)^{-\frac{1}{2}}
(1+z)^{-\frac{1}{2}} d_{L_{28}}^{\frac{1}{4}} 
A_{1p}^{-\frac{7}{8}}  \label{p12-4}
\end{eqnarray}
where
\begin{equation}
A_{1p} \equiv {\left(p-1\right) \over
\left(p-2\right)\left(3-p\right)},
\end{equation}
and we should emphasize that $t_\gamma$ is {\bf not} 
the burst duration -- it is the width of a single spike in the GRB 
prompt-lightcurve.

For a typical long duration GRB with $f_\gamma=1$mJy, $\nu_\gamma=100$keV, 
$t_\gamma=0.1$s, $z=1$, $d_{L_{28}} = 2$, and $t_a\sim t_\gamma$
-- henceforth we will refer to a GRB with these observed parameters 
as \grba~ -- the 5-parameters of the $\gamma$-ray source \fiveD  are 
obtained from equations \ref{p12-1}--\ref{p12-4} and are given by 
\begin{eqnarray}
&&\Gamma \,\gta\,\, 3.2 \times 10^3 Y^{-\frac{3}{16}} (1+Y)^{\frac{1}{4}} \\ \nonumber\\[-.5cm] 
&&\gamma_i \,\lta\,\, 6.0 \times 10^3 Y^{\frac{1}{16}} (1+Y)^{\frac{1}{4}} \\ \nonumber\\[-.5cm] 
&& B \,\lta\, 16 Y^{\frac{1}{16}} (1+Y)^{-\frac{3}{4}} \, {\rm Gauss} \\ \nonumber\\[-.5cm] 
&&\tau \,\gta\,\, 4.7 \times 10^{-9} Y^{\frac{7}{8}} (1+Y)^{-\frac{1}{2}}. 
\end{eqnarray}
In deriving these inequalities we took $p=2.5$, $\nu_c=\nu_\gamma=100$ keV, 
and $\nu_{i_5} < 0.1$. 

The dependence of $\Gamma$, $\gamma_i$, $B$ \& $\tau$ on $Y$ is weak,
so the coefficients in above expressions are reasonable estimates
for the $\gamma$-ray source basic physical parameters for \grba. We see
that the $\gamma$-ray source LF, $\Gamma$, is required to be rather
large -- $\Gamma\gta 3\times10^3$ -- if the radiation is to be produced
via the synchrotron process.  This large $\Gamma$ is not consistent with
afterglow modeling, which gives a value  of a few hundred or less
(Panaitescu \& Kumar, 2002). Furthermore, as shown below, the distance
of $\gamma$-ray source from the center of explosion turns out to be
larger than the deceleration radius for this large $\Gamma$ value,
unless $n_0$ is very small.  This suggests that the synchrotron solution
is internally inconsistent; after the deceleration radius $\Gamma$ is a
function of $N$, $\gamma_i$ and $n_0$ and is no longer an independent
parameter as considered in these derivations. The possibility that
$R_\gamma>R_d$ is also ruled out by early optical afterglow
data -- eg. GRBs 050801, 050820A, 060124, 060418, 060607A, 060614, 
060714 -- that show that $\gamma$-rays precede a rising afterglow flux
that is produced by a decelerating forward shock. Moreover, if
$R_\gamma \gta R_d$, then in this case of a decelerating source 
we should see an increasing GRB pulse duration with time, which is 
not observed. 

The distance of the $\gamma$-ray source from the center of explosion,
$R_\gamma\approx 2ct_\gamma\Gamma^2/(1+z)$, is calculated using 
eq. (\ref{p12-1}), and is given by
\begin{equation}
R_\gamma \approx 6.0 \times 10^{16} \quad
\nu_{\gamma_5}^{19-3p \over 16} f_\gamma^{\frac{3}{8}}
t_\gamma^{\frac{1}{4}} \nu_{i_5}^{3p-9 \over 16} 
t_a^{\frac{1}{2}} Y^{-\frac{3}{8}}
(1+Y)^{\frac{1}{2}} (1+z)^{-\frac{1}{2}} d_{L_{28}}^{\frac{3}{4}}
A_{1p}^{\frac{3}{8}} \,\, {\rm cm}
\label{p12r}
\end{equation}
or $R_\gamma \sim3\times10^{16}\,Y^{-\frac{3}{8}}(1+Y)^{\frac{1}{2}}\,
{\rm cm}$ for \grba.

\begin{figure}
\vskip -0.35in
\includegraphics[]{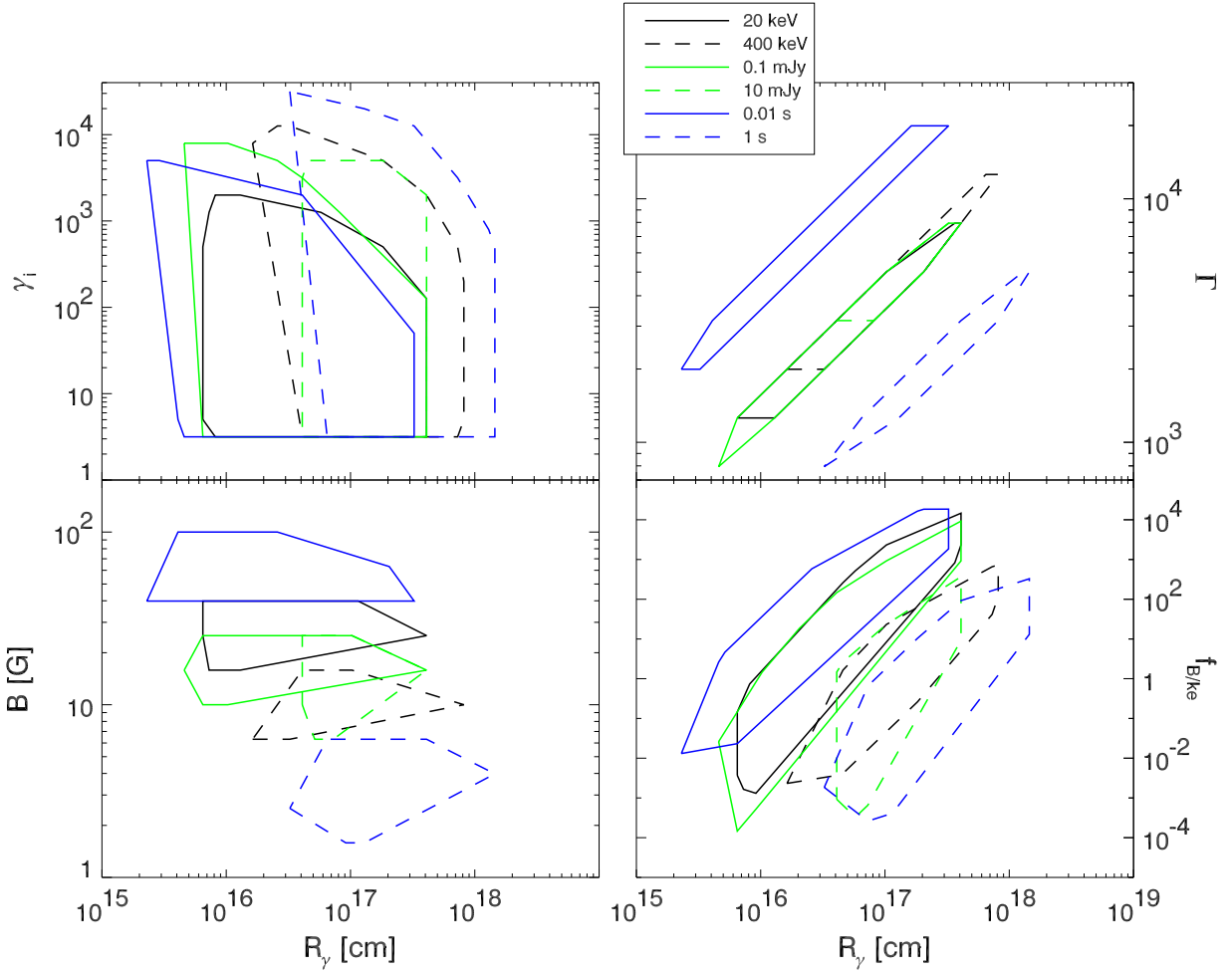}
\caption{Results of numerical calculation for the allowed synchrotron 
solution space when the spectrum below the peak of $\nu f_\nu$, at 
$\nu_\gamma$, is: $f_\nu\propto \nu^{-(p-1)/2}$ for $\nu<\nu_\gamma$. 
A point in the 5-D parameter space ($\gamma_i, \Gamma, B, N, \tau$) is 
considered an allowed solution for the observed GRB parameters 
($\nu_\gamma$, $f_\gamma$, $t_\gamma$, $\alpha$) provided 
that $\nu_\gamma$ is within a factor 2 of the observed value, the 
pulse duration ($t_\gamma$) \& flux at $\nu_\gamma$ ($f_\gamma$) are within 
a factor $1.5$ \& 3 of the observed value respectively; the larger tolerance
on flux is due to larger error in flux calculation.
The x-axis shows the distance of the $\gamma$-ray source from the
center of the explosion. The top left panel is $\gamma_i$ -- the
minimum LF of electrons in source comoving frame at the site where
they are accelerated (electron distribution function for $\gamma_e>\gamma_i$ 
is: $dn_e/d\gamma_e\propto\gamma_e^{-2.5}$ i.e. $p=2.5$). The top right panel 
shows the bulk LF of the source, the bottom left panel shows the comoving 
magnetic field in Gauss, and the bottom right panel shows the ratio of energy 
in the magnetic field and electrons. For all of the numerical calculations we 
took the burst redshift $z=1$.
Legend shows several different cases of GRBs corresponding to 
different observed values for $\nu_\gamma$, $f_\gamma$, and $t_\gamma$.
Only one observational parameter -- that noted in the legend -- is changed
at a time, all the remaining parameters are left unchanged; the base value
for the parameters is the same as we took for \grba, i.e. $\nu_\gamma=100$ keV,
$f_\gamma=1$mJy, $t_\gamma=0.1$s, and $t_a=t_\gamma$.  For instance,
for the 20 keV case, denoted by the solid black line, 
$\nu_\gamma=20$keV, and $f_\gamma$ \& $t_\gamma$ are same as for \grba~ i.e.
1 mJy and 0.1 s respectively.}
\label{fig:syncp12}
\end{figure}

We now compare these analytical estimates to the numerically computed
solution space for synchrotron radiation. A numerical search of the
allowed region of the 5-D parameter space that satisfies the
observational constraints ($\nu_\gamma$, $f_\gamma$ \& $t_\gamma$; the
same constraints that we used in the derivation of analytical
expressions), confirms that for synchrotron solutions $R_\gamma\gta
10^{16}$cm, $\Gamma\gta10^3$, and $10\lta\gamma_i\lta10^4$ (see fig.
\ref{fig:syncp12}).  We have considered a wide range of values of peak
frequency ($\nu_\gamma$), $\gamma$-ray flux at the peak, and pulse
duration, to see if we can find some viable synchrotron solutions for
any GRBs with $\alpha=-(p-1)/2$.  These solutions are shown in
Figure~\ref{fig:syncp12}.  We find that by decreasing any of the
observable parameters $R_\gamma$ decreases, but the dependence is weak
in agreement with the scaling given in equation (\ref{p12r}).
Furthermore, a decrease in $t_\gamma$ reduces $\Gamma$ as expected from
equation (\ref{p12-1}), but even for $t_\gamma=10$ ms, $\Gamma$ is still
$\gta 10^3$.  

\begin{figure}[h!]
\begin{center}
\includegraphics[height=3.9in,width=5.99in]{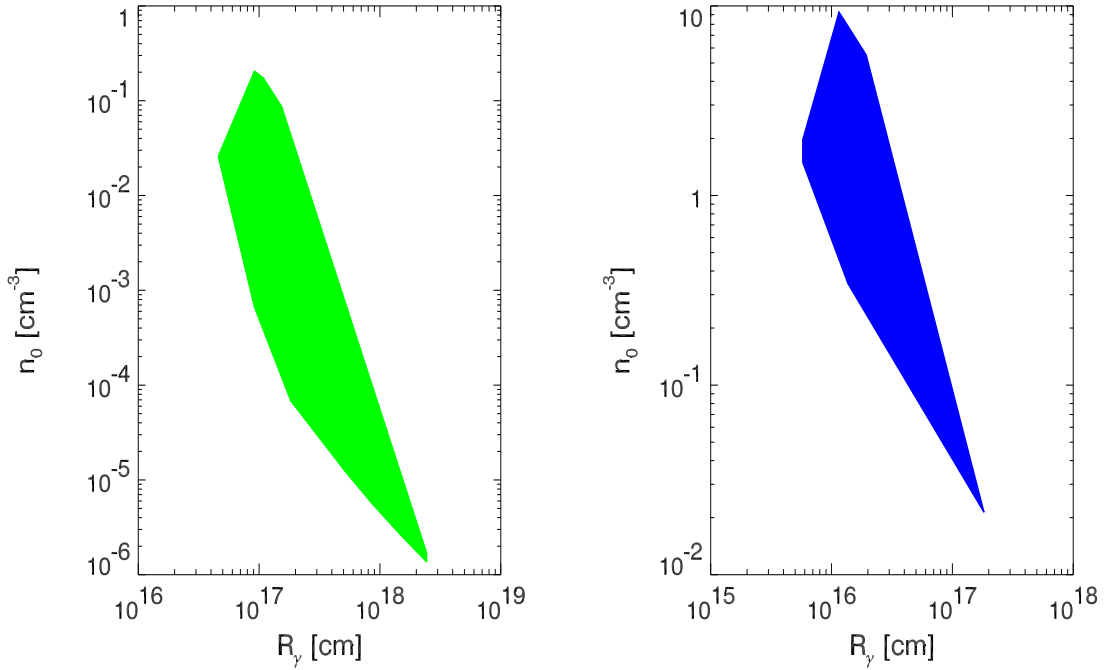}
\caption{\small
{\it Left panel:} the upper limit to the ISM density ($n_0$) for 
synchrotron solutions with $\alpha=-(p-1)/2$ for a burst with
$\nu_\gamma=100$ keV, $f_\gamma=1$ mJy, $t_\gamma=1$ s \& $t_a=t_\gamma$. 
{\it Right panel:} same as the left panel except that $t_a=t_\gamma/100$.
 Note that by decreasing the amount of time
electrons have to radiate away their energy before being re-accelerated
($t_a$) increases the $\max(n_0)$ roughly as $t_a^{-1}$. The $n_0$ upper 
limit decreases when any of the GRB parameters ($\nu_\gamma$, $f_\gamma$,
 $t_\gamma$) is increased;  $n_0\propa f_\gamma^{-1}$.}
\label{fig:n0maxp12}
\end{center}
\end{figure}

We next calculate the deceleration radius and compare it with $R_\gamma$
to ensure $R_\gamma<R_d$ for self consistent solutions.
The deceleration radius for GRB-ejecta is calculated using eq. (\ref{rd2})
and is given by 
\begin{equation}
R_d \approx \left\{ \begin{array}{ll}
\hskip -7pt 2.6 \times 10^{16} \, E_{53}^{1 \over 3} 
  n_0^{-{1 \over 3}} 
  \nu_{\gamma_5}^{3p-19 \over 48} f_\gamma^{-{1 \over8}}
  t_\gamma^{1 \over 4}  \nu_{i_5}^{3-p \over 16} 
t_a^{-{1\over 6}} Y^{1 \over 8} (1+Y)^{-{1 \over 6}} (1+z)^{-{1\over 6}}
d_{L_{28}}^{-{1\over4}} A_{1p}^{-{1 \over 8}} \,\,
   {\rm cm} & s=0 \\
   &  \\
\hskip -7pt 1.8\times 10^{13} \, E_{53} A_*^{-1}
\nu_{\gamma_5}^{3p-19\over 
  16} f_\gamma^{-{3\over8}} t_\gamma^{{3\over 4}}
\nu_{i_5}^{9-3p \over 16}  
t_a^{-{1 \over 2}} Y^{3 \over 8}
(1+Y)^{-{1 \over 2}} (1+z)^{-{1\over 2}} d_{L_{28}}^{-{3\over 4}} A_{1p}^{-{3\over 8}}\,\,{\rm cm} & s=2
\end{array} \right.
\end{equation} 
and the ratio of $R_\gamma$ and $R_d$ is:
\begin{equation}
{R_\gamma \over R_d} \approx  \left\{ \begin{array}{ll}
\hskip -7pt 2.3 E_{53}^{-{1\over3}} n_0^{1\over3}
\nu_{\gamma_5}^{19-3p\over12}
f_\gamma^{1\over2} \nu_{i_5}^{p-3 \over 4} 
t_a^{2\over3} Y^{-{1\over2}}
(1+Y)^{2\over3} (1+z)^{-{1 \over3}} d_{L_{28}} A_{1p}^{1\over2}\,\, & s=0 \\
  & \\
\hskip -7pt 3.3 \times 10^3 E_{53}^{-1} A_*
\nu_{\gamma_5}^{19-3p \over8}
f_\gamma^{3\over4} t_\gamma^{-{1\over2}} \nu_{i_5}^{3p-9
\over8} t_a Y^{-{3\over4}} (1+Y)
d_{L_{28}}^{3\over2}  A_{1p}^{3\over4} & s=2 
\end{array} \right.
\end{equation}
Substituting in the observable parameters for \grba~ into the above equation
and solving for $n_0$ \& $A_*$ such that $R_\gamma/R_d < 1$, we find
\begin{eqnarray}
 & n_0 < 0.057 E_{53} t_a^{-2} Y^{\frac{3}{2}} (1+Y)^{-2} 
    \,\, {\rm cm}^{-3} & s=0  \label{n01} \\ \nonumber\\[-.4cm] 
 & A_* < 5.5 \times 10^{-5} E_{53} t_a^{-1} Y^{\frac{3}{4}}
   (1+Y)^{-1} & s=2
\end{eqnarray}
Note that $n_0$ \& $A_*$ must be very small to ensure that
$R_\gamma<R_d$, especially for
$Y<1$ expected of synchrotron solutions. Figure (\ref{fig:n0maxp12}) shows 
the results of
numerical calculations which confirms these analytical estimates.
Moreover, if we want $R_\gamma/R_d \lta 0.5$, in order to have a clear
separation between internal and external shocks, then
$n_0\lta 10^{-2}$cm$^{-3}$. Therefore, self-consistent synchrotron solutions 
with $R_\gamma\lta R_d$ require very low density for the circumstellar 
medium compared with $n_0\sim 1$ cm$^{-3}$ obtained from afterglow
modeling \citep{pk01}.
 The limit on $n_0$ can be increased by decreasing $t_a$ (see eq. \ref{n01}).
Numerical result for the upper limit on $n_0$ when $t_a=t_\gamma/100$
is shown in fig. \ref{fig:n0maxp12}. It confirms the analytical result that 
$n_0\sim 1$ cm$^{-3}$ can give $R_\gamma< R_d$ provided that
$t_a\ll t_\gamma$. It should be noted that for systems involving 
shock heating of particles
we expect $t_a\sim t_\gamma$ because electrons are accelerated at the shock
front and there is no subsequent acceleration as particles travel downstream; in
magnetic reconnections or dissipation it is natural to expect $t_a\ll t_\gamma$.

We now estimate the ratio of energy in $e^\pm$ and magnetic field 
to find out if it is much less than unity or not when $t_a<t_\gamma$ 
(a small value for $E_\pm/E_B$ results in low efficiency for $\gamma$-ray 
generation). The ratio $E_\pm/E_B$ can be calculated
using eqs. (\ref{EB1}) \& (\ref{EP1}) and is given by
\begin{eqnarray}
{E_\pm\over E_B} \approx 1.5 (3-p) 
\left({\nu_{\gamma_5}\over\nu_{i_5}}\right)^{p-{5 \over 2}}
   \left[{t_a\over t_\gamma}\right] Y(1+Y),
\end{eqnarray}
for $2<p<3$ (numerical calculations take $p=2.5$). For the solution
space corresponding to $\alpha=-(p-1)/2$, $0.1\lta\nu_\gamma/\nu_i\lta
10^5$ and so $E_\pm>E_B$ even when $t_a/t_\gamma\sim10^{-2}$. Therefore,
small $t_a/t_\gamma$ solutions are fine from the point of radiative
efficiency; the above equation needs to be modified, when $t_a\ll
t_\gamma$, to include the total energy input in electrons during a GRB
pulse width of $t_\gamma$, which will further improve the radiative
efficiency when $t_a/t_\gamma$ is very small.

The reason that these synchrotron solutions have large $R_\gamma$ is not
hard to understand. It requires a certain minimum number of electrons to
produce the observed flux of $f_\gamma\sim 1$ mJy at $\nu_\gamma\sim100$
keV: $N\sim 10^{53}/(B\Gamma)$ -- see eq. \ref{s4}.  And in order to
keep the Compton-Y parameter, $Y\sim \tau \gamma_i \gamma_c$, less
$\sim10$ --- otherwise most of the energy will come out in IC-scattered
photons at $\nu\gg$ 1 MeV --- we must have large $R_\gamma$ for the
source. The solution offered by $t_a\ll t_\gamma$ is also easy to
understand. Frequent re-acceleration of charge particles makes it
possible to have larger magnetic field while keeping $\nu_c\gta 100$
keV.  This decreases the number of particles required to produce the
observed flux $f_\gamma$, and that in turn makes it possible to have a
smaller $R_\gamma$.

We conclude that the synchrotron process in a shock heated medium cannot 
account for the prompt $\gamma$-ray emission of long-duration GRBs with 
low energy spectrum $f_\nu\propto \nu^{-{p-1\over2}}$. However, synchrotron 
solutions appear to be viable when $t_a\ll t_\gamma$, i.e. when 
electrons are accelerated repeatedly, as might occur when magnetic field
is dissipated and the energy is deposited in e$^\pm$.

\subsection{Synchrotron solution when the low energy spectrum is
$\nu^\frac{1}{3}$}
\label{secsync13}

This is a special case of $\alpha=-(p-1)/2$
analyzed in the previous subsection (\S3.1) when $\gamma_i
\sim \gamma_c$; the solutions are a subset of those found
in \S3.1.  The analytical solutions for this case, obtained by substituting 
$\xi=\frac{p}{(p-2)}$ (see eq. \ref{xieq}), $\nu_{i_5} \sim \nu_{c_5}=
  \nu_{\gamma_5}$,
and $f_\gamma \sim f_{\nu_p}$, into equations \ref{s10}--\ref{s13} are
\begin{eqnarray}
&&\Gamma \approx 10^3 \nu_{\gamma_5}^{\frac{5}{16}} f_\gamma^{\frac{3}{16}}
t_\gamma^{-\frac{3}{8}} t_a^{\frac{1}{4}} Y^{-\frac{3}{16}}
(1+Y)^{\frac{1}{4}} (1+z)^{\frac{1}{4}} d_{L_{28}}^{\frac{3}{8}} 
\left[p\over p-2\right]^{\frac{3}{16}} \label{t1} \\ \nonumber\\[-.3cm] 
&&\gamma_i \approx 4.7 \times 10^4 \nu_{\gamma_5}^{\frac{9}{16}}
f_\gamma^{-\frac{1}{16}} t_\gamma^{\frac{1}{8}}
t_a^{\frac{1}{4}} Y^{\frac{1}{16}} (1+Y)^{\frac{1}{4}}
(1+z)^{\frac{1}{4}} d_{L_{28}}^{-\frac{1}{8}} 
\left[p\over p-2\right]^{-\frac{1}{16}}  \label{t2} \\ \nonumber\\[-.3cm] 
&&B \approx 4.0 \nu_{\gamma_5}^{-\frac{7}{16}} f_\gamma^{-\frac{1}{16}}
 t_\gamma^{\frac{1}{8}}  t_a^{-\frac{3}{4}} Y^{\frac{1}{16}}
(1+Y)^{-\frac{3}{4}}
(1+z)^{\frac{1}{4}} d_{L_{28}}^{-\frac{1}{8}}
\left[p\over p-2\right]^{-\frac{1}{16}} \,{\rm Gauss} \label{t3} \\ \nonumber\\[-.3cm] 
&&\tau \approx 3.3 \times 10^{-10} \nu_{\gamma_5}^{-\frac{9}{8}}
f_\gamma^{\frac{1}{8}} t_\gamma^{-\frac{1}{4}} t_a^{-\frac{1}{2}}
Y^{\frac{7}{8}} (1+Y)^{-\frac{1}{2}}
(1+z)^{-\frac{1}{2}} d_{L_{28}}^{\frac{1}{4}}
\left[p\over p-2\right]^{-\frac{7}{8}}  \label{t4}.
\end{eqnarray}
Substituting $f_\gamma=1$mJy, $\nu_{\gamma_5}=1$, $t_\gamma=0.1$s (the 
observed parameters for \grba), \& $t_a\sim t_\gamma$, in these equations, 
we find 
\begin{eqnarray}
&&\Gamma \sim 2.5 \times 10^3\, Y^{-\frac{3}{16}} (1+Y)^{\frac{1}{4}} \\  \nonumber\\[-.4cm] 
&&\gamma_i \sim 2 \times 10^4\, Y^{\frac{1}{16}} (1+Y)^{\frac{1}{4}} \\  \nonumber\\[-.4cm] 
&&B \sim 18\, Y^{\frac{1}{16}} (1+Y)^{-\frac{3}{4}} \,{\rm Gauss} \\  \nonumber\\[-.4cm] 
&& \tau \sim 6.3 \times 10^{-10}\, Y^{\frac{7}{8}} (1+Y)^{-\frac{1}{2}}   
\end{eqnarray}
and indeed, the solutions are a subset of the $\alpha=-(p-1)/2$ solution space
-- these have smaller $\tau$ and larger $\Gamma$ \& $\gamma_i$.  The distance
of the source from the center of the explosion is:
\begin{equation}
R_\gamma \approx 6.0\times10^{16}\, \nu_{\gamma_5}^{\frac{5}{8}} 
 f_\gamma^{\frac{3}{8}}  t_\gamma^{\frac{1}{4}} t_a^{\frac{1}{2}}
Y^{-\frac{3}{8}} (1+Y)^{\frac{1}{2}}
(1+z)^{-\frac{1}{2}} d_{L_{28}}^{\frac{3}{4}} 
\left[p\over p-2\right]^{\frac{3}{8}} \, {\rm cm}
\end{equation}
or $R_\gamma\sim2\times10^{16} \, Y^{-\frac{3}{8}} (1+Y)^{\frac{1}{2}}$ cm
for \grba.
This case has the same problems as $\alpha=-(p-1)/2$ case
discussed in \S3.1 i.e., large $R_\gamma$ and $\Gamma$, and requiring 
extremely small external density in order that $R_\gamma\lta R_d$.
Also, the conclusions drawn in \S3.1 regarding $t_a/t_\gamma\ll 1$ 
offering a way out of this problem apply here as well.

\begin{figure}[h!]
\begin{center}
\includegraphics[height=4.3in,width=5.7in]{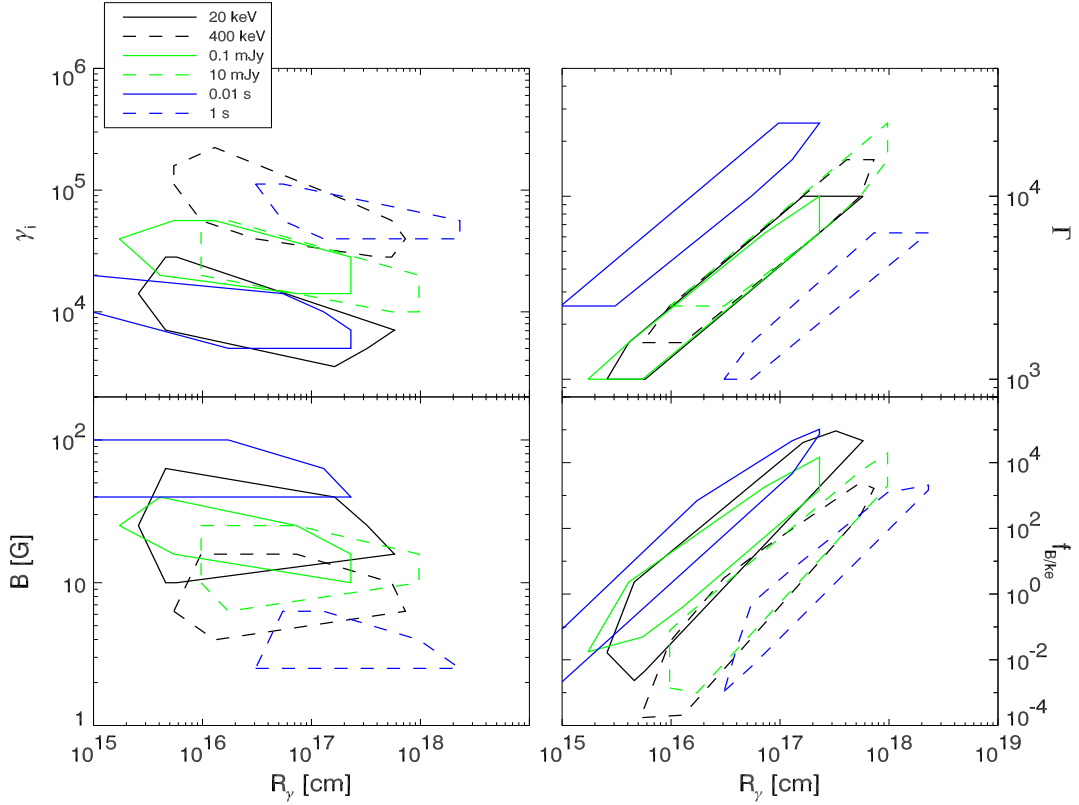}
\caption{ \small
 Synchrotron solution space when the spectrum below $\nu_\gamma$,
the peak of $\nu f_\nu$, is $f_\nu\propto\nu^{1/3}$, i.e. $\alpha=1/3$.
See Figure~\ref{fig:syncp12} caption for details. }
\label{fig:sync13}
\end{center}
\end{figure}

\begin{figure}[h!]
\begin{center}
\includegraphics[height=3.9in,width=5.99in]{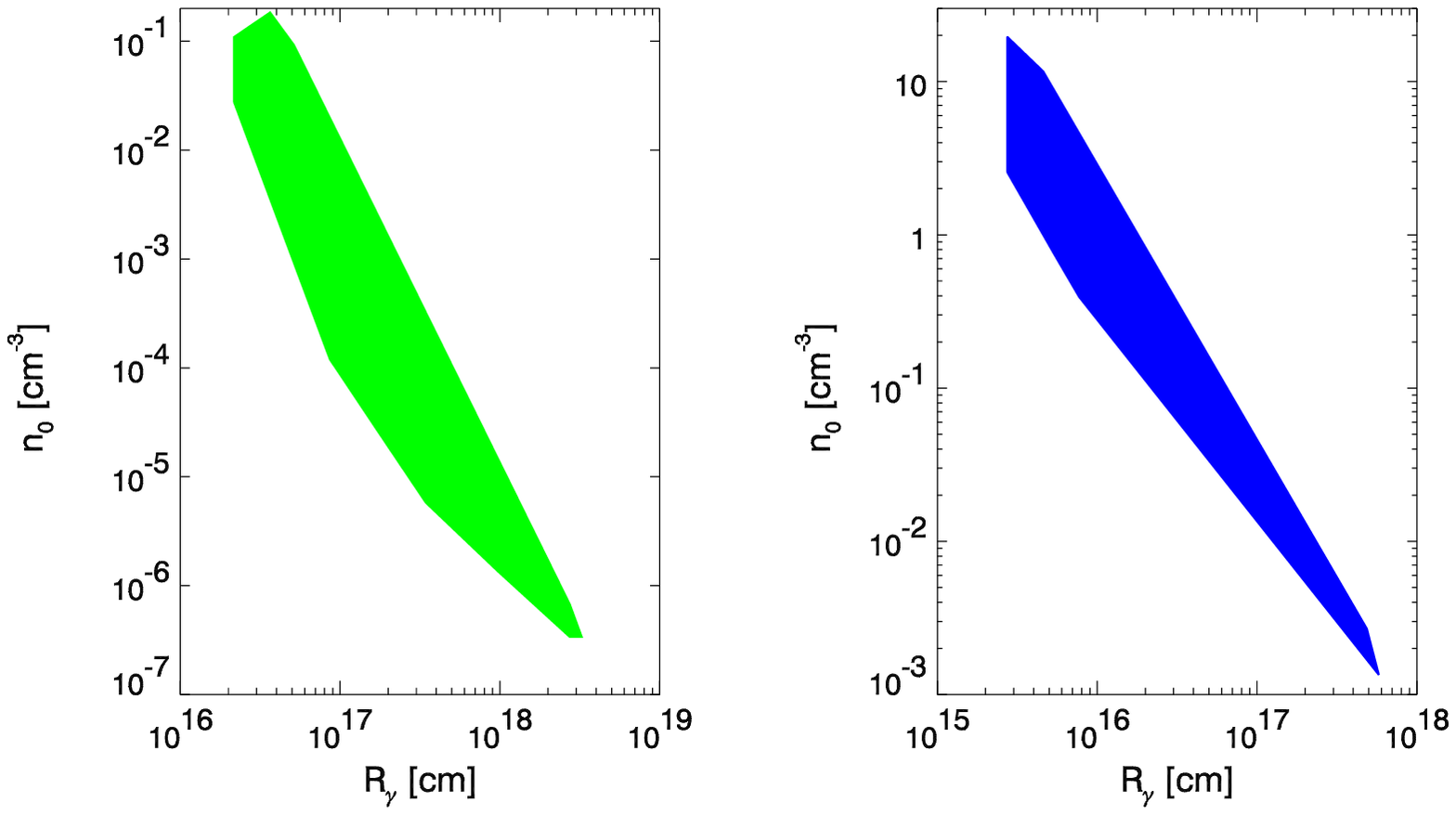}
\caption{\small
{\it Left panel:} the maximum density of the circum-bust-medium
so that $R_\gamma<R_d$ for synchrotron solutions with $\alpha={1/3}$
and for a burst with $\nu_\gamma=100$ keV, $f_\gamma=1$ mJy,
$t_\gamma=1$ s \& $t_a=t_\gamma$. {\it Right panel:} is same as 
the left panel except that $t_a=t_\gamma/100$; note that
by decreasing the amount of time electrons have to radiate away their
energy before being re-accelerated ($t_a$) increases the upper limit for
$n_0$ roughly as $t_a^{-1}$.  The $n_0$ upper limit is weakly dependent on
$\nu_\gamma$, and decreases with increasing $t_\gamma$ \&
$f_\gamma$ being most sensitive to $f_\gamma$ -- $\max(n_0)\propa
f_\gamma$.
}
\label{fig:n0max13}
\end{center}
\end{figure}

The numerical calculation of the hypersurface in 5-D parameter space allowed
by GRB observations -- $\nu_\gamma$, $f_\gamma$ \& $t_\gamma$ -- for \grba~ 
finds $\gamma_i\gta10^4$, $\Gamma \gta 10^3$, $200\lta\gamma_i/\Gamma\lta700$,
source radius ($R_\gamma$) 
$10^{16}-10^{18}$ cm, and $B$ between 1 and 10$^2$Gauss for the entire
solution space (see fig. \ref{fig:sync13}) -- which is in very good
agreement with analytical estimates. For a wide range of values for the
three observable parameters we find the GRB source to be located between
$\sim10^{15}$ cm \& $10^{18}$ cm, $\gamma_i \gta 3 \times 10^3$, 
and $\Gamma \gta 10^3$ (fig. \ref{fig:sync13}). In order that 
$R_\gamma/R_d <1$, the density of the surrounding medium ($n_0$) is 
required to be less than $\sim 0.1$ cm$^{-3}$ which is much smaller 
than the value inferred from late time afterglow modeling for long
duration GRBs. The density requirement is relaxed if $t_a\ll t_\gamma$
(see fig. \ref{fig:n0max13}).

In conclusion, the synchrotron process, in a shock heated medium, has 
serious problems accounting for prompt $\gamma$-ray emission for those 
bursts that have spectrum below the peak frequency ($\nu_\gamma$) scaling 
as $f_\nu\propto\nu^{1/3}$ or $\nu^{-(p-1)/2}$. 
A possible resolution is provided if electrons are
more or less continuously accelerated while they are radiating 
$\gamma$-ray photons during the entire time period of a spike in 
the observed GRB lightcurve; in other words $t_a\ll t_\gamma$.
It should be pointed that $t_a\sim t_\gamma$ in shocks whereas 
continuous acceleration might be possible in regions of magnetic
reconnection/dissipation.

% add explanation of spaces for plot?

%%%%%%sync -1/2 solutions *******************************
\subsection{Synchrotron solution when the low energy spectrum is $\nu^{-1/2}$}

Substituting $f_{\nu_p} = f_\gamma\left(\nu_{i_5}/\nu_{c_5}\right)^
{\frac{1}{2}}$, $\nu_{\gamma_5}=\nu_{i_5}$, and $\xi$ from equation 
(\ref{xieq}) for the case where
$\nu_c<\nu_i$, into eqs.  \ref{s10}--\ref{s13} we find the allowed part of 
the 5-D parameter space when the spectrum below $\nu_\gamma$ is 
$f_\nu\propto\nu^{-1/2}$
\begin{eqnarray}
&&\Gamma \approx 10^3 \nu_{\gamma_5}^{\frac{3}{16}}
f_\gamma^{\frac{3}{16}} t_\gamma^{-\frac{3}{8}} \nu_{c_5}^{\frac{1}{8}}
t_a^{\frac{1}{4}} Y^{-\frac{3}{16}} (1+Y)^{\frac{1}{4}}
(1+z)^{\frac{1}{4}} d_{L_{28}}^{\frac{3}{8}}
A_{2p}^{\frac{3}{16}} \label{s12-1}\\  \nonumber\\[-.4cm] 
&&\gamma_i \approx 4.7 \times 10^4 \nu_{\gamma_5}^{\frac{7}{16}}
f_\gamma^{-\frac{1}{16}} t_\gamma^{\frac{1}{8}} \nu_{c_5}^{\frac{1}{8}}
t_a^{\frac{1}{4}} Y^{\frac{1}{16}} (1+Y)^{\frac{1}{4}}
(1+z)^{\frac{1}{4}} d_{L_{28}}^{-\frac{1}{8}} 
A_{2p}^{-\frac{1}{16}}  \label{s12-2} \\  \nonumber\\[-.4cm] 
&&B \approx 4.0 \nu_{\gamma_5}^{-\frac{1}{16}} f_\gamma^{-\frac{1}{16}}
t_\gamma^{\frac{1}{8}} \nu_{c_5}^{-\frac{3}{8}} 
 t_a^{-\frac{3}{4}} Y^{\frac{1}{16}} (1+Y)^{-\frac{3}{4}}
(1+z)^{\frac{1}{4}} d_{L_{28}}^{-\frac{1}{8}} 
A_{2p}^{-\frac{1}{16}} \quad {\rm Gauss} \label{s12-3} \\  \nonumber\\[-.4cm] 
&&\tau \approx 3.3 \times 10^{-10} \nu_{\gamma_5}^{-\frac{3}{8}}
f_\gamma^{\frac{1}{8}} t_\gamma^{-\frac{1}{4}} \nu_{c_5}^{-\frac{3}{4}}
t_a^{-\frac{1}{2}} Y^{\frac{7}{8}} (1+Y)^{-\frac{1}{2}}
 (1+z)^{-\frac{1}{2}} d_{L_{28}}^{\frac{1}{4}}
A_{2p}^{-\frac{7}{8}},  \label{s12-4}
\end{eqnarray}
where 
\begin{equation}
A_{2p} \equiv {p-1 \over p-2}.
\end{equation}
For \grba ~ --- $t_\gamma=0.1$s, $\nu_\gamma=100$keV, $f_\gamma=1$mJy \& 
$t_a=t_\gamma$ --- and taking $\nu_{c_5} < 10^{-4}$ (in agreement with
the numerical calculation) we find 
$\Gamma \lta 720 Y^{-\frac{3}{16}}$,  $\gamma_i \lta 6.7 \times 10^4$, \& 
$B \gta 590 (1+Y)^{-\frac{3}{4}}$ Gauss.
In contrast to the previous two cases considered in \S3.1 \& \S3.2, 
$\Gamma < 1000$ in this case.
We find the $\gamma$-ray source distance, $R_\gamma$, to be
\begin{equation}
R_\gamma \approx 6.0 \times 10^{16} \, \nu_{\gamma_5}^{\frac{3}{8}}
f_\gamma^{\frac{3}{8}} t_\gamma^{\frac{1}{4}} \nu_{c_5}^{\frac{1}{4}} 
t_a^{\frac{1}{2}} Y^{-\frac{3}{8}}
(1+Y)^{\frac{1}{2}} (1+z)^{-\frac{1}{2}} d_{L_{28}}^{\frac{3}{4}} 
A_{2p}^{\frac{3}{8}}\, {\rm cm}
\end{equation}
which is $R_\gamma \sim 2 \times 10^{15} Y^{-\frac{3}{8}}
(1+Y)^{\frac{1}{2}}$ cm for \grba.  
We compare this radius to the deceleration radius, $R_d$, which is obtained
from eq. (\ref{rd2}) and is given by
\begin{equation}
R_d \approx \left\{ \begin{array}{ll}
\hskip -7pt 2.6 \times 10^{16}\, E_{53}^{1\over3} n_0^{-{1\over3}}
\nu_{\gamma_5}^{-{1\over8}}  f_\gamma^{-{1\over8}}
 t_\gamma^{1\over4}  \nu_{c_5}^{-{1\over12}}
t_a^{-{1\over6}} Y^{1\over8} (1+Y)^{-{1\over6}} (1+z)^{-{1\over6}}
d_{L_{28}}^{-{1\over4}} A_{2p}^{-{1\over8}}\,\, {\rm cm} & s=0 \\
   & \\
\hskip -7pt 1.8\times 10^{13} \, E_{53} A_*^{-1} \nu_{\gamma_5}^{-{3\over8}}
f_\gamma^{-{3\over8}} 
t_\gamma^{{3\over4}} \nu_{c_5}^{-{1\over4}} t_a^{-{1\over2}} Y^{3\over8} 
  (1+Y)^{-{1\over2}} (1+z)^{1\over2} d_{L_{28}}^{-{3\over4}} A_{2p}^{-{3\over8}} \,\, {\rm cm} & s=2
\end{array} \right.
\end{equation}
and for the ratio $R_\gamma/R_d < 1$, we find that $n_0 \lta 8\,E_{53} 
Y^{\frac{3}{2}} (1+Y)^{-2}$ cm$^{-3}$ and $A_* < 0.014 E_{53} 
Y^{\frac{3}{4}} (1+Y)^{-1}$ if $\nu_{c_5}\sim0.1$; the limits on $n_0$ \&
$A_*$ are much higher for $\nu_{c_5}\ll0.1$ and poses no problem for 
synchrotron solutions in a shock heated source.

If the synchrotron solutions were to arise in a shock heated medium,
we can calculate the LF of the shock front wrt the unshocked fluid, 
$\Gamma_{sh}$, using equation \ref{gamsh}:
\begin{equation}
\Gamma_{sh} \approx 26\, \epsilon_e^{-1} \nu_{\gamma_5}^{\frac{7}{16}} 
f_\gamma^{-\frac{1}{16}} t_\gamma^{\frac{1}{8}} \nu_{c_5}^{\frac{1}{8}}
t_a^{\frac{1}{4}} Y^{\frac{1}{16}}
(1+Y)^{\frac{1}{4}}  (1+z)^{\frac{1}{4}} d_{L_{28}}^{-\frac{1}{8}} 
 A_{2p}^{\frac{15}{16}}.
\end{equation}
or $\Gamma_{sh} \lta 32 Y^{\frac{1}{16}} (1+Y)^{\frac{1}{4}}$ for \grba 
$\,$ assuming that electrons receive half of the shock energy and that there 
are no $e^\pm$ pairs. Note that as long as $\nu_{\gamma_5} \sim 1$, 
$\Gamma_{sh}$ is pretty high ($\sim 20$), and it is insensitive to 
$\nu_{c_5}$ (and the other quantities).  In
order to produce $\Gamma_{sh} \sim 20$ in internal shocks, we need 
the relative LF of the two colliding shells $\Gamma_{rel} \sim 2 
\Gamma_{sh}^2 (n_1/n_2) ^{\frac{1}{2}}$ \citep{pk04,sp95}, where $n_1$ and $n_2$ are the 
comoving densities of the two colliding shells (see appendix A for a discussion
of how we calculate $\Gamma_{rel}$).  For $\Gr\lta5$, so that the ratio
of the LFs of the colliding shells is not larger than 10, $n_1/n_2\lta
\sim10^{-5}$ is required (see appendix A); $\Gamma_{rel}>5$ is an unlikely 
situation to be realized in nature.

\begin{figure}[h!]
\begin{center}
\includegraphics[height=4.1in,width=5.4in]{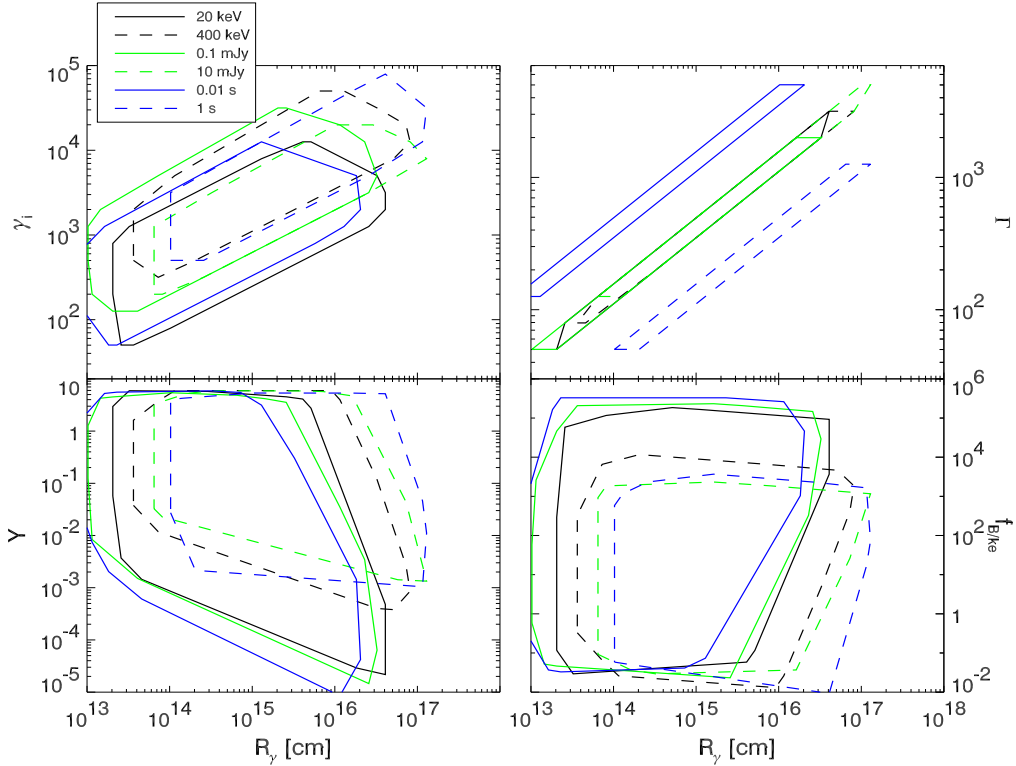}
\caption{\small
Synchrotron solution space when the spectrum below the peak of
$\nu f_\nu$ is $f_\nu\propto \nu^{-1/2}$ for $\nu < \nu_\gamma$.
$\gamma_i$, $\Gamma$, $Y$, and $f_{{B}/{ke}}$ are plotted against 
the distance of the source from the center of explosion ($R_\gamma$). 
The solution spaces for several different values of $\nu_\gamma$, $f_\gamma$,
and GRB pulse duration ($t_\gamma$) are shown in the four panels. Please
see the Figure~\ref{fig:syncp12} caption for details.  }
\label{fig:synchalf}
\end{center}
\end{figure}

We calculate the total energy in electrons ($E_\pm$) and magnetic field
($E_B$) to determine the efficiency for synchrotron radiation ---
if there is a lot more energy in magnetic field than that in the
electrons, the efficiency for $\gamma$-ray radiation would be small.  
 The magnetic and electron energies for the case of $\nu^{-1/2}$ spectrum
are obtained from equation (\ref{energies}) and the solutions for $B$, 
$\Gamma$, $\gamma_i$ and $R_\gamma$ derived above, and are given by --
\begin{equation}
E_B = B^2 R_\gamma^3/6 \approx 6.0 \times 10^{50} \, \nu_{\gamma_5} f_\gamma 
t_\gamma Y^{-1} (1+z)^{-1}  d_{L_{28}}^2 A_{2p}\,
  {\rm ergs},
\end{equation}
and
\begin{equation}
E_\pm = N(p-1)m_e c^2 \gamma_i\Gamma/(p-2)\approx 8.7 \times 10^{50} \,
  \nu_{\gamma_5} f_\gamma t_a (1+Y) (1+z)^{-1} d_{L_{28}}^2 A_{2p}\, {\rm ergs}.
\end{equation}
The ratio $E_B/E_\pm$ is
\begin{equation}
f_{B/ke} \equiv {E_B \over E_\pm} \approx  0.68 t_\gamma t_a^{-1} Y^{-1}
(1+Y)^{-1}. 
\label{s12-fBke}
\end{equation}
For $t_a\ll t_\gamma$
the above expression for the ratio $f_{B/ke}$ would need to be modified
to include the total energy deposited in $e^\pm$s as a result of 
multiple acceleration episodes during a GRB pulse time period of $t_\gamma$;
for $t_\gamma/t_a\sim 1$ the expression for $f_{B/ke}$ reduces to the 
familiar form that depends only on the Compton-$Y$.
For $Y \ll 1$ most of the energy is in the magnetic field and for 
these solutions the radiative efficiency to produce a GRB is very small.

\begin{figure}[h!]
\begin{center}
\includegraphics[height=3.9in,width=5.4in]{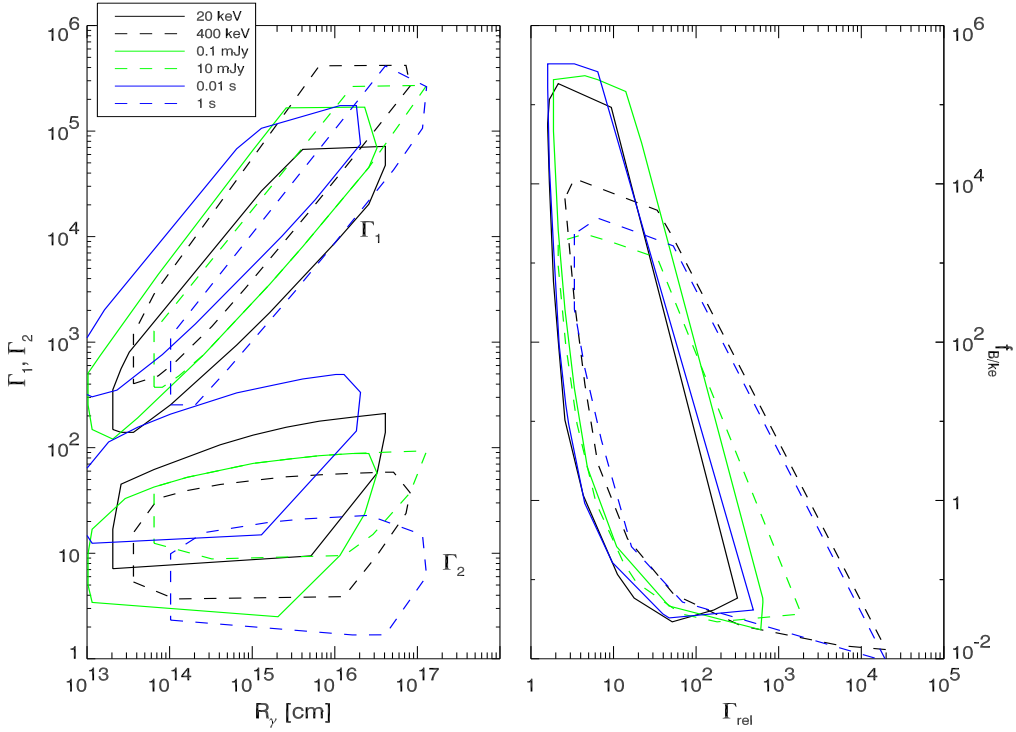}
\caption{\small
This figure, unlike all the previous ones, shows {\it model dependent}
results. For each point in the allowed region of the 5-D parameter space,
corresponding to synchrotron solutions for $\alpha=-1/2$ and for given
$\nu_\gamma$, $f_\gamma$ \& $t_\gamma$,
we calculate parameters for the popular internal shock model for GRBs
(see appendix A for details). The {\it left panel} shows the LFs $\Gamma_1$
\& $\Gamma_2$ of the two colliding shells in the internal shock model
as a function of $R_\gamma$, the distance from the center where the
shells collide.  The {\it right panel} shows
$\Gamma_{rel}=\Gamma_1\Gamma_2\left(1-v_1v_2\right)$ vs. $f_{{B}/{ke}}$
(the ratio of energy in magnetic field and electrons). Different
$\nu_\gamma$, $f_\gamma$ and $t_\gamma$ cases are displayed as described
in Figure~\ref{fig:syncp12}. Note that $\Gr$ is correlated with $f_{B/ke}$;
for high $\Gr$, $f_{B/ke}$ is small ($< 1$), and for low $\Gr$,
$f_{{B}/{ke}}\gg1$.  Low $\Gr$ \& $f_{{B}/{ke}}\gg1$ solutions might
correspond to a highly magnetized outflow.  }
\label{fig:syncshells}
\end{center}
\end{figure}

We numerically search the 5-D parameter space subject to the three 
observational constraints ($\nu_\gamma$, $f_\gamma$, $t_\gamma$) and 
find solutions with $10^{13}$ cm $<R_\gamma< 10^{17}$ cm, $10$ 
Gauss $<B<10^7$Gauss, $100 <\gamma_i<3 \times 10^4$, 
$80 <\Gamma<3000$, and $10^{-3}<Y<7$ (see fig. \ref{fig:synchalf})
--- all in good agreement with analytical estimates presented above. 
 We find $2 < \Gamma_{sh} < 100$ and $10^{-2} <
f_{B/ke} < 10^4$. So it would seem that we have solutions with 
$f_{B/ke} \sim1$ and $\Gamma_{sh}$ of order a few -- however
it turns out that for $f_{B/ke}\lta10$, $\Gamma_{sh} \gta 5$.  
 For $\Gamma_{sh} \gta 5$, $10^{-5}<\frac{n_1}{n_2}<0.1$, and
the ratio of the LFs of two colliding shells, $\Gamma_{rel}$, to produce 
this $\Gamma_{sh}$ is greater than 20 (see fig. \ref{fig:syncshells}) --- 
 fluctuations in the LF of the outflow with $\Gamma_{rel}$ on the order 
of a few are typically expected in internal shocks.

Numerical solutions for the allowed part of the 5-D space for a range of 
observable parameters are shown in Figure~\ref{fig:synchalf}. An increase in
$\nu_\gamma$ leads to a slight increase of $\gamma_i$ and $R_\gamma$ whereas
$\Gamma$ is quite insensitive to it. These behaviors are consistent with our
analytical calculations (eqs. \ref{s12-1}--\ref{s12-4}). The decrease of
$f_{{B}/{ke}}$ with $\nu_\gamma$ (fig. \ref{fig:synchalf}) is due to an 
increase of $Y$. An increase of $f_\gamma$ has little effect on $\gamma_i$
(for the allowed solution space), $\Gamma$ \&
$Y$ increase a little, and $f_{B/ke}$ decreases;  these parameters have 
a very weak dependence on $f_\gamma$ (see eqs. \ref{s12-1}--\ref{s12-4}).
  And finally, when $t_\gamma$ is increased,
$\gamma_i$ and $Y$ increase, and $\Gamma$ and $f_{{B}/{ke}}$ decrease. This
is again in agreement with the analytical estimates -- $\gamma_i \propto
t_\gamma^{\frac{1}{2}}$. The large decrease in $\Gamma$ with $t_\gamma$
is due to an increase of $Y$ with $t_\gamma$ -- $\Gamma \propto 
t_\gamma^{-\frac{1}{8}} Y^{-\frac{3}{16}}$.

We have looked at the variation of $\Gamma_{sh}$, $\Gamma_{rel}$ and
$f_{{B}/{ke}}$ with $\nu_\gamma$, $f_\gamma$ \& $t_\gamma$. The results
are shown in Figure~\ref{fig:syncshells}.  The minimum value of
$\Gamma_{rel}$ has a weak dependence on $\nu_\gamma$, $f_\gamma$, and
$t_\gamma$; $\Gamma_{rel}$ is between 5 and 20 for $f_{{B}/{ke}} \lta
10$; $\Gamma_{rel} \sim 5$ solutions are only present when $t_\gamma
\lta 0.01$s, $f_\gamma \lta 0.1$ mJy, or $\nu_\gamma < 20$ keV (note
that we only alter one of the three at a time, i.e. for $t_\gamma \sim
0.01$s, $\nu_\gamma \sim 100$ keV and $f_\gamma \sim 1$ mJy). $\Gamma_{rel}$
may be small enough, then, that the synchrotron mechanism can produce
GRBs with $\nu^{-\frac{1}{2}}$ spectra if it has very short pulse
duration, small peak frequency, or small flux.

\subsubsection{X-ray flux during the GRB when $\alpha=-\frac{1}{2}$} 

So far, we have only considered prompt $\gamma$-ray emission due to the
synchrotron process. We now calculate emission in other wavelengths,
particularly the X-ray and optical, that should accompany $\gamma$-ray
photons. In this subsection, and in \S3.3.2, we relate the solutions we
found in \S3.3 to the internal shock model for GRBs (Rees \& Meszaros,
1994; see Piran, 1999, for complete references) according to which
shells of material ejected in the explosion undergo collisions and the
resulting shocks convert part of the kinetic energy of the outflow to
radiation.  Throughout this section we assume that the $\gamma$-ray
emission is produced in shell `1' which is taken to be the faster of the
two shells.  The results are essentially identical if we assume that the
GRB is produced in the outer, slower, shell, which we shall refer to as
shell `2'.   The x-ray flux from shell `1', however, is independent of
the internal shock model, and is expected to accompany the prompt
synchrotron $\nu^{-\frac{1}{2}}$ $\gamma$-ray emission.

The x-ray and optical flux from shell `1' lie on the $\nu^{-1/2}$
extrapolation of the $\gamma$-ray flux and therefore it is
straightforward to calculate these using $f_\gamma$ and the information
that $\nu_c\lta 2$eV \& $\nu_a\lta 5$eV for the entire solution
sub-space of the 5-D parameter space. The calculation of emission from
shell `2' is more involved and also a bit uncertain. We provide here
(and in \S3.3.2) a lower limit to the x-ray and optical flux from shell
`2' for each point in the 5-D space that satisfies the three
observational constraints ($\nu_\gamma$, $f_\gamma$, $t_\gamma$). The
calculation of flux from shell `2' requires the knowledge of the LF of
the shock front moving into this shell as well as the ratio of densities
(${n_1}/{n_2}$). The calculation for these quantities is described in
appendix A. The synchrotron injection frequency in shell `2',
$\nu_{i_2}$, is smaller than that in shell 1 by a factor of
$\left(\Gamma_{s_2}-1\right)^2/\left(\Gamma_{s_1}-1\right)^2$;
$\Gamma_{s_1}$ \& $\Gamma_{s_2}$ are shock front LFs into shell `1' \&
`2' wrt unshocked gas in shells `1' \& `2' respectively. This factor is
approximately $\simeq \frac{n_1}{n_2}$ for $\Gamma_{s_1},\Gamma_{s_2}
\gg 1$, but the approximation breaks down for $ \frac{n_1}{n_2} \lta
10^{-2}$ (see Figure~\ref{fig:n12}) since the shock in shell `2' becomes
mildly relativistic (we note that this approximation is not used in our
numerical calculations). The peak flux of the synchrotron spectrum at
$\nu=\min(\nu_{i_2},\nu_{c_2})$ in shell `2', $f_{\nu_{p_2}}$, is larger
than $f_{\nu_{p_1}}$ by a factor of $\left( \frac{n_2}{n_1}\right)^{
\frac{1}{2}}$.  The magnetic field is assumed to be the same in the two
shells, and therefore the difference between the cooling frequencies in
the shells is due to different $Y$-parameters; since synchrotron
dominates over SSC here by design, the difference between $\nu_{c_2}$
and $\nu_{c_1}$ ends up being very small. The shell `2' synchrotron self
absorption frequency, $\nu_{a_2}$ is larger than that in shell 1 by a
factor of $\left( \frac{n_1}{n_2}\right)^{0.2}$, or a factor of a few.

\begin{figure}[h!]
\begin{center}
\includegraphics[height=4.1in,width=5.4in]{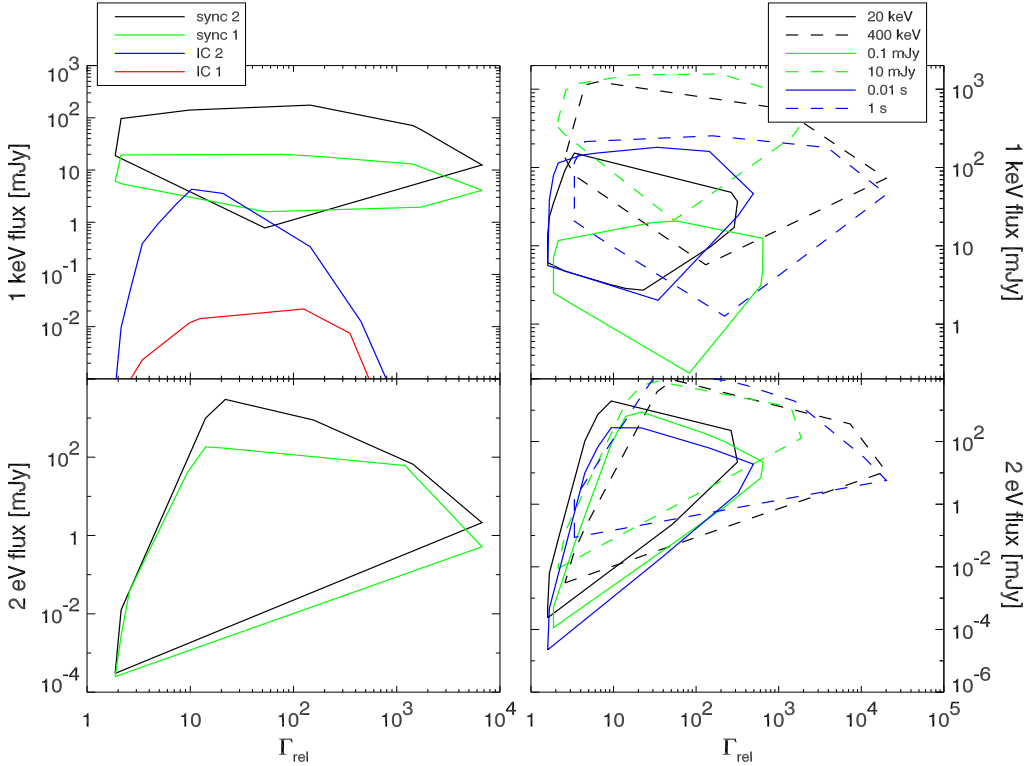}
\caption{\small
Prompt x-ray [1 keV] and optical [2 eV] flux associated with points in the 
solution sub-space of 5-D parameter space when cast in terms of the internal
shock model or two colliding shells; the solutions are for synchrotron 
radiation with low energy spectrum $f_\nu\propto\nu^{-\frac{1}{2}}$.
Top left: contributions of SSC and synchrotron to x-ray flux at 1 keV
from shells `1' and `2' (see legend) for \grba.  Top right: sum of all 
contributions is shown for the cases that have been described in 
Figure~\ref{fig:syncp12}.  Bottom left: contributions
of synchrotron emission from shells `1' and `2' to the optical flux; IC
contributions are negligible from either shell.  Bottom right:  sum
of all contributions in the optical, for the same cases as the top right
panel.  The results shown here assume that the $\gamma$-ray flux is
produced in shell `1'; production of the GRB in shell `2' does not change
the results significantly.
}
\label{fig:synchalfxo}
\end{center}
\end{figure}

The 1 keV synchrotron and SSC flux from shells `1' and `2' for \grba~ is shown 
in the  top left panel of Figure~\ref{fig:synchalfxo}.
 The shell `2' synchrotron flux contributes the
most to the prompt x-ray flux, and the shell 1 synchrotron flux
contributes a slightly smaller amount.  SSC flux from either
shell is negligible. There is a weak dependence of x-ray flux on $\Gr$.  
The shell `1' synchrotron flux at 1 keV is about 10 mJy.  This is simply the
extrapolation of the 100 keV flux back to 1 keV, with a spectral index
of $-\frac{1}{2}$ or $\left(\frac{1}{100}\right)^{-\frac{1}{2}} 1
 \mathrm{mJy} \sim 10 $mJy for \grba.  

Shell `2' synchrotron x-ray flux ranges from 1 to 100 mJy.
We expect the flux from shell `2' at 1 keV to be
\begin{equation}
f_{2_x} \sim f_{\gamma} {f_{\nu_{p_2}} \over f_{\nu_{p_1}}} \left({1
  \mathrm{keV} \over \nu_{i_2}}\right)^{-\frac{p}{2}} \left({\nu_{i_2} \over
  100 \mathrm{keV}}\right)^{-\frac{1}{2}} \sim  f_{\gamma} \left({n_1
  \over n_2 }\right)^{\frac{1}{4}} \left(\frac{\mathrm{1
keV}}{\nu_\gamma}\right)^{-\frac{5}{4}}\, {\rm mJy}
\label{eq:xcheck}
\end{equation}
using the above arguments and assuming that $\nu_{i_2} < 1$ keV (valid
when $\frac{n_1}{n_2} \lta 0.01$ -- satisfied by roughly half of the 
solution points in the 5-D space), $\nu_{c_1}\sim\nu_{c_2}$, and $p=2.5$.  
From this equation, we estimate that the flux at 1 keV should be between
$\sim20$ and 300 mJy (for $f_{1_{\gamma}}=1$ mJy and $\nu_\gamma \sim
100$ keV) -- in agreement with
the numerical results in Figure~\ref{fig:synchalfxo}.

The shell `2' SSC flux at 1 keV ranges from $10^{-5}$ to almost 10 mJy.  
The SSC peak frequency ($\sim \nu_{i_2} \gamma_{i_2}^2$) ranges from 
about 100 eV to very high values, so over a large part of the solution 
space, the expected SSC flux at 1 keV in comparison to the synchrotron 
100 keV flux from shell `1' is (assuming that $\nu_{i_2} > \nu_{c_2}$) 
\begin{equation}
{f^{ic}_{2_x} \over f_{\gamma}} \approx {f_{\nu_{p_2}} \tau_{2}
\over f_{\nu_{p_1}}}
\left({1 \mathrm{keV} \over \gamma_{c_2}^2 \nu_{c_2} }\right)^{-\frac{1}{2}}
\left({\nu_{c_1} \over 100 \mathrm{keV}}\right)^{-\frac{1}{2}} \sim
10 \left({n_2 \over n_1}\right) \tau_{1} \gamma_{c_2}
\end{equation}
which, after substituting in the solutions for $\tau$ and $\gamma_c$, is
\begin{equation}
f^{ic}_{2_x} \approx 1.5\times10^{-4}\, \left({n_2 \over n_1}\right)
\nu_{\gamma_5}^{-\frac{7}{16}} f_\gamma^{\frac{17}{16}} 
t_\gamma^{-\frac{1}{8}} \nu_{c_5}^{-\frac{1}{8}}
t_a^{-\frac{1}{4}} Y^{\frac{15}{16}} (1+Y)^{-\frac{1}{4}}
(1+z)^{-\frac{1}{4}} d_{L_{28}}^{\frac{1}{8}}
A_{2p}^{-\frac{15}{16}}\, {\rm mJy};
\end{equation}
with $Y \sim 1$ and $1 < {n_2}/{n_1} < 10^5$, the range for the x-ray flux 
obtained from the above equation is in agreement with the numerical solutions. 

The sum of synchrotron and SSC contributions to flux at 1 keV from both
shells for various values of $\nu_\gamma$, $f_\gamma$, and $t_\gamma$ are
shown in the upper right hand panel of Figure~\ref{fig:synchalfxo}.  The
1 keV flux ranges from 0.1 to a few thousand mJy, and is most sensitive
to $f_\gamma$ and $\nu_\gamma$ -- in agreement with equation
(\ref{eq:xcheck}), $f_{2_x} \propto f_\gamma \nu_\gamma^{\frac{5}{4}}$,
since synchrotron emission dominates. 

The early 0.2-10 keV x-ray flux as observed by the Swift x-ray telescope
ranges from $10^{-12}$ to $10^{-8}$ ergs cm$^{-2}$
s$^{-1},$\footnote{url{http://swift.gsfc.nasa.gov/cgi-bin/swift/grb\_table/grb\_table.py}} which corresponds to 1 keV flux of about
$10^{-4}$ to a few mJy (assuming $\nu^{-\frac{1}{2}}$ in the x-ray band).  
These observations are made at roughly 100s after the GRB trigger.  The x-ray
light curve from the $\gamma$-ray source should peak at about the same time 
as the GRB light curve.
After the peak, assuming that the outflow opening angle is greater than
$1/\Gamma$, the emission should be dominated by off-axis emission and
the light curve should fall off as $t^{-2+\beta}$ \citep{kp00}, which 
in this case is $t^{-2.5}$ since $f_\nu\propto\nu^{-1/2}$. Extrapolating the
observed 1 keV flux of $\sim 10^{-4}$--1 mJy backwards in time from 100s
to 10s, we find the x-ray flux during the GRB to be consistent with
values shown in the upper right panel of Figure~\ref{fig:synchalfxo}.  

\subsubsection{Prompt optical emission when the GRB index $\alpha=-\frac{1}{2}$}

In the bottom left panel of Figure~\ref{fig:synchalfxo}, the R band (2
eV) flux from shell `1' and `2' are plotted against $\Gamma_{rel}$.
Optical flux from the $\gamma$-ray source can be pretty bright during
the burst for these solutions ranging from $10^{-3}$ and 100 mJy (24th
to 11th magnitude in the R band). The synchrotron flux is smaller for
smaller $\Gr$ solutions. The SSC makes negligible contribution to the
optical flux compared to the synchrotron process, because $\nu_{a}
\gamma_c^2$ is well above the optical. 

If we extrapolate the shell `1' 100 keV flux back to 2 eV using the
spectral index $\alpha=-\frac{1}{2}$, we expect $f_{1R} \sim 225
\nu_{\gamma_5}^{1/2} f_{\gamma} \sim 225$mJy for \grba  whereas for
most of the solution space the optical flux for shell `1' falls below 10
mJy (fig. \ref{fig:synchalfxo} bottom left panel) --- this is because
the synchrotron self absorption frequency is larger than the R-band
frequency by a factor $\sim10$ or more.  The range of shell `2' R-band
flux is higher than shell `1' by a factor of
$\frac{f_{\nu_{p_2}}}{f_{\nu_{p_1}}} \sim \sqrt{\frac{n_2}{n_1}}\sim
30$.

In the bottom right panel of Figure~\ref{fig:synchalfxo}, we show
the affect of varying $\nu_\gamma$, $f_\gamma$, and $t_\gamma$ on the
prompt optical flux; the total flux -- obtained by adding the contributions 
for the two colliding shells -- ranges from $10^{-4}$ to $10^3$ mJy, or 
$R$ magnitude of 26 to 9. The optical flux increases with $\nu_\gamma$, 
$f_\gamma$, and $t_\gamma$ -- longer GRB pulses with higher peak frequency 
and/or flux tend to be brighter in the optical band. Synchrotron self 
absorption is larger at smaller $\Gr$, and that makes the optical 
flux smallest at the minimum of $\Gr$.  There are $\Gr \sim 5$ 
solutions with $f_{{B}/{ke}} \sim 1$, that have 
small enough optical flux to be in accord with the observed upper limits, 
especially for smaller
$\nu_\gamma$, $f_\gamma$, and $t_\gamma$.  Although $\nu_{a} > 2$ eV for much
of the solution space, the optical light curve should peak at
the same time as the GRB light curve, since $\nu_{c} < 2$ eV.  After the
peak, the light curve should fall off -- if dominated by off-axis emission
--  as $t^{-2.5}$.

\subsection{Synchrotron solutions for $f_\nu\propto\nu^{-\frac{p}{2}}$}

If the observed spectral index is steep and consistent with 
$\nu^{-\frac{p}{2}}$ and no break is detected in the observed energy band of
15-150 keV, for instance, then this is a special case of either 
$\nu^{-\frac{1}{2}}$ or $\nu^{-\frac{(p-1)}{2}}$ low energy spectrum discussed
in \S3.1 \& 3.3 -- with $\nu_{\gamma_5} = \max(\nu_{\gamma_5},\nu_{c_5}) 
\lta 0.15$.  
The allowed solution space for this situation should be close to the
$\nu_\gamma=20$ keV case in Figures~\ref{fig:syncp12} and \ref{fig:synchalf};
Specifically, $\gamma_i \lta 10^3$, $\Gamma > 100$, and $2\lta\Gr\lta200$.

%************

\section{Synchrotron-self-Compton -- SSC -- solutions}

In this section, we present solutions for the prompt $\gamma$-ray
emission to be produced via the 
synchrotron-self-inverse-Compton radiation or the SSC process.  
The basic approach is same as in section 3. We determine the hypersurface
in the 5-D parameter space ($\gamma_i, \Gamma, B, N, \tau$) that has
SSC emission consistent with the three observational constraints 
$\nu_\gamma$, $f_\gamma$ \& $t_\gamma$.
Since different cases of low energy spectral index have different
ordering for the characteristic synchrotron frequencies $(\nu_i, \nu_c, 
\nu_a)$, we do not
consider a general SSC solution, but describe analytical and numerical
solutions for the positive low energy spectral index case i.e., $f_\nu
\propto \nu^{\alpha}$ with $\alpha>0$ for $\nu<\nu_\gamma$, and the negative 
index case i.e., $\alpha<0$, separately in several subsections below.
 
\subsection{SSC solutions: Positive low energy spectral index}

This section is broken up in two subsections. One dealing with the
special case of $f_\nu\propto \nu^{1/3}$ is discussed below. All the
other cases of $\alpha>1/3$ are discussed in \S4.1.2.

\subsubsection{SSC solutions: $\alpha\approx+$1/3} 

The SSC spectrum ($\nu f_\nu^{ic}$) peaks at $\nu_\gamma \sim 4
\max(\nu_i,\nu_c) \max(\gamma_i,\gamma_c)^2$; where $\nu_i$ \& $\nu_c$ are
the injection and cooling frequencies of the underlying synchrotron
radiation, $\gamma_i$ is the minimum LF of electrons in the source
comoving frame and $\gamma_c$ is the LF of electrons that cool on
time scale $t_a$ available since last accelerated. For the spectrum below
$\nu_\gamma$ to be $\sim\nu^{1/3}$, we must have $\gamma_i \sim \gamma_c$, 
and in that case $\nu_\gamma\sim 4\nu_i\gamma_i^2$. The IC flux at $\nu_\gamma$
is $f_\gamma \sim f_{\nu_p} \tau$ ($f_{\nu_p}$ is the synchrotron flux 
at $\nu_i$). The equations for pulse duration $t_\gamma$ and Compton-$Y$ 
are same as in \S3. The equations for $\nu_i\sim\nu_c$, 
the peak IC frequency $\nu_\gamma$, the IC flux $f_\gamma$ at $\nu_\gamma$,
and the Compton Y-parameter are given below --
\begin{eqnarray}
&&B^2 \Gamma \gamma_i \approx 7.7 \times 10^8 (1+z) t_a^{-1} (1+Y)^{-1} 
\label{e13} \\  \nonumber\\[-.4cm] 
&&\ B \gamma_i^4 \Gamma \approx 2.3 \times 10^{12} \nu_{\gamma_5} (1+z) \label{e14} \\  \nonumber\\[-.4cm] 
&&B \Gamma^5 \tau^2 \approx 1.6 \times 10^6 f_\gamma t_\gamma^{-2} (1+z)
d_{L_{28}}^2 \label{e15} \\  \nonumber\\[-.4cm] 
&&\quad\ \tau \gamma_i^2 \approx {3 \over 4} Y \left({p-1 \over p-2}\right)^{-1}
\label{e16}
\end{eqnarray}
We first eliminate $\tau$ from equation (\ref{e15}) using (\ref{e16}) to get
\begin{equation}
B \Gamma^5 \gamma_i^{-4} \approx 2.8 \times 10^6 f_\gamma t_\gamma^{-2}
(1+z) d_{L_{28}}^2 Y^{-2} \left({p-1\over p-2}\right)^2 .
\label{e17}
\end{equation}
Next, we divide equations (\ref{e13}) and (\ref{e14}) to eliminate $\Gamma$:
\begin{equation}
B \gamma_i^{-3} \approx 3.3 \times 10^{-4} t_a^{-1} (1+Y)^{-1}
\nu_{\gamma_5}^{-1},
\label{e18}
\end{equation}
divide equations (\ref{e14}) and (\ref{e17}) to eliminate $B$:
\begin{equation}
\gamma_i^8 \Gamma^{-4} \approx 8.2 \times 10^5 \nu_{\gamma_5} f_\gamma^{-1}
t_\gamma^2 d_{L_{28}}^{-2} Y^2 \left({p-1\over p-2}\right)^{-2},
\label{e19}
\end{equation}
and combine equations (\ref{e13}) and (\ref{e17}) to obtain:
\begin{equation}
B \Gamma \approx 4.6 \times 10^4 t_a^{-\frac{4}{9}} (1+Y)^{-\frac{4}{9}} f_\gamma^{\frac{1}{9}}
t^{-\frac{2}{9}} (1+z)^{\frac{5}{9}} d_{L_{28}}^{\frac{2}{9}} Y^{-\frac{2}{9}} \left({p-1\over
p-2}\right)^{\frac{2}{9}}.
\label{e20}
\end{equation}
Equations (\ref{e18}) and (\ref{e20}) give
\begin{equation}
\Gamma \gamma_i^3 \approx 1.4 \times 10^8 \nu_{\gamma_5} t_a^{\frac{5}{9}} (1+Y)^{\frac{5}{9}}
f_\gamma^{\frac{1}{9}} t_\gamma^{-\frac{2}{9}} (1+z)^{\frac{5}{9}} d_{L_{28}}^{\frac{2}{9}} Y^{-\frac{2}{9}}
\left({p-1\over p-2}\right)^{\frac{2}{9}},
\label{e21}
\end{equation}
and substituting this into equation (\ref{e19}), we find the solution for
$\gamma_i$ to be
\begin{equation}
\gamma_i \approx 84 \nu_{\gamma_5}^{\frac{1}{4}}
f_\gamma^{-\frac{1}{36}} 
 t_\gamma^{\frac{1}{18}} t_a^{\frac{1}{9}}Y^{\frac{1}{18}}
(1+Y)^{\frac{1}{9}} 
d_{L_{28}}^{-\frac{1}{18}} (1+z)^{\frac{1}{9}} A_{2_p}^{-\frac{1}{18}}.
\label{e22}
\end{equation}
Note that the electron LF $\gamma_i$ has a very weak dependence on the
observed quantities as well as the Compton-$Y$ parameter, and therefore
$\gamma_i\sim80$ for the entire SSC solution space. By plugging equation
(\ref{e22}) back into equations (\ref{e18}), (\ref{e19}), and
(\ref{e16}),
we find the remaining parameters
\begin{eqnarray}
&& B \approx 200 \nu_{\gamma_5}^{-\frac{1}{4}} f_\gamma^{-\frac{1}{12}} 
t_\gamma^{\frac{1}{6}}
t_a^{-\frac{2}{3}}Y^{\frac{1}{6}}(1+Y)^{-\frac{2}{3}} 
  d_{L_{28}}^{-\frac{1}{6}} (1+z)^{\frac{1}{3}} 
A_{2_p}^{-\frac{1}{6}} 
  \, {\rm Gauss}\label{e23} \\  \nonumber\\[-.4cm] 
&& \Gamma \approx 240 \nu_{\gamma_5}^{\frac{1}{4}}
f_\gamma^{\frac{7}{36}} 
t_\gamma^{-\frac{7}{18}}
t_a^{\frac{2}{9}}Y^{-\frac{7}{18}}(1+Y)^{\frac{2}{9}}
d_{L_{28}}^{\frac{7}{18}} (1+z)^{\frac{2}{9}} A_{2_p}^{\frac{7}{18}}
\label{e24} \\  \nonumber\\[-.4cm] 
&& \tau \approx 1.1\times10^{-4}
\nu_{\gamma_5}^{-1/2}f_\gamma^{\frac{1}{18}}
  t_\gamma^{-\frac{1}{9}} t_a^{-\frac{2}{9}}
  Y^{\frac{8}{9}} (1+Y)^{-\frac{2}{9}} 
  d_{L_{28}}^{\frac{1}{9}} (1+z)^{-\frac{2}{9}} 
  A_{2_p}^{-\frac{8}{9}} \label{e25}.
\end{eqnarray}
All of these parameters are weakly dependent on the three observable
quantities viz. $\nu_{\gamma_5}$, $f_\gamma$ \& $t_\gamma$.
Substituting the {\it observed} values for \grba, i.e. $\nu_{\gamma_5}
 =1$, $f_\gamma =1$ mJy, and $t_a \sim t_\gamma = 0.1$ s, $z=1$, \&
taking p=3.2, we find that $\gamma_i \sim 58 Y^{\frac{1}{18}} 
(1+Y)^{\frac{1}{9}}$, $B \sim 1.1 \times 10^3 Y^{\frac{1}{6}}
(1+Y)^{-\frac{2}{3}}$Gauss, $\Gamma \sim 110 Y^{-\frac{7}{18}}
(1+Y)^{\frac{2}{9}}$, and $\tau \sim 1.3 \times 10^{-4} Y^{\frac{8}{9}}
(1+Y)^{-\frac{2}{9}}$.

The distance of the $\gamma$-ray source from the center of the
explosion,
$R_\gamma$, is given by
\begin{equation}
R_\gamma \approx 3.3\times10^{15} \, \nu_{\gamma_5}^{\frac{1}{2}}
f_\gamma^{\frac{7}{18}} t_\gamma^{\frac{2}{9}}
t_a^{\frac{4}{9}}
Y^{-\frac{7}{9}} (1+Y)^{\frac{4}{9}}
(1+z)^{-\frac{5}{9}}  d_{L_{28}}^{\frac{7}{9}}
A_{2_p}^{\frac{7}{9}} \,{\rm cm}\label{e26},
\end{equation}
or $R_\gamma \sim 1.3\times 10^{15} Y^{-\frac{7}{9}}
(1+Y)^{\frac{4}{9}}$cm  for \grba.
This distance is smaller than the deceleration radius for a homogeneous
or a
wind external medium, unlike the situation when $\gamma$-rays are
produced
via the synchrotron process (see \S3).

One constraint that we have not yet considered is that the SSC self absorption
frequency, $\nu_a^{ic} \sim 4 \nu_a \gamma_i^2$, must be smaller than $\sim$20
keV otherwise the low energy spectral index, obtained by Band function
fit to the BATSE or \emph{Swift}/BAT data, would be steeper than $\alpha=1/3$ 
we are considering in this subsection. The expression for $\nu_a^{ic}$,
valid for $\nu_a < \nu_i, \nu_c$, is
\begin{equation}
\nu_a^{ic} \sim {1.7 \times 10^{-14} \gamma_i^2 \Gamma^{6\over5} \over 
 (1+z)} \left({f_p' \nu_i^{-\frac{1}{3}} \over 2 \gamma_i m_e}
   \right)^{\frac{3}{5}} 
\end{equation}
where $f_p' \equiv \sqrt{3} q^3 B \tau/ \sigma_T m_e c^2$ is the comoving
synchrotron flux. Substituting for $\gamma_i$, $\Gamma$, $B$ \& $\tau$ 
using equations \ref{e22}--\ref{e25} we find
\begin{equation}
\nu_a^{ic} \sim 2.2 \times 10^5 \, \nu_{\gamma_5}^{\frac{1}{10}}
f_\gamma^{\frac{1}{6}} t_\gamma^{-\frac{1}{3}} t_a^{-\frac{1}{15}}
Y^{\frac{4}{15}} (1+Y)^{-\frac{1}{15}}
(1+z)^{-\frac{2}{3}} d_{L_{28}}^{\frac{1}{3}}
A_{2_p}^{-\frac{4}{15}}\,{\rm eV} \label{e27}
\end{equation} 
which is very insensitive to \emph{all} of the observed quantities and for 
a wide range of observables $\nu_a^{ic}\sim100$ keV which is
 too large to produce an SSC spectrum with $f_\nu^{ic}\propto\nu^{\frac{1}{3}}$ 
below the peak.

We now try relaxing one of the constraints we had imposed 
to simplify the analytical calculation i.e., $\gamma_i \sim \gamma_c$.  
This approximation was guided by the observational result that
the observed spectrum for $\nu>\nu_\gamma$ is almost always $\sim \nu^{-1.5}$
for GRBs with $\alpha\sim1/3$. This result suggests $\gamma_i \sim \gamma_c$
provided that $p\approx3$. However, if the electron distribution is steeper,
$p \sim 4$, then $\nu_c$ can be much greater than $\nu_i$, and a
high energy spectrum of $\nu^{-(p-1)/2} \sim \nu^{-1.5}$ would be
consistent with observations. We now investigate this possibility and
determine if letting $\nu_c>\nu_i$ would allow for a smaller
$\nu_a^{ic}$ and hence $\alpha=1/3$ solutions. Note that the opposite 
arrangement of frequencies ($\nu_c < \nu_i$) is uninteresting, since the 
low energy index is -1/2 in this case.

For $\nu_c > \nu_i$ equation (\ref{e13}) is modified to read
\begin{equation}
\gamma_i B^2 \Gamma = {7.7 \times 10^8 (1+z) \over \eta_i t_a (1+Y)}
\end{equation}
where $\eta_i \equiv\gamma_c/\gamma_i$. We also need to use the appropriate
expression for $Y$ when $\gamma_c \gg \gamma_i$ and $p>3$ (see eq. ~\ref{s5})
\begin{equation}
Y = {4 \over 3} {\left(p-1\right) \over \left(p-2\right)
\left(p-3\right)} \tau \gamma_i^2,
\end{equation}
which apart from a factor $(p-3)$ is same as equation (\ref{e16}).
We solve the above two equations together with eqs. (\ref{e15}) \&
(\ref{e16}) to find that $\nu_a^{ic} \propto \eta_i^{-\frac{1}{15}}$;
 so $\nu_a^{ic}$ does not decrease by much even if we take $\gamma_c$
to be larger than $\gamma_i$ by many orders of magnitude, and therefore there
are no SSC solutions with low energy spectrum of $\nu^{1/3}$ between 
$\sim15$ keV and 200 keV. 

The above analytical calculation is based on a number of approximations for
the SSC spectrum and flux. We check the validity of analytical results
using numerical calculations and by searching the 5-D parameter
space for SSC solutions with low energy spectral index $\alpha\approx1/3$.
 It turns out that numerically also we find no solutions --- $\nu_a^{ic}$ is
indeed too high to produce a GRB with a low energy spectrum of $\nu^{1/3}$
in the $\sim$15--200 keV band.

The only other possibility is that the $\alpha\approx1/3$ index is transitory, 
i.e.  the spectrum is changing continuously from $\alpha\approx1$ at
$\nu\sim 20$ keV to $\alpha\approx-1$ at $\nu=\nu_\gamma$ and that 
$\alpha\sim 1/3$ is realized at some intermediate frequency.
 This might, however, pose a problem for those
GRBs with $\nu_\gamma\gta100$keV, since the spectrum would be steeper than 
$\nu^{1/3}$ near 15 keV and therefore a Band function fit to the
spectrum will yield $\alpha>1/3$.

\subsubsection{SSC solutions: Spectral index $1/3<\alpha\le1$} 

\begin{figure}[h!]
\begin{center}
\includegraphics[height=3.9in,width=5.4in]{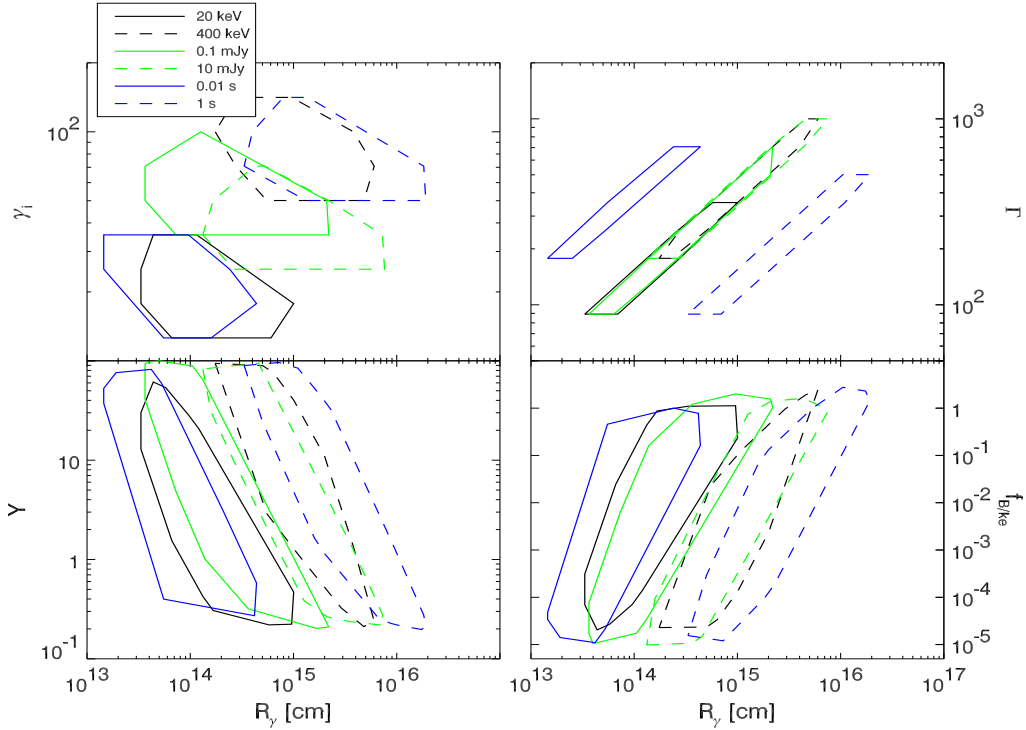}
\caption{\small
The SSC solution space when the low energy spectrum is $f^{ic}_\nu
\propto\nu^\alpha$, for $\nu<\nu_\gamma$, with $1/3<\alpha\le1$. 
The different cases in legend are as described in Figure~\ref{fig:syncp12}.  }
\label{fig:icpos}
\end{center}
\end{figure}

The analytical solution for this case is similar to the SSC $\alpha=1/3$
case analyzed in \S4.1.1. We take $\gamma_i \sim \gamma_c$ in order that the
high energy spectrum is $\propto \nu^{-p/2}\sim\nu^{-1.5}$. The equations
we solve are for $\nu_\gamma$, $f_\gamma$, Compton-Y, and $\gamma_i \sim 
\gamma_c$:
\begin{eqnarray}
&& B^2 \Gamma \gamma_i \approx 7.7 \times 10^8 (1+z) t_a^{-1} (1+Y)^{-1} 
\label{p13} \\  \nonumber\\[-.4cm] 
&&\ B \gamma_i^4 \Gamma \approx 2.3 \times 10^{12} \nu_{\gamma_5} (1+z)
   \eta_a^{-2} \label{p14} \\  \nonumber\\[-.4cm] 
&& B \Gamma^5 \tau^2 \approx 1.6 \times 10^6 f_\gamma t_\gamma^{-2} (1+z)
d_{L_{28}}^2 \eta_a^p \label{p15} \\  \nonumber\\[-.4cm] 
&&\quad \tau \gamma_i^2 \approx {3 \over 4} Y \left({p-1 \over p-2}\right)^{-1}
\label{p16}
\end{eqnarray}
where $\eta_a \equiv \gamma_a/\gamma_i\gta1$ since $\nu_a \gta \nu_c,\nu_i$.
The above equations are solved in the exact same way
that we solved them in section 4.1.1, and we find that the solutions are:
\begin{eqnarray}
&& \gamma_i \approx 84 \nu_{\gamma_5}^{{1\over4}} f_\gamma^{-{1\over36}}
   t_\gamma^{{1\over18}} t_a^{{1\over9}} Y^{{1\over18}}
   (1+Y)^{{1\over9}}
   d_{L_{28}}^{-{1\over18}} (1+z)^{{1\over9}} 
   A_{2p}^{-{1\over18}} \eta_a^{-\left(p+18\right) \over 36}
\label{pos1} \\
&& B \approx 200 \nu_{\gamma_5}^{-{1\over4}} f_\gamma^{-{1\over12}} 
  t_\gamma^{{1\over6}} 
   t_a^{-{2\over3}} Y^{{1\over6}} (1+Y)^{-{2\over3}}
  d_{L_{28}}^{-{1\over6}} (1+z)^{{1\over3}} 
   A_{2p}^{-{1\over6}} \eta_a^{6-p \over 12}\,{\rm Gauss} \label{pos2} \\
&& \Gamma \approx  240 \nu_{\gamma_5}^{{1\over4}} f_\gamma^{{7\over36}}
  t_\gamma^{-{7\over18}}
   t_a^{{2\over9}} Y^{-{7\over18}} (1+Y)^{{2\over9}}
   d_{L_{28}}^{{7\over18}} (1+z)^{{2\over9}} 
  A_{2p}^{{7\over18}} \eta_a^{7p -18 \over 36} \label{pos3} \\
&& \tau \approx 1.1 \times 10^{-4} \nu_{\gamma_5}^{-{1\over2}}
f_\gamma^{{1\over18}} t_\gamma^{-{1\over9}} 
   t_a^{-{2\over9}} Y^{{8\over9}} (1+Y)^{-{2\over9}}
  d_{L_{28}}^{{1\over9}} (1+z)^{-{2\over9}} 
   A_{2p}^{-{8\over9}} \eta_a^{p+18\over18}. \label{pos4}
\end{eqnarray}
For $p=3.2$, $t_a\sim t_\gamma$, $\eta_a\gta 1$, and parameters 
corresponding to \grba~ we find from the above equations that
$\Gamma \gta 110 Y^{-\frac{7}{18}}$, $\gamma_i \lta 100$, and
$\tau \gta 1.3 \times 10^{-4} Y^{\frac{8}{9}} (1+Y)^{-\frac{2}{9}}$.
These analytical results are roughly consistent with numerical determination
of the allowed region in 5-D parameter space (see fig. \ref{fig:icpos});
we also find $\eta_a\approx1$ numerically.

The distance of the $\gamma$-ray source from the center of the explosion is 
shown in fig. (\ref{fig:icpos}) for various GRB parameters and is greater
than $\sim10^{14}$cm for $Y\lta10$.  The ratio of magnetic to
electron energy is small -- $f_{B/ke} < 0.1$ for the entire solution space
(fig. \ref{fig:icpos}) since $Y\gta1$. 

The SSC solutions we have found can be related to the internal shock model.
The relative LF of collision between shells -- obtained from $\gamma_i$ (see 
eq. \ref{gamsh} and appendix A) -- is found to be between 2 and 10, which
is significantly less than what we were finding for synchrotron solutions.
The LFs of shells before collision (assuming that $\gamma$-rays are produced
in the inner, faster, shell) is $300 \lta 
\Gamma_1 \lta 5000$ and $100 \lta \Gamma_2 \lta 1000$; the ratio 
$\Gamma_1/\Gamma_2>2$ and the efficiency for producing $\gamma$-rays is
$\gta$10\% for the allowed 5-D parameter space for SSC. The bulk
LF of post-shock gas $100 < \Gamma < 1000$ is compatible with late time 
afterglow modeling.

We now calculate the x-ray and optical emissions accompanying the 
$\gamma$-ray pulse.

\bigskip
\centerline{\it 4.1.2a. X-ray emission for $1/3<\alpha\le1$ SSC solutions}
\medskip

\begin{figure}[h!]
\begin{center}
\includegraphics[height=4.1in,width=5.4in]{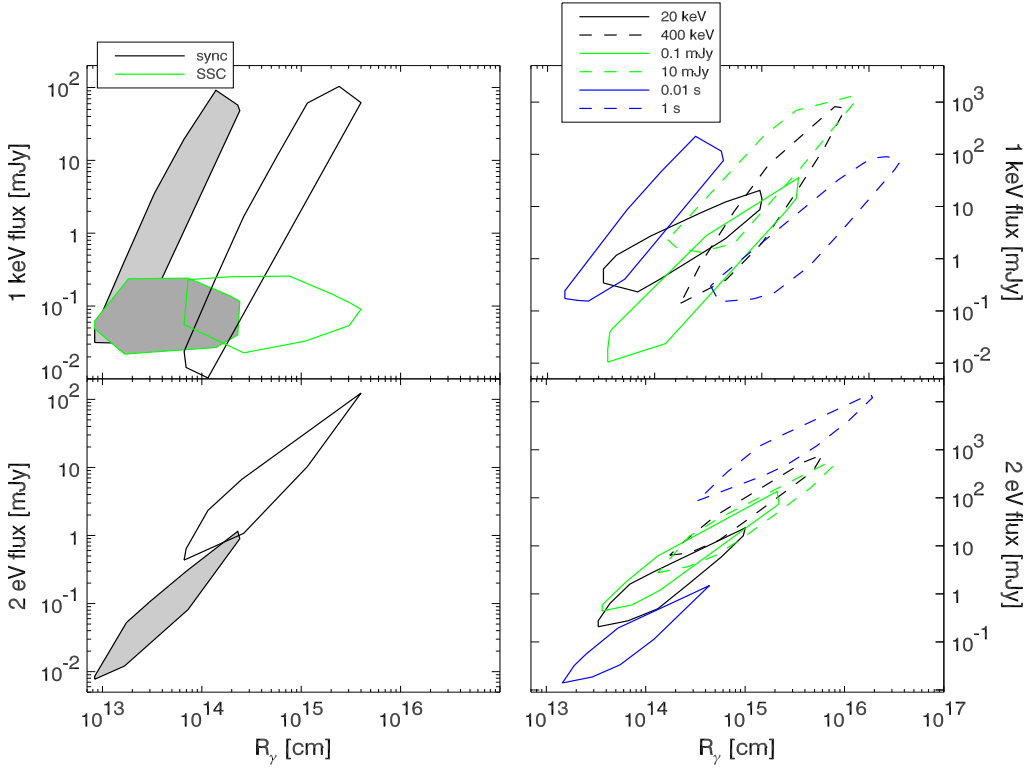}
\caption{\small    X-ray (1 keV) and optical (2eV) synchrotron 
and SSC flux accompanying the prompt GRB emission for SSC solutions 
with $1/3<\alpha\le1$.
{\bf Right panels:} The sum of SSC and synchrotron contributions for several
different sets of $\nu_\gamma$, $f_\gamma$ \& $t_\gamma$ as described in
Figure~\ref{fig:synchalfxo}. {\bf Left two panels:} the areas shaded in
gray are the x-ray and optical flux for \grba~ (a GRB that has
$\nu_\gamma=100$keV, $f_\gamma=1$mJy \& $t_\gamma=0.1$s) and with $t_a=
 t_\gamma/100$, and the unshaded areas are for \grba~ with $t_a=t_\gamma$.
}
\label{fig:icposxo}
\end{center}
\end{figure}

The 1 keV prompt emission from SSC and synchrotron processes is shown 
in the top two panels of figure~\ref{fig:icposxo}. The contributions of
SSC
\& synchrotron to 1 keV flux is shown separately in the top left panel
for \grba, and the sum of the two for a variety of GRB parameters can
be found in the top right panel. 

The x-ray flux can be estimated analytically using the expression for
synchrotron flux $f_x = f_{\nu_p} (1 \mathrm{keV}/\nu_i)^{-p/2}$, since $\nu_i
\sim \nu_c \sim \nu_a$; $f_x$ can also be expressed as $f_x \sim
f_\gamma \tau^{-1} \gamma_i^{-p} (25)^{p/2} \nu_{\gamma_5}^{p/2}$,
or in terms of observable parameters
\begin{equation}
f_x \sim {9\times10^3\over 17^p} \nu_{\gamma_5}^{p+2 \over 4}
f_\gamma^{34 + p \over 36} t_\gamma^{2-p \over 18} t_a^{2-p \over 9}
Y^{-{16+p \over 18}} (1+Y)^{2-p \over 9}
d_{L_{28}}^{p-2 \over 18} (1+z)^{2-p \over 9}
A_{2p}^{16+p \over 18}
\eta_a^{\left(p-2\right)\left(p+18\right) \over 36}\,{\rm mJy}, 
\label{icposfx}
\end{equation}
or $f_x \sim 9 Y^{-\frac{16}{15}} (1+Y)^{-\frac{2}{15}}$mJy for \grba~
with $\eta_a > 1$, $p=3.2$ and $t_a\sim t_\gamma$; this is roughly
consistent with the numerically calculated flux shown in 
fig. \ref{fig:icposxo}.  

In the top right panel of Figure~\ref{fig:icposxo}, the total 1 keV
prompt flux is shown for a number of different values of $\nu_\gamma$,
$f_\gamma$, and $t_\gamma$. The dependence of the x-ray flux on these
quantities agree with equation (\ref{icposfx}), which gives $f_x\propa
\nu_\gamma^{\frac{13}{10}} f_\gamma^{\frac{31}{30}}
t_\gamma^{-\frac{1}{15}}$ for $p=3.2$: an increase in $\nu_\gamma$ or
$f_\gamma$ leads to an increase of $f_x$, and an increase in $t_\gamma$
has little effect on the x-ray flux (fig.  \ref{fig:icposxo}). The 1 keV
flux for all of these cases ranges from 0.01 to $10^3$ mJy during the
burst; the flux at 100s, the time when the x-ray telescope aboard the
\emph{Swift} satellite starts looking at the burst, would be smaller by
a factor of $\sim10$--10$^3$ depending on GRB pulse duration.  So, the
x-ray flux accompanying the $\gamma$-ray radiation, for the SSC model of
GRBs, is consistent with the observed data.

\bigskip
\centerline{\it 4.1.2b. Optical emission for $1/3<\alpha\le1$ SSC solutions}
\medskip

The prompt optical flux accompanying $\gamma$-rays, in the
SSC model, is shown in the bottom left panel of Figure~\ref{fig:icposxo}.  
Analytically we find the prompt R-band flux due to the synchrotron component
underlying the SSC model to be: $f_R\sim f_{\nu_p}\sim f_\gamma\tau^{-1}$, or 
\begin{equation}
f_R \sim 9\times10^3 
\nu_{\gamma_5}^{\frac{1}{2}} f_\gamma^{\frac{17}{18}}
t_\gamma^{\frac{1}{9}} t_a^{\frac{2}{9}} Y^{-\frac{8}{9}}
(1+Y)^{\frac{2}{9}}
d_{L_{28}}^{-\frac{1}{9}} (1+z)^{\frac{2}{9}} A_{2_p}^{\frac{8}{9}}
\eta_a^{17p-18 \over 18} \mathrm{mJy}
\label{icposfr}
\end{equation}
which for \grba~ reduces to $f_R \sim (8\times10^3$ mJy) 
$Y^{-\frac{8}{9}}(1+Y)^{\frac{2}{9}}$. There are, however, many numerical 
solutions corresponding to $Y\gta10$ for which 
$f_R \sim 70$ mJy.  
It turns out that $\nu_a > 2$ eV for the low radius
solutions, by up to a factor of 3.  If $\nu_i < 2$ eV$ < \nu_a$, then
the expression for optical flux is
\begin{equation}
f_R \sim 50 \nu_{\gamma_5}^{-\frac{3}{4}} f_\gamma^{\frac{29}{36}}
t_\gamma^{\frac{7}{18}} t_a^{\frac{7}{9}} Y^{-\frac{11}{18}}
(1+Y)^{\frac{7}{9}}
d_{L_{28}}^{-\frac{7}{18}} (1+z)^{\frac{7}{9}}
A_{2_p}^{\frac{11}{18}}
\eta_a^{-{\left(7p+126\right) \over 36}} \mathrm{mJy}
\label{icposfrnua}
\end{equation}
giving $f_R \lta 7 Y^{-\frac{11}{18}} (1+Y)^{\frac{7}{9}}$ mJy for \grba.

The optical flux, obtained by numerical calculations, is shown in the
bottom right panel of Figure~\ref{fig:icposxo}, for several sets of
($\nu_\gamma$, $f_\gamma$, $t_\gamma$). The results are consistent with
the dependences found in equations (\ref{icposfr}) or (\ref{icposfrnua})
when $\nu_a>2$eV.  The reason that the self absorbed $f_R$ increases
with $\nu_\gamma$ numerically while equation (\ref{icposfrnua}) shows a
decrease is that $\eta_a \sim \nu_\gamma^{-1/2}$ (confirmed
numerically), and the huge dependence of $f_R \propto \eta_a^{4.1}$
gives positive dependence of $f_R \sim \nu_\gamma^{1.3}$.  The
dependence of optical flux on the duration of the GRB pulse is due to
the fact that longer pulses have larger $R_\gamma$ and $\nu_a<2$eV.  The
range of optical flux for SSC solutions is between 0.01 mJy to a few
times $10^4$ mJy ($R$ magnitude from 21- to 6-mag).  There is an
approximately linear relationship between $f_R$ and $R_\gamma$.
Solutions with $R$-magnitude of above 9-mag (1 Jy) are most likely ruled
out.  In particular, this rules out the 1 s pulse duration solutions
that have $R_\gamma\gta2\times10^{15}$cm. If the pulse width were 10 s,
the SSC solutions would have prompt optical of between 1 and 4 Jy, or $R
\sim$ 7 mag, which is too bright to have been missed in optical follow
up observations.  We note that if GRB dissipation radius is
$\sim10^{16}$cm as found in \citet{kumar07}, then bright optical flux of
$R > 9$ mag is expected in every GRB produced via SSC.  Since this
bright optical emission is not seen, this may pose major problems for
the SSC process to produce GRBs with positive $\alpha$.  

If, however, electrons are accelerated multiple times during the course
of a $\gamma$-ray pulse in the GRB light-curve, i.e. $t_a\ll t_\gamma$,
the optical flux can be reduced significantly.  The dependence $f_R
\propto t_a^{2/9}$ (eq. \ref{icposfr}) itself does not reduce $f_R$ by
much, but since $\nu_a \propto t_a^{-{4\left(p+3\right) \over 9
\left(p+5\right)}} \sim t_a^{-0.3}$ as well, a smaller $t_a$ gives a
larger $\nu_a$ and that reduces the optical flux by an additional factor
of $\sim t_a^{-0.6}$ and results in $f_R \propa t_a^{7/9}$;  for $t_a =
t_\gamma/100$, the optical flux is reduced by a factor of $\sim10^2$
compared with the case where $t_a\sim t_\gamma$, in agreement with the
numerical results found in the lower left hand panel of
Figure~\ref{fig:icposxo}.  We've numerically searched the whole range of
observable parameters and find that the scaling $f_R\propa t_a^{7/9}$ is
valid through the entire solution space, and even the highest optical
flux levels of 10 Jy (for the $t_\gamma \sim 1 $s case) is reduced below
0.1 Jy, or $R > 11$ mag, if $t_a\lta t_\gamma/10^2$. Multiple
acceleration episodes for electrons is, therefore, a possible way of
reducing the excessive optical flux that otherwise necessarily
accompanies SSC solutions for $\gamma$-ray emission.

The optical flux should peak at the same time the GRB light curve peaks,
since $\nu_{c}$ is on the order of 2 eV.  The temporal decay of optical
flux in this case is dictated by the curvature or the off-axis emission
(Kumar \& Panaitescu, 2000) --- the optical light curve should fall off
as $t^{-2-p/2} \sim t^{-3.6}$ as long as $\nu_a$ is below the R band. At
first glance, it might seem that if $\nu_a > 2$ eV, and the synchrotron
spectrum $\propto\nu^{5/2}$ or $\nu^2$ in the optical band, the optical
LC would be flat $\sim t^0$ or even rise as $t^{1/2}$. This behavior,
however, lasts for a very short time since ($\nu_a/2$eV)$\lta$ a few,
and $\nu_a\propto t^{-1}$ for off-axis emission; once $\nu_a$ drops
below 2eV the optical flux would start falling off as $\sim t^{-3.6}$.
The upper limit of $V\sim 18.5$-mag (0.2 mJy) at 100s for many Swift
detected bursts \citep[e.g.][]{roming06} is a lot smaller than the flux
expected during the burst for the SSC model. If the pulse occurs at
$\sim 1$s post-trigger then the optical flux at 100s would be smaller
than the prompt optical flux by a factor $\sim10^7$ and that is quite
consistent with observational upper limits for almost the entire
solution space for the SSC-model.

We emphasize that a bright optical flash ($R\lta14$-mag) concurrent with
$\gamma$-ray emission is a generic prediction of the SSC-model for GRBs
with positive low energy index. Bright, prompt, optical radiation has
been reported for a few bursts with positive $\alpha$ -- e.g. 990123
\citep{akerlof99,briggs99}, 061007 \citep{gcn5722,gcn5724}, and the
second emission episodes of 050401 \citep{gcn3179} and 050820A
\citep{gcn3858} -- however, if future observations fail to detect prompt
optical with $R\lta14$-mag then that will suggest that one of the
assumptions of the model developed in this work has to be abandoned --
the most likely possibility, in our view, is to discard the assumption
that $t_a\sim t_\gamma$ and that suggests that $\gamma$-rays are not
generated in a shock heated medium.

%*****************************************
\subsection{SSC solutions for negative low energy spectral index}

In this section we consider synchrotron-self-Compton solutions when the 
spectrum below $\nu_\gamma$, the peak of $\nu f^{ic}_\nu$, is 
$f^{ic}_\nu\propto\nu^\alpha$ with $\alpha<0$. There are two different class 
of solutions in this case -- 
those with the underlying seed synchrotron spectrum having $\nu_c < \nu_i$ 
and vice versa. We treat the two cases separately analytically, but plot the 
numerical solutions for both cases together in Figure~\ref{fig:ichalf}. We 
use one vital piece of information gained from the numerical solutions 
to simplify our analytical calculations: the synchrotron 
self absorption frequency, $\nu_a$, is larger than $\max(\nu_i, \nu_c)$, 
and therefore $\nu_\gamma \sim 4 \nu_a\max(\gamma_i,\gamma_c)^2$; note that
even though $\nu_a > \max(\nu_i, \nu_c)$, the spectral index below 
$\nu_\gamma$ is negative down to the frequency $\sim \nu_a
\min(\gamma_i,\gamma_c)^2 < 10$ keV. 

\subsubsection{$\nu_c < \nu_i$ case}

The equations that are solved for this case can be cast in a form very
similar to the SSC $\alpha\approx1/3$ case (considered in \S4.1.1) by 
introducing two variables:  $\eta_i \equiv \gamma_i/\gamma_c$ and 
$\eta_a \equiv \gamma_a/\gamma_i$. The equations for $\nu_c$, $\nu_\gamma$,
$f_\gamma$, and Compton-$Y$ expressed in terms of $\eta_i$ \& $\eta_a$ are:
\begin{eqnarray}
&& B^2 \Gamma \gamma_i \sim 7.7 \times 10^8 \eta_i (1+z) t_a^{-1} (1+Y)^{-1}
\\  \nonumber\\[-.4cm] 
&& B \gamma_i^4 \Gamma \sim 2.3 \times 10^{12} \nu_{\gamma_5} (1+z) \eta_a^{-2} \\  \nonumber\\[-.4cm] 
&& B \Gamma^5 \tau^2 \sim 1.6 \times 10^6 f_\gamma t_\gamma^{-2} (1+z)
d_{L_{28}}^2 \eta_a^p \eta_i^2 \\  \nonumber\\[-.4cm] 
&& \tau \gamma_i^2 \sim 0.75 Y \left({p-1 \over p-2}\right)^{-1} \eta_i.
\end{eqnarray}
These equations are solved the same way as 
outlined in the previous section, and we find the solutions to be:
\begin{eqnarray}
&& \gamma_i \sim 84 \nu_{\gamma_5}^{1/4} f_\gamma^{-1/36} 
t_\gamma^{1/18} t_a^{1/9} Y^{1/18} (1+Y)^{1/9}
d_{L_{28}}^{-1/18} (1+z)^{1/9} A_{2p}^{-1/18}
\eta_i^{-\frac{1}{9}} \eta_a^{-\left(p+18\right) \over 36} \label{icneg1}\\
&& B \sim 200 \nu_{\gamma_5}^{-1/4} f_\gamma^{-1/12} 
t_\gamma^{1/6} t_a^{-2/3} Y^{1/6} (1+Y)^{-2/3}
d_{L_{28}}^{-1/6} (1+z)^{1/3} A_{2p}^{-1/6}
\eta_i^{\frac{2}{3}} \eta_a^{6-p \over 12}\,{\rm Gauss} \label{icneg2} \\
&& \Gamma \sim 240 \nu_{\gamma_5}^{1/4} f_\gamma^{7/36} 
t_\gamma^{-7/18} t_a^{2/9} Y^{-7/18} (1+Y)^{2/9}
d_{L_{28}}^{7/18} (1+z)^{2/9} A_{2p}^{7/18}
\eta_i^{-\frac{2}{9}} \eta_a^{7 p -18 \over 36} \label{icneg3} \\
&& \tau \sim 1.1 \times 10^{-4} \nu_{\gamma_5}^{-1/2} f_\gamma^{1/18}
t_\gamma^{-1/9} t_a^{-2/9} Y^{8/9} (1+Y)^{-2/9}
d_{L_{28}}^{1/9} (1+z)^{-2/9}  A_{2p}^{-8/9}
\eta_i^{\frac{11}{9}} \eta_a^{p+18 \over 18}.
\label{icneg4}
\end{eqnarray}
The dependence of $B$, $\Gamma$, $\gamma_i$ \& $\tau$ on the observables is 
same as in equations~\ref{e22}-\ref{e25} -- the difference is in the
dependence on $\eta_a$ and $\eta_1$.  The distance of the $\gamma$-ray
source from the center of the explosion is $R_\gamma \propto t_\gamma\Gamma^2 
\propto \eta_i^{-\frac{4}{9}} \eta_a^{7 p -18 \over 18}$.  We use the
constraint that $\nu_a^{ic} \sim 4 \nu_a \gamma_c^2 \sim 10^5 \nu_{\gamma_5}
\eta_i^{-2}  < 10$ keV to infer that $\eta_i\gta 3.2$ for these solutions; 
we numerically find that $\eta_i \gta 10$.

\begin{figure}[h!]
\begin{center}
\includegraphics[height=3.9in,width=5.4in]{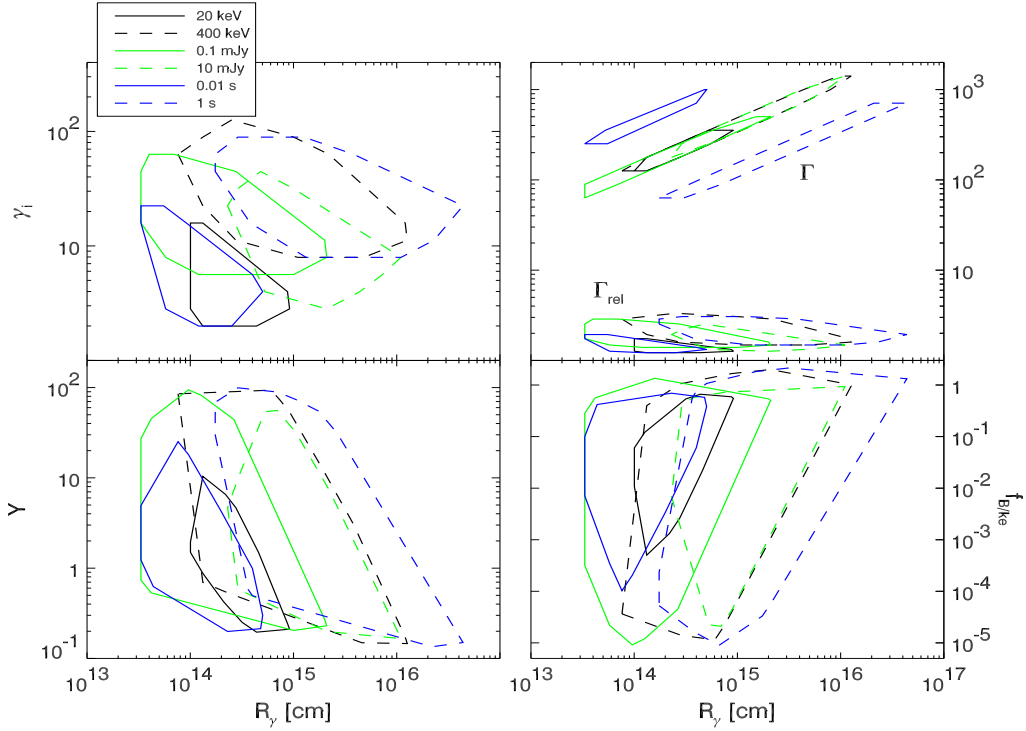}
\caption{\small
Synchrotron-Self-Compton solutions when the spectral index, $\alpha$, below
$\nu_\gamma$ (the peak of $\nu f^{ic}_\nu$) is less than 0. There are two 
branches of the solutions both of which are shown in the figure. For one 
of these branches
$\nu_c < \nu_i$, and for the other $\nu_c > \nu_i$;  $\gamma$-ray sources
corresponding to the $\nu_c < \nu_i$ branch lie at a smaller distance from
the center of explosion ($R_\gamma$) than the other branch. Allowed regions
for the 5-D parameter space is shown for a number of different sets of
GRB observable parameters (see figure ~\ref{fig:syncp12} caption for details).
The top right panel shows the LF of the $\gamma$-ray source (one of the
five basic parameters we use to describe the source) as well as the $\Gr$ --
the relative LF of collision of two shells obtained by mapping the 5-D 
parameter solution to internal shocks (see appendix A).
}
\label{fig:ichalf}
\end{center}
\end{figure}

Numerical calculations of the allowed region in the 5-D parameter space 
for \grba~ give $\gamma_i \lta 50$, $B \gta 200$ Gauss, 
$\Gamma \lta 300$, and $\tau \gta 3 \times 10^{-3}$.  There is good
agreement between the numerical and analytical  $\gamma_i$ and $\Gamma$
solutions (see fig. \ref{fig:ichalf}), although analytical and numerical 
solutions for $B$, $\tau$ \& $R_\gamma$ can differ by a factor of a few
due to the difference of a factor of a few in the analytical and
numerical value of $\eta_i$.

In figure~\ref{fig:ichalf}, we show the numerical solutions for the allowed
region in the 5-D space for the SSC-model for a wider range of observables.  
Note that results for both $\nu_i > \nu_c$ \& $\nu_i < \nu_c$ 
(to be considered analytically in the next subsection) are plotted 
together in fig. \ref{fig:ichalf}. The allowed range for electron LF
$\gamma_i$ is 3--200 (fig. \ref{fig:ichalf}) which is characteristic
of mildly relativistic shocks. If we cast the 5-D parameter solutions
in terms of colliding shells as described in appendix A, we find the relative
LF of collision between shells to be less than a few.
For $\Gamma \sim 100$ \& $\Gr \sim$ a few, there is little chance of an 
external forward-shock origin for these SSC photons. However, the 5-D
solutions we find appear to be consistent with an internal shock;
$R_\gamma$ is smaller than the deceleration radius and $f_{B/ke} \lta 1$
for the entire SSC solution space.

The dependence of the solution space on $\nu_\gamma$, $f_\gamma$, and
$t_\gamma$ is in agreement with analytical estimates; for example, 
we have verified the analytical dependence, $\gamma_i \propto 
\nu_\gamma^{\frac{1}{4}} f_\gamma^{-\frac{1}{36}} t_\gamma^{\frac{1}{6}}$
(for $t_a = t_\gamma$), given in equation (\ref{icneg1}), 
 and $\Gamma \propto \nu_\gamma^{\frac{1}{4}} f_\gamma^{\frac{7}{36}} 
  t_\gamma^{-\frac{1}{6}}$ (eq. \ref{icneg3}).
We find that SSC can produce a wide variety of GRBs with the low
energy spectral index $\alpha\sim -1/2$ without requiring extreme 
parameter values.  

In order for these solutions to remain viable, the prompt optical and
x-ray flux needs to be in accord with observations and/or upper limits.
We next look at the prompt x-ray and optical emission from these
solutions.

\bigskip
\centerline{\it 4.2.1a.  Prompt x-ray emission for SSC solutions with $\alpha
  \le-1/2$}
\medskip

\begin{figure}[h!]
\begin{center}
\includegraphics[height=3.9in,width=5.4in]{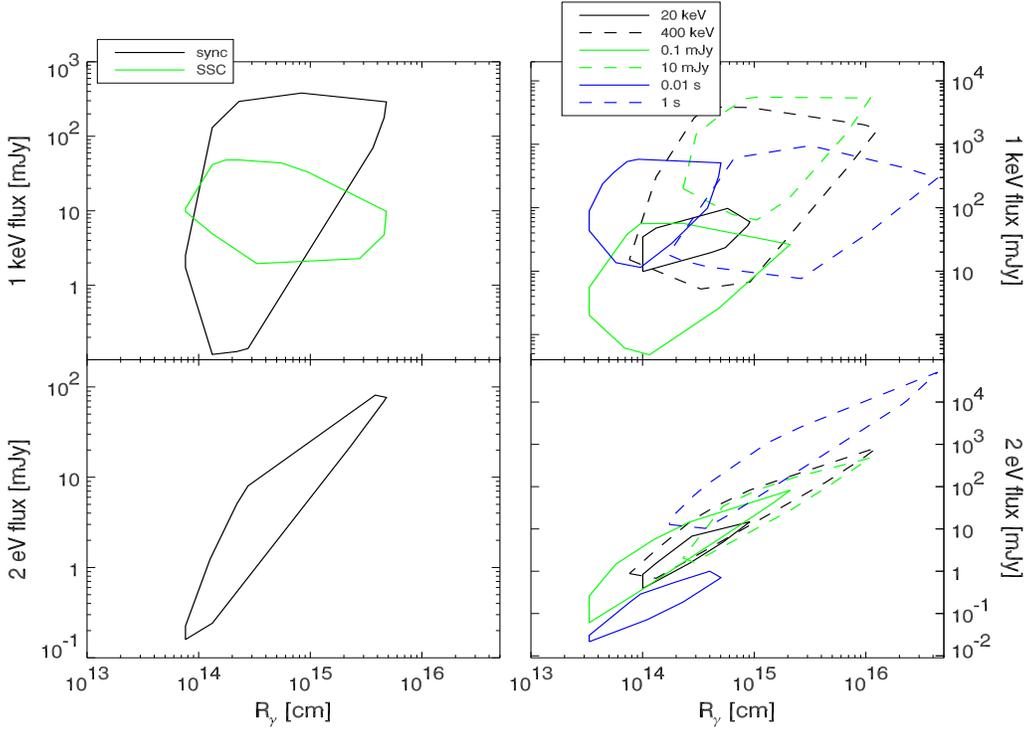}
\caption{\small
X-ray (1 keV) and optical flux simultaneous with the GRB emission for
the SSC solutions with $\alpha<0$ are plotted against the $\gamma$-ray
emission radius. The top left panel shows the
synchrotron \& SSC contributions to the 1 keV flux for \grba, and the
bottom left panel shows the optical (2 eV) flux associated with \grba-SSC
 solutions. The right panels show the
1 keV flux (top right) and 2 eV flux (bottom right) by summing up 
contributions of synchrotron and SSC for various sets of observed 
parameters for GRBs as described in the caption for fig. \ref{fig:syncp12}.}
\label{fig:ichalfxo}
\end{center}
\end{figure}

In the top two panels of Figure~\ref{fig:ichalfxo}, we've plotted the
synchrotron and SSC contributions to the x-ray (1 keV) flux accompanying 
the $\gamma$-ray emission during the burst. In the top left panel, we see 
that the 1 keV flux ranges from 0.1 to over 100 mJy for \grba~and much of 
it is due to the underlying synchrotron emission.

There are at least a few solutions for each value of $\nu_\gamma$,
$f_\gamma$ \& $t_\gamma$ we have considered with x-ray flux less than 10
mJy (fig. \ref{fig:ichalfxo}) with the exception of the $f_\gamma \sim
10$ mJy case. The high end of this range of x-ray flux is above the value
typically observed by \emph{Swift} at 100 seconds following the burst
($10^{-4}$ to a few mJy); but we know from early x-ray observations that
this flux is initially falling off very steeply, $\sim t^{-3}$, and
therefore x-ray flux of $\sim10^3$ mJy during the burst,
$t_\gamma\lta10$s, would be less than 1 mJy at 100s, or within the
observed flux range of the x-ray telescope aboard \emph{Swift}.

We expect the x-ray flux  to be
\begin{equation}
f_x = f_{\nu_p} \left({1 \mathrm{keV} \over \nu_i}\right)^{-\frac{p}{2}}
\left({\nu_i \over \nu_c}\right)^{-\frac{1}{2}},
\end{equation}
(since synchrotron dominates). In terms of the observable quantities the 
flux is
\begin{equation}
f_x \approx {9\times10^3\over 17^p} \nu_{\gamma_5}^{p+2 \over 4}
f_\gamma^{34 + p \over 36} 
t_\gamma^{2-p \over 18} t_a^{2-p \over 9} Y^{-{16+p\over 18}}
(1+Y)^{2-p \over 9}
d_{L_{28}}^{p-2 \over 18} (1+z)^{2-p \over 9} A_{2p}^{16+p \over 18}
\eta_a^{\left(p-2\right)\left(p+18\right) \over 36}
\eta_i^{p-2 \over 9}\, {\rm mJy},
\label{icfx}
\end{equation}
so for $p=2.5$ and $t_a \sim t_\gamma$, $f_x \propto \nu_\gamma^{\frac{9}{8}}
f_\gamma^{\frac{73}{72}} t_\gamma^{-\frac{1}{12}}$. This analytical formula
is in good agreement with numerical solutions shown in fig. \ref{fig:ichalfxo}.

The upper limit to the synchrotron flux can be understood from the limit
we place on the synchrotron 10 keV flux.  We restrict synchrotron flux
at 10 keV to be less than the 10 keV SSC flux, in order that the GRB
spectrum is SSC dominated.  The 1 keV synchrotron flux is then
$<f_{{\gamma}}(10 \mathrm{keV}/\nu_\gamma)^{(p-1)/2} \sim 55$ mJy for
\grba, in agreement with Figure~\ref{fig:ichalfxo}.

\bigskip
\centerline{\it 4.2.1b. Prompt optical emission for SSC solutions with $\alpha
  \le -1/2$}
\medskip

In the bottom two panels of Figure~\ref{fig:ichalfxo}, we plot the
optical emission accompanying the $\gamma$-ray radiation for the SSC
solutions. In the optical (2 eV), we look only at the synchrotron
emission, since the SSC flux is highly suppressed as $\nu_a^{ic}\gg2$eV.
The optical flux is between 0.2 and 100 mJy, or between $R \sim $ 18
and 11 mags for \grba~(fig. \ref{fig:ichalfxo}).
  For the range of $\nu_\gamma$, $f_\gamma$, and $t_\gamma$ we have
considered, we find the optical flux to be between 0.02 mJy and
 50 Jy, or $R \sim $ 20 to 4 mags (bottom right panel of fig. 
\ref{fig:ichalfxo}). Observational limits on $R$ band flux, 100s after 
the burst, are $\gta$18.5 mag, or $\lta$0.2 mJy, for approximately half 
of Swift bursts \citep{roming06}. 
For small values of $\nu_\gamma$, $f_\gamma$, and $t_\gamma$ there is 
no problem satisfying this upper limit, but large $\nu_\gamma$, $f_\gamma$, 
or $t_\gamma$ might exceed the observed optical flux limit. Since the prompt 
optical flux falls off very rapidly with time as $t^{-(4+p)/2}$, for
$t>t_\gamma$, SSC solutions with hundreds of mJy optical flux during the
burst are consistent with the upper limit of $\sim 18$mag at 100s. 
 The solutions with optical flux greater than 100 mJy or so, however, 
can be ruled out, since prompt optical flux this bright is very rare.

Analytically, the R-band flux, dominated by synchrotron photons, is 
\begin{equation}
f_R \approx \left({2 \mathrm{eV} \over \nu_a}\right)^{\frac{5}{2}} 
\left({ \nu_a \over \nu_i}\right)^{-\frac{p}{2}} 
\left({\nu_i \over \nu_c}\right)^{-\frac{1}{2}} f_{\nu_p},
\end{equation}
provided that $\nu_a>2$eV, and $\nu_c<\nu_i<\nu_a$. In terms of the 
observed quantities the flux is given by
\begin{equation}
f_R \sim 2\times 10^3 \,\nu_{\gamma_5}^{-\frac{3}{4}} 
f_\gamma^{\frac{29}{36}} t_\gamma^{\frac{7}{18}} t_a^{\frac{7}{9}}
Y^{-\frac{11}{18}} (1+Y)^{\frac{7}{9}}
  d_{L_{28}}^{-\frac{7}{18}} (1+z)^{\frac{7}{9}} A_{2p}^{11\over 18}
\eta_a^{-{7\left(p+18\right)\over 36}} \eta_i^{-\frac{7}{9}}\,{\rm mJy}.
\end{equation}
If we assume that $t_a \sim t_\gamma$, 
we find $f_R \propto \nu_\gamma^{-\frac{3}{4}} f_\gamma^{\frac{29}{36}}
t_\gamma^{\frac{7}{6}}$.  This agrees with what we find numerically for
$f_\gamma$ and $t_\gamma$, but not $\nu_\gamma$.  Numerically, $f_R$
increases with $\nu_\gamma$, while this expression shows a decrease.
This difference is caused by the sensitivity of $f_R \sim \eta_a^{-4}$
-- numerically we find that $\eta_a \propto \nu_\gamma^{-1/2}$, changing
the above dependence on $\nu_\gamma$ to be $f_R \sim
\nu_\gamma^{5\over4}$, in accord with results shown in
Figure~\ref{fig:ichalfxo}.

For \grba, $f_R \lta 120 Y^{-\frac{11}{18}} (1+Y)^{\frac{7}{9}}$ mJy. 
This estimation for $f_R$ is larger by a factor $\sim10$ than what we find
numerically (see fig. ~\ref{fig:ichalfxo}). This factor of 10 difference is
due to the fact that $\gamma_c \sim 1$ for many of these
solutions. When $\gamma_c < 2$, we have a population of electrons
that don't radiate synchrotron emission and we need to reduce the 
number of radiating particles. This is done by using $\nu_c \propto 
(\gamma_c - 1)^2$.  Since $f_R \propto \nu_c^{\frac{1}{2}}$, the correct
value of $f_R$, is a factor of $\gamma_c/(\gamma_c-1)$ smaller than the 
crude analytical estimate above. The smallest value of $\gamma_c$ that
we find numerically is 1.08, and therefore the analytical expression
for $f_R$ overestimates the true flux, calculated numerically, by a factor 
of about $\sim14$. This indeed reconciles the analytical and numerical 
results. 

The R-band flux increases with increasing $t_\gamma$. This is because
increasing the pulse duration increases the radius at which the GRB
emission is produced, and the synchrotron self absorption frequency is
smaller at larger radii, and therefore brighter optical flux is
observed.  Thus, a prediction of the SSC model is that brighter optical
flux accompanies wider GRB pulses --- very spiky light curves (with
short variability time scale) will have small optical flux that can
escape detection, but those with wide pulses should have bright early
optical afterglows.  If a pulse duration were to be 10 s, we should
expect prompt optical flux of 100 mJy or larger (R-band magnitude
smaller than 11$^{th}$).   If this optical emission is not detected, it
will point to one of the assumptions in our model for the SSC-emission
being violated -- most likely $t_a\not\sim t_\gamma$, i.e. electrons are
not accelerated just once, but multiple times, during the course of a
pulse duration of $t_\gamma$.

\subsubsection{$\nu_i < \nu_c$ case}

Analytically this case is very similar to the case of $\nu_c < \nu_i$
discussed in \S4.2.1. We require this time that $\gamma_i/\gamma_c = \eta_i
 < 1$ and $\eta_a$ is still $ > 1$. Since the ratio $\gamma_a/\gamma_c =
\eta_a \eta_i$, we will see that some of the solutions can be expressed in
terms of $\eta_a\eta_i$ instead of using $\eta_i$ and $\eta_a$ separately.  

The equations that we are solving for, in this case are 
\begin{eqnarray}
&&\ B^2 \Gamma \gamma_i \sim 7.7 \times 10^8 \eta_i (1+z) t_a^{-1} (1+Y)^{-1}
\\  \nonumber\\[-.4cm] 
&&\ B \gamma_i^4 \Gamma \sim 2.3 \times 10^{12} \nu_{\gamma_5} (1+z) \eta_a^{-2}
\eta_i^2 \\   \nonumber\\[-.4cm] 
&& B \Gamma^5 \tau^2 \sim 1.6 \times 10^6 f_\gamma t_\gamma^{-2} (1+z)
d_{L_{28}}^2 \eta_a^p \eta_i^{2-p} \\   \nonumber\\[-.4cm] 
&& \tau \gamma_i^2 \sim 0.75 Y \left({p-1 \over \left(p-2\right)
\left(3-p\right)}\right)^{-1} \eta_i^{3-p}.
\end{eqnarray}
These equations are almost identical to the SSC $\nu^\frac{1}{3}$ case
of \S4.1.1, with exception of the dependence on the variables $\eta_i$
and $\eta_a$ and a slight change in the $Y$ expression with an
additional factor of ($3-p$).  These equations are solved same way as
outlined in  \S4.1.1. We find that the solutions have the same 
dependence on $\nu_\gamma$, $f_\gamma$ \& $t_\gamma$ as in equations 
\ref{icneg1}--\ref{icneg4}. And the dependence on $\eta_i$ and $\eta_a$ are:
$\gamma_i \propto \eta_a^{-{(p+18)\over 36}} \eta_i^{18-p \over 36}$,
$B \propto \left(\eta_i \eta_a\right)^{6-p \over 12}$,
$\Gamma \propto \left(\eta_i \eta_a\right)^{7 p -18 \over 36}$,
and $\tau \propto \eta_i^{36-17 p \over 18} \eta_a^{p+18 \over 18}$. 
The distance of $\gamma$-ray 
source $R_\gamma\propto\Gamma^2\propto\left(\eta_i \eta_a\right)^{7 p -18\over
18}$. We constrain $\eta_i$ by requiring $\nu_a^{ic} \sim 4 \nu_a
\gamma_i^2 \sim \nu_\gamma \eta_i^2 \lta 10$ keV, which suggests that 
$\eta_i \lta 0.3$, in accord with the numerical calculation.

Numerically, we find $\gamma_i \lta 20$, $B \lta 2\times10^3$ Gauss,
$\Gamma \lta 10^3$, and $\tau \gta 10^{-3}$ for \grba~ (fig. \ref{fig:ichalf}). 
The $\gamma$-ray source radius $R_\gamma \lta 2 \times 10^{15}
Y^{-\frac{7}{9}} (1+Y)^{\frac{4}{9}}$ cm is in good agreement with 
$R_\gamma \lta 5 \times 10^{15}$cm obtained by numerical calculations. 
 Note that $R_\gamma$ for this case is a factor of a few higher than the
$\nu_c < \nu_i$ case discussed in \S4.2.1.

The numerical results for the allowed region of 5-D space, for $\nu_c>\nu_i$,
are also shown in Figure~\ref{fig:ichalf}. The solutions corresponding
to $\nu_i > \nu_c$ are those at larger radius -- the right hand side of
each solution contour, or bubble, shown in the figure.  These solutions
have smaller $\gamma_i$, larger $\Gamma$, smaller $B$ \& $\tau$,
and a little bit smaller $Y$.  Since these solutions too seem viable
for shock models, we explore below the x-ray and optical flux accompanying
$\gamma$-ray emission.

\bigskip
\centerline{\it 4.2.2a. Prompt X-ray and optical flux}
\medskip

The analytical expression for the x-ray flux is almost identical to that 
found for the $\nu_c < \nu_i$ case (eq. \ref{icfx}) -- the only
difference is in the dependence on $\eta_a$ and $\eta_i$.  We find that --
\begin{equation}
f_x \propto \left( \eta_i \eta_a \right)^{\left(p-2\right) \left(p+18\right)
\over 36},
\end{equation}
and the lower limit on the x-ray flux is $\sim 50Y^{-\frac{37}{36}}
 (1+Y)^{-\frac{1}{18}}$mJy; this is in agreement with numerical results
shown in fig. (\ref{fig:ichalfxo}). 

The R-band flux, dominated by synchrotron emission, is
\begin{equation}
f_R \sim \left({2 \mathrm{eV} \over \nu_a}\right)^{2} 
\left({ \nu_a \over \nu_c}\right)^{-\frac{p}{2}} 
\left({\nu_c \over \nu_i}\right)^{-\frac{p-1}{2}} f_{\nu_p}
\end{equation}
which is, in terms of the observed quantities, given by
\begin{equation}
f_R \sim 3 \times10^3 \, \nu_{\gamma_5}^{-\frac{1}{2}} f_\gamma^{\frac{5}{6}} 
t_\gamma^{\frac{1}{3}} 
 t_a^{\frac{2}{3}} Y^{-\frac{2}{3}} (1+Y)^{\frac{2}{3}}
  d_{L_{28}}^{-\frac{1}{3}} (1+z)^{\frac{2}{3}} 
A_{1p}^{\frac{2}{3}}
\left(\eta_a \eta_i\right)^{-{p+18\over 6}}\, {\rm mJy}
\end{equation}
with the value of $f_R \lta 1700$ mJy for \grba.  This is higher than 
the numerical results that give $f_R \lta 200$ due to the sensitivity of
$f_R$ on $\eta_a \eta_i \gta 2$.  
 If we assume that $t_a \sim t_\gamma$, as has
been done numerically, we find $f_R \propto \nu_\gamma^{-\frac{1}{2}} 
f_\gamma^{\frac{5}{6}} t_\gamma$. Numerically we find that $f_R$ increases
with $\nu_\gamma$, $f_\gamma$, and $t_\gamma$, and is most sensitive to
$t_\gamma$.  The increase of $f_R$ with $\nu_\gamma$ is again due to the
high sensitivity on $\eta_a \eta_i$ which numerically we find is $
\propto \nu_\gamma^{-1/2}$.  

We note that the optical flux for this case ($\nu_i < \nu_c$) is larger
than when $\nu_c < \nu_i$ -- 10$^2$-- 10$^3$mJy -- since $R_\gamma$ is
larger for these solutions.  For $t_a/t_\gamma=10^{-2}$ $f_R$ is reduced
by a factor of $\sim50$. This brings the highest flux for the $t_\gamma
\sim 1$ s solutions down to about 140 mJy -- still a bit bright, but
there are many other solutions for this case that have flux smaller than
about 10 mJy, and probably in accord with observations and upper limits.
Thus, bright optical flux is a generic prediction of the SSC model for
$\gamma$-ray generation whether the low energy spectral index is
positive or negative; the optical is particularly bright for $\alpha>0$.
This is a problem for the SSC model if $R_\gamma \sim 10^{16}$ cm, as
found in \citet{kumar07}, as the brightest optical flux is produced at
the larger $R_\gamma$.  One of the, possibly only, ways to avoid bright,
prompt, optical emission is if $t_a\ll t_\gamma$.

\section{GeV photon signal for synchrotron and SSC solutions}

Detection of GeV photons by the Gamma Ray Large Area Space Telescope
(GLAST) \citep{mcenery06} will be a useful piece of evidence to use to
determine if GRBs are produced by synchrotron or SSC emission
mechanisms.  The IC scattering of $\gamma$-ray photons produced by
synchrotron will peak above the GLAST band, $\gg 100$ GeV, while the
second IC scattered SSC photons will peak at $\sim$ 1 GeV.  

For the synchrotron cases $\alpha = 1/3$ and $-(p-1)/2$,
$\max(\gamma_i,\gamma_c) \sim 10^4$ and $\tau \lta 10^{-6}$, giving the
peak of IC scattered flux $\nu f_\nu$ at $\nu_G \sim \min[\nu_\gamma
\max(\gamma_i,\gamma_c)^2, \max(\gamma_i,\gamma_c) \Gamma m_e c^2/(1+z)]
 \sim 10^4$ GeV and flux $\nu f_\nu \sim \tau
f_\gamma \nu_G \sim 10^{-7}$ erg s$^{-1}$ cm$^{-2}$. We need to ensure
that such high energy photons can escape the source region and are not
converted to electron-positron pairs. This effect is incorporated in our 
numerical calculations and is discussed below. Morever, photons of energy
larger than $\sim 1$TeV are converted to $e^\pm$ by collision with infrared
photons and therefore we would not see $\gta$1TeV photons from GRBs at cosmological
distances.

The SSC $\alpha >0 $ solutions have $\gamma_i \sim \gamma_c \lta 10^2$
and $\tau \gta 10^{-4}$ giving the second scattering peak of $\nu_G \sim
4 \nu_\gamma \gamma_i^2 \sim 1$ GeV and the flux $\nu f_\nu \sim
10^{-7}$ erg s$^{-1}$ cm$^{-2}$.  The SSC $\alpha <0$ solutions have
very similar $\nu_G$ and $\nu f_\nu$.  The SSC signals are well
above the GLAST threshold
so we expect to see a GeV signal from GRBs produced by the SSC process.
For synchrotron solutions, however, the IC flux might be below the GLAST threshold.  
The spectral shape will also be different -- $\nu_G$ is well above the
GLAST band for the synchrotron case, while SSC should peak at the low
end of the GLAST band.

Using analytical results for the synchrotron $\alpha = 1/3$ case (Equations
\ref{t1}--\ref{t4}), we find that the IC scattered signal peaks at
\begin{equation}
\nu_G^s \sim \gamma_i\Gamma m_e c^2/(1+z)\sim 2\times 10^4 \nu_{\gamma_5}^{1 
  \over 4} f_\gamma^{1 \over
8} t_\gamma^{-{1 \over 4}} t_a^{1 \over 2} Y^{-{1 \over 8}} (1+Y)^{1
\over 2} d_{L_{28}}^{1 \over 4} (1+z)^{-{1 \over 2}} \left[{
p \over p-2}\right]^{1 \over 8} \mathrm{GeV};
\end{equation}
This is due to the Klein-Nishina reduction to cross-section above electron rest
frame photon energy of $m_e c^2$.

 The frequency above which the high energy spectrum is attenuated due to 
$\gamma \gamma \to e^-e^+$ within the GRB source is $\nu_\pm
\sim (\Gamma m_e c^2)^2/\nu_{\tau \sim 1}$, where $\nu_{\tau \sim 1}$ is
the frequency of the synchrotron photon at which the optical depth
to pair production with $\nu_\pm$ photons is 1.  The optical depth 
to pair production is $\tau_{\gamma \gamma} \sim \sigma_T
n'_{\nu'} R_\gamma/ \Gamma$, where $n'_{\nu'} \sim L_{iso}(\nu)/4
\pi \Gamma c R_\gamma^2$ is the comoving number density of photons
between $\nu'$ and $2 \nu'$,  $R_\gamma /\Gamma$ is the comoving shell
width, and $L_{iso}(\nu)$ is the observed isotropic luminosity per unit
frequency.  To find $\nu_{\tau \sim 1}$, we set $\tau_{\gamma \gamma}
\sim 1$ and solve the equation using the observed $\gamma$-ray spectrum;
$\nu_{\pm}$ is calculated from $\Gamma$ \& $\nu_{\tau\sim1}$.
In terms of the observable parameters for the synchrotron $\alpha =1/3 $
case, we find
\begin{equation}
\nu_\pm \sim 4 \times 10^3 \nu_{\gamma_5}^{1 \over 4} f_\gamma^{1 \over
4} t_\gamma^{-1} t_a Y^{-{3 \over 4}} (1+Y) d_{L_{28}}^{1 \over 2}
(1+z)^{ 1 \over 2} \left[{p \over p-2}\right]^{3 \over 4} \,\mathrm{GeV}
\end{equation}
where we have assumed that the synchrotron GRB spectrum is $L_{iso}(\nu)\propto
\nu^{-2}$ for $\nu > \nu_\gamma$.  Since $\nu_\pm < \nu_{G}^s$ (calculated
above), and the spectrum falls off very steeply above $\nu_\pm$, the IC 
spectrum will peak at $\nu_\pm$ for many of the synchrotron solutions.

The flux at $\nu_G^s$, with appropriate Klein-Nishina cross section, is
\begin{equation}
[\nu f_{\nu}]^{s} \sim 8 \times 10^{-10} \nu_{\gamma_5}^{-{3 \over 2}}
f_\gamma^{3 \over 2} t_\gamma^{-1} Y^{1 \over 2}  d_{L_{28}} (1+z)^{-2}
\left[{p \over p-2}\right]^{-{1 \over 2}} \, \mathrm{erg} \, \mathrm{s}^{-1}
\mathrm{cm}^{-2}.
\label{nufnus13}
\end{equation}
This flux is probably just at the GLAST threshold for detection.  If
$\nu_\pm < \nu_G^s$, the $\nu f_\nu$ spectrum peaks at $\nu_\pm$, and
the flux at this frequency will be smaller than that in
equation (\ref{nufnus13}); the attenuation of $\gta$TeV photons as they propagate
through the inter-galactic medium would further reduce the observed flux.  
The results are very similar for the $\alpha =
-(p-1)/2$ case, since the $\alpha = 1/3$ case is a subset of the $\alpha
= -(p-1)/2$ solutions with $\gamma_i \sim \gamma_c$.  

\begin{figure}[h!]
\begin{center}
\includegraphics[height=4.9in,width=5.4in]{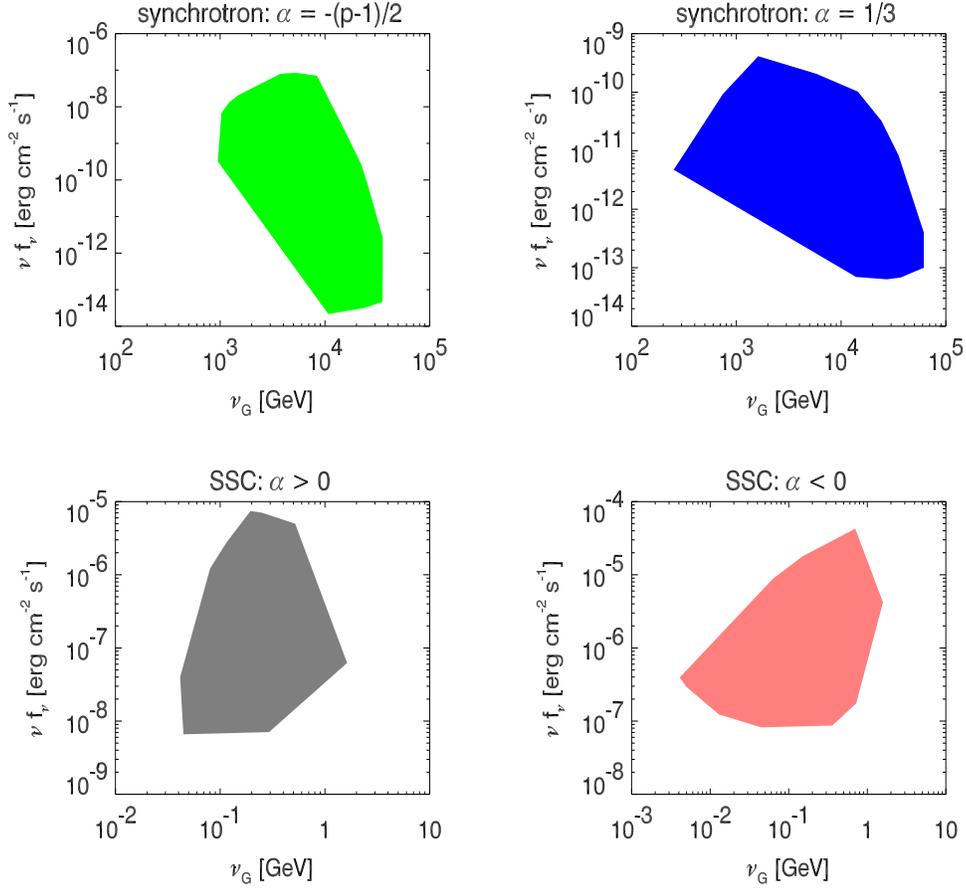}
\caption{\small
Numerical results for the IC scattering of prompt $\gamma$-ray photons in the
source. The four panels show the peak of IC spectrum ($\nu_G$) and the flux 
($\nu f_\nu$) at the peak when the $\gamma$-ray emission is produced via 
the synchrotron process (top two panels) and via the SSC process 
(bottom two panels) for  $\alpha>0$ \& $\alpha<0$ cases; for all of these
cases we took the underlying $\gamma$-ray spectrum with $\nu_\gamma=100$keV 
\& $f_\gamma=0.1$mJy. The IC signal for many of the synchrotron solutions 
are affected by photon-photon pair production within the GRB source, and that is 
included in these numerical calculations;  however the conversion of $\gta$1TeV
photons to $e^\pm$ due to collision with infrared photons in the inter-galactic
medium is not shown in the figure (no photon above 1 TeV can reach us from a
GRB source at cosmological distances). The effect of pair
production is very small in the SSC cases. The Klein-Nishina cross-section 
has been used in these calculations, and it significantly affects the
IC peak flux when $\gamma$-rays are generated via the synchrotron process.
}
\label{fig:GeV}
\end{center}
\end{figure}

For the SSC case, using the $\alpha > 0$ analytical results in Equations
\ref{pos1} -- \ref{pos4}, the 2$^{nd}$ IC peak is
\begin{equation}
\nu_G^{ic} \sim 3 \nu_{\gamma_5}^{3 \over 2} f_\gamma^{-{1\over18}}
t_\gamma^{1 \over 9} t_a^{2 \over 9} Y^{1 \over 9} (1+Y)^{2 \over 9}
d_{L_{28}}^{-{1 \over 9}} (1+z)^{2 \over 9} A_{2_p}^{-{1 \over 9}}
\eta_a^{-{p+18 \over 18}} \, \mathrm{GeV}
\end{equation}
and the flux at this peak is
\begin{equation}
[\nu f_{\nu}]^{ic} \sim 7 \times 10^{-7} \nu_{\gamma_5} f_\gamma Y 
A_{2_p}^{-1} \, \mathrm{erg} \, \mathrm{ s}^{-1} \mathrm{cm}^{-2}. 
\end{equation}
$\nu_G$ is smaller than $\nu_\pm$ for the majority of the SSC solution 
space, so the 2nd inverse Compton scattering spectrum will indeed peak 
at $\nu_G$. For the SSC $\alpha < 0$ case, the expressions are very similar
-- the only difference is the dependence on $\eta_a$, $\eta_i$, and $p$:
\begin{equation}
\nu_G^{ic} \sim 3 \nu_{\gamma_5}^{3 \over 2} f_\gamma^{-{1\over18}}
t_\gamma^{1 \over 9} t_a^{2 \over 9} Y^{1 \over 9} (1+Y)^{2 \over 9}
d_{L_{28}}^{-{1 \over 9}} (1+z)^{2 \over 9} A_{1_p}^{-{1 \over 9}}
\eta_a^{-{p+18 \over 18}} \eta_i^{18-p \over 18} \, \mathrm{GeV}
\end{equation}
and the flux at this peak is
\begin{equation}
[\nu f_{\nu}]^{ic} \sim 7 \times 10^{-7} \nu_{\gamma_5} f_\gamma
Y \eta_i^{3-p}
A_{1_p}^{-1} \, \mathrm{erg} \, \mathrm{ s}^{-1} \mathrm{cm}^{-2}. 
\end{equation} 

We show numerical results for $\nu_G$ and $\nu f_\nu$ for 4 cases in
Figure~\ref{fig:GeV} -- the two synchrotron cases of $\alpha = -(p-1)/2$
and $1/3$ and the SSC cases of $\alpha <0$ and $>0$. These numerical
results are in rough agreement with our analytical estimates.

In summary, IC scattering of prompt $\gamma$-ray photons, when the
GRB emission is produced via the synchrotron process, gives rise to a
spectrum that peaks at $\sim$ TeV,
the flux is of order 10$^{-9}$ erg cm$^{-2}$ s$^{-1}$ (fig. 13), the
spectrum below the peak is between $\nu^{1/3}$ or $\nu^{-(p-1)/2}$,
and the spectrum above the peak is expected to be sharply cutoff due
to pair production. On the other hand, if the GRB emission is produced 
via the SSC process then the spectrum of 2$^{nd}$ IC scattered  
photons should peak at $\sim$ 1 GeV, with a flux of $\sim10^{-6}$ 
erg cm$^{-2}$ s$^{-1}$ (fig. 13), and the spectrum above the peak should be 
$f_\nu\propto\nu^{-1.5}$. This signal should be detected by GLAST. If GRBs are 
not detected in the GLAST band, that would suggest that GRB prompt
emission is not generated via the SSC process.
We note that the GLAST band flux cannot be reduced in the case of 
repeated acceleration of electrons during a single GRB pulse.

\section{Comparison with prior work on $\gamma$-ray generation mechanism}

We provide a brief comparison with published work on $\gamma$-ray generation
in the internal shock model \citep{rm94,paczynskixu94,papmeszaros96,
meszarosrees97,saripiran97b} and the formalism/results of this paper.
There is a fairly extensive literature on this topic and this is not
the place to provide a general review. What we wish to do is to 
describe the main difference between previous approaches and the work 
presented here.

We should note that $\gamma$-ray generation in the  external shock model
has also been looked at by a number of people eg. Rees \& Meszaros
(1992), Meszaros \& Rees (1993), Piran et al. (1993), Dermer et al.
(1999), Dermer \& Mitman (1999), McMahon et al. (2004), Ramirez-Ruiz \&
Granot (2006); the  issue of variability in external shocks is discussed
in Sari \& Piran (1997a), Dermer \& Mitman (1999), Nakar \& Granot
(2006).  We don't have anything particularly enlightening to say
regarding the external shock model that has not already been mentioned
by one of these authors; the general problem with shocks is discussed 
in \S3 \& \S4.

The main difference between previous works and our approach is that
previous works considered the forward problem i.e., starting with a
parameterization of the properties of colliding shells and resulting
shocks the emergent radiation field was calculated, whereas our approach
is to start with the minimum number of physical parameters needed (five)
to calculate the observed flux and spectrum -- at one instant in time or
for one pulse in a multi-pulse GRB lightcurve -- and determine these
using the observed data.  For synchrotron \& IC radiations the
parameters needed are $\gamma_i, \Gamma, B, N$, and $\tau$,  which are
determined by the observational data $\nu_\gamma$, $f_\gamma$,
$t_\gamma$, and $\alpha$ for a pulse in GRB LC.  The 5 parameters in
turn are used to provide constraint on the nature of GRB source.  The
{\it old} {\it forward approach} is wedded to a particular model --
either internal or external shock -- whereas the method used in this
work is relatively model independent. 

The parameters of the internal shock model can be mapped into the five
parameters $(\gamma_i$, $\Gamma$, $B$, $N$, $\tau)$ in a straightforward
manner -- this in fact is done implicitly in all the {\it forward
approach} papers in order to calculate the emergent radiation. The
converse of this is not true, however, since the internal shock model
has more than 5 independent parameters (appendix A describes how to go
from the five parameters to providing a limit on some of the internal
shock parameters such as the initial LFs of colliding shells and their
comoving densities).

\citet{papmeszaros96} and \citet{saripiran97b} carried out a fairly
detailed analysis of prompt emission in the internal shock model. These
authors addressed a set of questions such as the ability of
synchrotron/SSC in internal shocks to produce a spectral peak near 100
keV, the observed flux in the $\gamma$-ray band, and short time scale
variability.  \citet{papmeszaros96} and \citet{saripiran97b} realized
that the cooling time scale for electrons (compared to the dynamical
time) for internal shocks is short and although \citet{saripiran97b}
don't explicitly say, their work applies to GRBs with $\alpha=-1/2$.
\citet{papmeszaros96} look at composite synchrotron/SSC spectra, however
there are many free parameters and little comparison to observed
properties of GRBs.  Had these authors investigated the self-consistency of
synchrotron internal-shock model for the case of $\alpha=1/3$, they
would have discovered the problem reported in this paper using their
{\it forward modeling} approach.

Ghisellini et al. (2000) did in fact worry about synchrotron solutions
when $\alpha>0$ and concluded that it is impossible to account for it in
shock based models (this is paraphrasing their actual wordings).  They
pointed out that electron cooling time, during prompt emission, is much
smaller than the dynamical time [as was reported in
\citet{papmeszaros96} and \citet{saripiran97b}] and therefore the GRB
spectrum below the peak should be always $\nu^{-1/2}$ if the radiation
is produced via synchrotron process.  Ghisellini et al. (2000)
considered the possibility that a lower strength magnetic field would
avoid the excessive cooling of electrons, and rejected it based on the
argument that IC emission would dominate in this case i.e., $Y\gg1$, and
that IC spectrum too would be falling of as $\nu^{-1/2}$ or steeper due
to IC cooling. We find that a smaller magnetic field can avoid excessive
cooling, so that $\alpha=1/3$, and at the same time Compton $Y\sim 1$.
The reason for these different conclusions is that we don't impose any 
restriction on the source distance that {\it forward calculations} based 
on internal shocks do. The most serious problem with synchrotron 
$\alpha=1/3$ case is that $R_\gamma>R_d$ unless
$n_0\ll1$ cm$^{-3}$ (see \S3.2). 

Ghisellini et al. (2000) correctly pointed out that re-acceleration
of electrons, in shock based models, would not work because it requires
too much energy; one has a continuous stream of electrons crossing the
shock front -- and to accelerate all of these electrons to their
original energy distribution, so that $\alpha=1/3$, while they are
rapidly losing energy to radiation will indeed require much more energy
than is available in the shock. The re-acceleration invoked in this work
is not shock based.  It in fact requires abandoning shock models and
considering a scenario where particles are NOT being added to the
``system" -- the source for $\gamma$-ray photons -- continuously (as in
shock based models) but where the source has a fixed number of particles
that are being continuously accelerated; there is no excessive energy
problem in this scenario.

\section{Discussion}

In this paper we have investigated the generation of $\gamma$-rays in
gamma-ray bursts via synchrotron or synchrotron-self-inverse-Compton
(SSC) emissions in a relativistic outflow. The SSC radiation from a
relativistic source can be fully described by a set of 5 parameters
($\gamma_i$, $\Gamma$, $B$, $N$, $\tau$); see table 1 for definition of
symbols used in the paper. For each possible low energy spectral index,
we have analytically and numerically determined the region of the 5-D
parameter space that is consistent with a set of GRB observations --
$\nu_\gamma$, $f_\gamma$, \& $t_\gamma$.  For these allowed regions --
or {\it solution} sub-space -- we calculate the x-ray and optical
fluxes that should be seen concurrent with the $\gamma$-ray radiation
to further narrow down the properties of $\gamma$-ray sources. The set
of five parameters also allows us to determine the distance of the source
from the center of explosion ($R_\gamma$).

We find that if $\gamma$-ray emission were to be produced via the
synchrotron process, the required set of parameters and burst radius
have extreme values that are not internally consistent and are in conflict
with afterglow data.  In particular, when the low energy spectrum is
$f_\nu\propto\nu^{1/3}$ or $\nu^{-{p-1\over 2}}$, the Lorentz factor of
the source is required to be larger than $\sim10^3$, in disagreement
with afterglow modeling, and the source distance ($R_\gamma$) is larger
than the deceleration radius even when the density of the medium is as
small as $\sim0.1$ cm$^{-3}$.  The requirement on the magnetic field
strength is also very stringent; the comoving field strength is required
to lie in a very narrow range of about $\sim 10$--30 Gauss in order to
explain the radiation for a typical burst as synchrotron emission.
Allowing for the possibility of multiple electron acceleration episodes,
i.e. $t_a\ll t_\gamma$, alleviates the problem of large $R_\gamma$ \&
$\Gamma$; in this scenario the synchrotron process could account for the
prompt $\gamma$-ray radiation for GRBs (see \S 3.1 \& 3.2) although $R_\gamma$
is still larger than the source distance determined for a subset of bursts
detected by Swift (Kumar et al. 2007).

The reason that synchrotron solutions with $\alpha=-(p-1)/2$ \& 1/3 require
large $R_\gamma$ \& $\Gamma$ is easy to understand. The number of
electrons needed to produce the observed
flux via the synchrotron process is $\sim 10^{53}f_\gamma/(B\Gamma)$.
And in order to keep the Compton-Y parameter ($Y\sim \tau \gamma_i
 \gamma_c$) less than $\sim10$ --- otherwise most of the energy will come out
in IC-scattered photons at $\nu\gg$ 1 GeV --- the source must have small
$\tau$ or large $R_\gamma$. Moreover, since
$t_\gamma\sim R_\gamma/(4c\Gamma^2)$, large $R_\gamma$ solutions also
have large $\Gamma$ for a given GRB pulse width of $t_\gamma$. The reason that
$t_a\ll t_\gamma$ offers a way out of this problem is also easy to understand.
Frequent re-acceleration of charge particles makes it
possible to have larger magnetic field while keeping $\nu_c\gta 100$ keV.
This decreases the number of particles required to produce the observed flux
($f_\gamma$), and that in turn makes it possible to have a smaller $R_\gamma$.

 For $f_\nu\propto\nu^{-1/2}$, the allowed region of the 5-D parameter
space for the synchrotron process is quite reasonable. However,
interpreting these solutions in terms of the internal shock model
requires the ratio of LFs of the two colliding shells to be rather large
($\gta 10-20$) when the ratio of magnetic energy to electron energy is
$\lta 10$, i.e. if we want the energy fraction in electrons to be not
too small -- otherwise $\gamma$-ray production would be very
inefficient (see \S3.3).

The SSC process provides viable solutions for the prompt emission of a large 
fraction of GRBs. We have considered almost all different possibilities of the
low energy spectrum for GRBs: $f^{ic}_\nu\propto\nu^\alpha$ below the peak of 
$\nu f^{ic}_\nu$ with $-1\lta\alpha\le 1$. The solution space (a hypersurface 
in the 5-D parameter space) is quite large for $\alpha<0$ and $0.5\lta\alpha
\le1$. However, there are no SSC solutions when $\alpha\sim 1/3$; the reason 
is that the synchrotron characteristic and cooling frequencies should be equal 
($\nu_i\sim\nu_c$) in order that $\alpha=1/3$, and in that case the 
synchrotron-self-absorption frequency is shown to be roughly equal 
to $\nu_i$ as well (see \S 4.1.1), and therefore the low energy spectral
index is $\sim 1$ and NOT $\sim 1/3$ as desired.

The SSC solution space for $\alpha<0$ and $0.5\lta\alpha\le1$ has
source LF of order 100, the minimum electron energy $\lta 10^2 m_e c^2$
(characteristic of mildly relativistic shocks), and
$10^{14}\lta R_\gamma \lta 10^{16}$cm is smaller than the
deceleration radius.  These solutions are accompanied by bright optical
synchrotron flux of $\sim 10$ mJy (14$^{th}$ mag) to several 
hundred mJy for bursts at $z\sim2$ -- brightest for bursts with 
$\alpha\sim 1$ and those bursts with pulse duration on the order of 
$\gta 1$s. Moreover, the optical flux is correlated with $R_\gamma$ 
-- the flux is larger for larger $R_\gamma$ -- and for 
$R_\gamma\gta 2\times10^{15}$ (cf. Kumar et al. 2007), the optical 
flux is $\gta$100mJy ($<12^{th}$ mag). Bright optical flux contemporaneous 
with $\gamma$-rays is a prediction of the SSC model that is in conflict with
prompt optical follow up observations of a large number of bursts detected by 
Swift (Roming et al.  2006).\footnote{The simultaneous optical and $\gamma$-ray observations for a 
few bursts show bright optical flares. For example, GRB 041219a was observed
in optical and IR simultaneously with $\gamma$-rays
\citep{vestrand05,blake05}; the optical flux peaked at 13.7 magnitude at
approximately the same time as the first main pulse of the GRB
lightcurve (as expected for the SSC model), and IR measurements 
\citet{blake05} show evidence of rapid variability with the last spikes
in the GRB light curve. GRB 990123, with a positive $\alpha$, had a 
peak optical flux during the burst of $\sim 1$Jy \citep{akerlof99}, 
however the optical lightcurve was not correlated with the $\gamma$-ray 
lightcurve. }
We note that the large optical flux accompanying
$\gamma$-rays can be reduced only if each radiating electron is
accelerated numerous times ($\gta100$) in time period of order the
duration of a pulse in the GRB lightcurve, i.e. $t_a\lta t_\gamma/10^2$.
GRB models based on converting kinetic energy to radiation via shocks
have $t_a\sim t_\gamma$, where particles are accelerated at the
shock-front and not down-stream, but continuous particle acceleration
might work for some alternate scenarios such as magnetic field
reconnection/dissipation.
Magnetic outflow model for GRBs has been proposed/investigated by a 
number of people cf. \citet{usov92,usov94}, \citet{thompson94},
\citet{katz97}, Meszaros \& Rees (1997), Wheeler et al. (2000 \& 2002),
Vlahakis \& Konigl (2001), Spruit et al. (2001), Lyutikov \&
Blandford (2003), Thompson (2006). However, the model has not been 
developed to the extend where it can be tested with GRB observations.

The data from the upcoming high-energy mission GLAST should be able to settle 
the question whether GRBs are produced via synchrotron or the SSC process (see 
\S5); see Gupta \& Zhang (2007), Granot et al. 2007 (and references therein)
for recent work on how GLAST would help our understanding of GRBs.

To summarize our main conclusions, we find that the synchrotron process
has serious difficulty accounting for the prompt emission in GRBs. The
SSC offers reasonable solutions for all GRBs except those with spectral
index of $\sim 1/3$ below the peak. SSC solutions predict very bright
optical emission ($>10$mJy or 14-mag for $z\sim2$) accompanying $\gamma$-ray
lightcurves which is in conflict with a number of well observed bursts. 
A possible solution to this problem might be to drop the assumption that 
$t_a \sim t_\gamma$. The assumption of one shot acceleration of electrons i.e.,
$t_a\sim t_\gamma$, is motivated by shock based physics for GRBs and
it may have to be replaced with an alternate scenario in which all
electrons that radiate in the $\gamma$-ray band are accelerated
continuously throughout the duration of a $\gamma$-ray pulse. In that
scenario there are viable synchrotron solutions when $\alpha\sim 1/3$ 
 -- a case that otherwise cannot be explained by the SSC mechanism.

\section*{Acknowledgments}
PK dedicates this work to Bohdan Paczynski, a mentor, a friend and an excellent
scientist from whom he has learned much about GRBs. This paper is the 
culmination of work over a long period of time and the authors are much
indebted to Tsvi Piran for many enlightening discussions during this
period, and Craig Wheeler for discussions about magnetic outflows and
for his encouragement. We thank Savannah Kumar and Monica Kidd for help 
with fig. 1, and Alin Panaitescu \& Bing Zhang for useful comments on 
the paper. This work is supported in part by grants from NSF (AST-0406878) 
and NASA Swift-GI-program.

\appendix

\section{Determination of the relative Lorentz factor of colliding shells}

\begin{figure}[h!]
\begin{center}
\includegraphics[height=3.9in,width=5.4in]{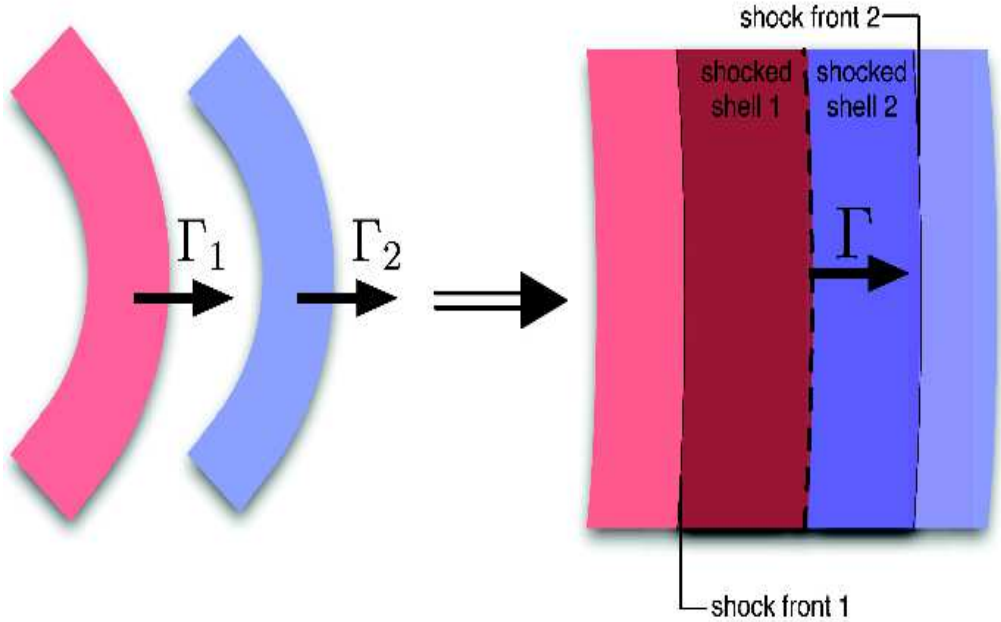}
\caption{Pictorial representation of two colliding shells --- internal shock 
  model.}
\label{fig:twoshells}
\end{center}
\end{figure}

We can relate solutions ~\fiveD ~ we find for a set of observables
to parameters for any model for GRBs. In this appendix we relate our
solutions to the parameters for the internal shock model where
$\gamma$-rays are produced by collision of two shells, as
shown in Figure~\ref{fig:twoshells}. Internal shocks are produced by a
collision between fast ejecta catching up with slower ejecta 
\citep{rm94,paczynskixu94} which in
the discrete version is modeled as the collision between two homogeneous
shells moving with LFs $\Gamma_1$ and $\Gamma_2$ ($\Gamma_1$ is LF of
the faster, inner shell). [The external shock can be thought of as a 
special case of internal shocks which results from collision
between a stationary circum-stellar medium ($\Gamma_2 =1$) and the
ejecta from the burst.] 

This appendix is devoted to describing the method we use to place a
lower limit to the relative LF of the colliding shells from $\Gamma_{sh}$,
the LF of the shock front in the $\gamma$-ray producing shell, 
which is directly related to $\gamma_i$ (one of the 5 parameters) via
equation (\ref{gamsh}), and by varying the ratio of densities in the 
inner and outer shells ($n_1/n_2$) subject to the condition that the 
efficiency for $\gamma$-ray production is not less than $\sim$10\%; as
a by product we find the comoving density ratio for the shells.

The LF of unshocked inner shell, which is moving faster and lies a bit closer
to the center of explosion than the outer shell, can be determined from
the addition of LFs $\Gamma_{sh}$ and $\Gamma$ --
\begin{equation}
\Gamma_{1} = \Gamma_{sh} \Gamma(1 + v_{sh} v),
\end{equation}
provided that the $\gamma$-ray emission we observe is produced in the inner
shell when it is shock heated ($v$ \& $v_{sh}$ are speeds corresponding 
to LFs $\Gamma$ -- the LF of the $\gamma$-ray source -- \& $\Gamma_{sh}$). 
On the other hand if it is the outer shell that produces the observed 
$\gamma$-ray emission then its LF factor, before it was shocked, is given by
\begin{equation}
\Gamma_{2} = \Gamma_{sh} \Gamma(1 - v_{sh} v).
\end{equation}
The LF of the shell that does not contribute significantly
to the observed $\gamma$-ray emission cannot be determined uniquely.
The problem is that in the absence of any observed emission that
can be identified with this shell we cannot say anything
directly about the LF of the shock front moving into this shell.
However, we can still constrain its LF by requiring that the comoving
density of this shell is such that the efficiency for thermal energy produced
in the collision is no less than $\sim$10\% (most GRBs for which
we have good afterglow data and we can determine the kinetic energy of the
ejecta show that the efficiency for converting kinetic energy to $\gamma$-rays
is about 10\% or more).  

The efficiency of $\gamma$-ray production is used to constrain $n_1/n_2$,
and that in turn provides a lower limit for the relative LF $\Gr$. 
The fraction of kinetic energy
converted to thermal energy when two shells of mass $m_1$ and $m_2$
collide with a relative LF of $\Gr$ is (Kumar, 1999; Piran, 1999)
\begin{equation}
f_r = 1 - \left[ 1 + {2(m_2/m_1)(\Gr - 1)\over [1+m_2/m_1]^2}
      \right]^{-1/2}
\label{eq:fR}
\end{equation}
For equal mass shells a $\Gr=1.5$ collision converts 10\% of the
kinetic energy of shells to thermal energy, and for $\Gr>5$ the
conversion efficiency is more than 40\%.  We need, however, to figure 
out the fraction of energy produced in a collision that is radiated in
the typical $\gamma$-ray observing band of 15--400 keV during
the time interval $t_\gamma$, in two shells that are not of equal mass.  

The ratio of the mass of the two shells is the mass swept up by
the two shocks in the time equal to the shock transit time for the shell
that produces the observed $\gamma$-ray emission. This is given by
\begin{equation}
{m_1\over m_2} = {n_1 (4\Gamma_{s_1} + 3) v_{ss_1}\over
      n_2 (4\Gamma_{s_2} + 3) v_{ss_2}  }. 
\end{equation}
where $n_1$ and $n_2$ are the densities of shell 1 \& 2, $\Gamma_{s_1}$
and $\Gamma_{s_2}$ are the shock LFs in the frame of each unshocked
shell (if the GRB emission is predominantly from shell 1, then
$\Gamma_{s_1} = \Gamma_{sh}$), and $v_{ss_1}$ and $ v_{ss_2}$ are the
speeds of the shock fronts with respect to each unshocked shell.

The shock front speeds are determined from the following cubic equation
obtained from the continuity of energy, momentum and particle number
fluxes across the shock front (Landau \& Lifshitz, 1980):
\begin{equation}
\Gamma_{ss}^3 + \left(1  - {2\over \alpha}\right)
\Gamma_{su}\Gamma_{ss}^2
  - \left(1 - {1\over \alpha^2}\right) \Gamma_{ss}  - \Gamma_{su}
    \left(1  - {1\over \alpha}\right)^2 = 0.
\label{eq:cubic}
\end{equation}
where $\Gamma_{su}$ \& $\Gamma_{ss}$ are the LF of the shock front wrt
to the
unshocked and shocked fluid respectively, and $\alpha$ depends on the
equation of state of the shocked gas and is approximated by
\begin{equation}
\alpha = {4\Gamma_0 + 1\over \Gamma_0 + 1},
\end{equation}
which provides a smooth interpolation between the sub-relativistic value
for $\alpha=5/2$ and the highly relativistic value of 4.

The shock LF wrt the unshocked fluid in the shell not dominating the GRB
emission ($\Gamma_{s_2}$, if we assume that the GRB is being produced in
shell 1) can be determined from the condition of pressure balance across
the surface of discontinuity separating the shocked fluids in the two
shells-
\begin{equation}
n_1(4\Gamma_{s_1} + 3)(\Gamma_{s_1} - 1) = n_2(4\Gamma_{s_2} + 3)
   (\Gamma_{s_2} - 1). 
\end{equation}

For a given density ratio $n_1/n_2$ and $\Gamma_{s_1}$ we can solve the
above
equation to determine $\Gamma_{s_2}$, which in turn is used to determine
shock front speed wrt the unshocked fluid for the outer/inner shell
using 
\begin{equation}
\Gamma_{s_2} = \Gamma_{ss_2}\Gamma_{su_2}\left[ 1 - v_{ss_2}
v_{su_2}\right],
\end{equation} 
and the cubic equation \ref{eq:cubic} for $\Gamma_{ss_2}$.  These pieces
together give us $m_1/m_2$ and $\Gr$, which are used to calculate 
$f_r$ -- the fraction of the kinetic energy of the two shells converted 
to thermal energy in shell collision (eq. A3).

The ratio of the internal energy of the shocked gas in these two shells is
\begin{equation}
{E_1\over E_2} = {n_1 (4\Gamma_{s_1} + 3)(\Gamma_{s_1} -1) v_{ss_1}\over
      n_2 (4\Gamma_{s_2} + 3) (\Gamma_{s_2} -1 )v_{ss_2}  }.
\end{equation}
The fraction of the total internal energy of the shocked gas in shell `1' is 
then $f_1 = {E_1\over E_2}/\left({E_1\over E_2}+1\right)$.  We assume that
the majority of this energy is indeed being radiated in the GRB band.  For
shell `2', $f_2 = 1-f_1$, and we find the fraction of the total
radiation contributing to the GRB band, $f_{GRB}$;  $f_{GRB}$ is found
from the ratio $\nu_{i_1} f_{\nu_2}\left(\nu_{i_1}\right)/\nu_{i_2}
f_{\nu_2}\left(\nu_{i_2}\right)$, where $\nu_{i_1}/\nu_{i_2} =
\left(\gamma_{i_1}/\gamma_{i_2}\right)^2$ (assuming that the magnetic
field is equal in both shells).  The radiation efficiency in the GRB band 
for the shell collision is then $f_R\left(f_1 + f_2 f_{GRB}\right)$.

When considering synchrotron radiation as the primary source of emission
in the $\gamma$-ray band we need to take into account the energy
fraction that is lost to very high energy photons ($\nu\gg 200$ keV)
produced via the inverse-Compton process.  The fraction of energy
radiated via the synchrotron emission is $1/(1+Y)$, therefore, the total
efficiency for energy production in shell collisions must be larger than
the desired 10\% by a factor of $1+Y$ -- this effectively restricts
solutions to $Y\lta 10$. $Y\gta0.1$ if inverse-Compton emission is the
main source for the observed $\gamma$-ray emission and also $Y$ must not
be greater than $\sim 10^3$ otherwise most of the radiative energy will
be in the 2nd inverse-Compton photons of much higher energy.  In the
same sense, we need to make sure the ratio of magnetic energy to that in
electrons is $\lta 1$, in order for the electrons to radiate
efficiently.

\begin{figure}[h!]
\begin{center}
\includegraphics[height=3.9in,width=5.95in]{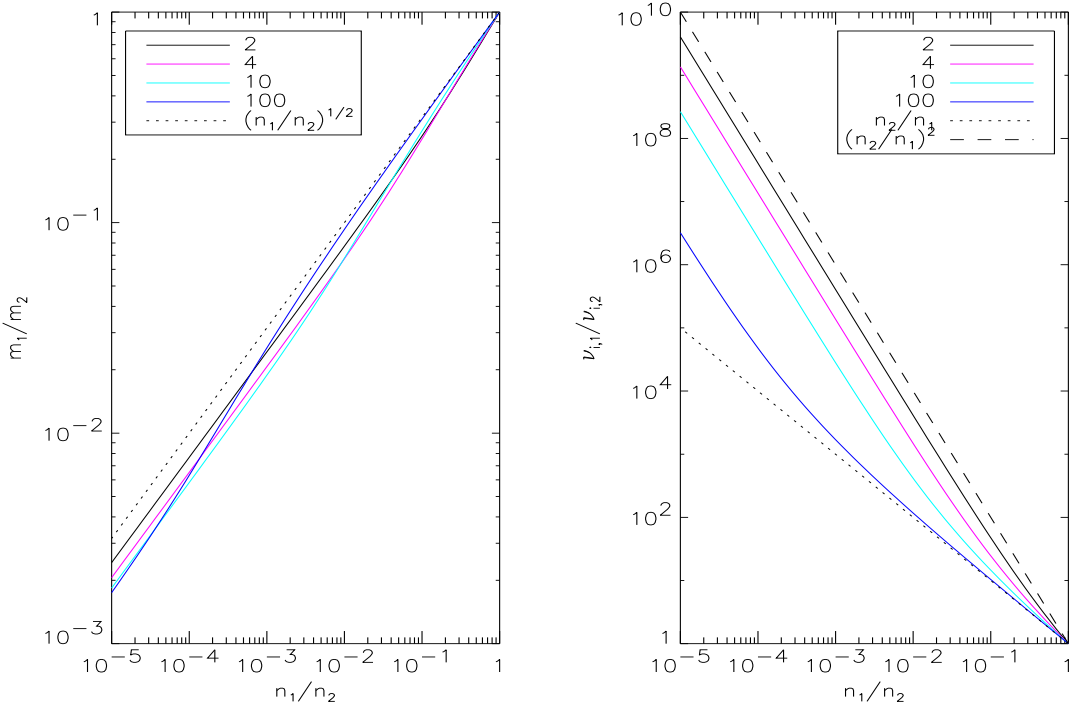}
\caption{\small Left panel: Dependence of the ratio of the two shell masses,
$m_1/m_2$, on the ratio of densities, $n_1/n_2$.  The colored lines show
the relation for different relative Lorentz factor, $\Gr$, as shown
in the legends.  The dotted and dashed lines show the applicable
dependence on the density ratio.  For high $\Gr$ and $n_1/n_2 >
0.01$, $m_1/m_2 \propto \left(n_1/n_2\right)^{1/2}$, however the deviation from
this relation for small $\Gr$ and $n_1/n_2$ is small -- about a
factor of 2.  Right panel: Dependence of the ratio of synchrotron injection
frequencies ($\nu_i$) in the two shells on $n_1/n_2$.  For high $\Gr$ and
$n_1/n_2 > 0.01$, $\nu_{i,1}/\nu_{i,2} \propto n_2/n_1$.  At low
$\Gr$ and small $n_1/n_2$, the ratio of injection frequencies in
the two shells significantly deviates from this, and for mildly
relativistic $\Gr$, can be better approximated by
$\nu_{i,1}/\nu_{i,2} \propto \left(n_2/n_1\right)^{2}$, shown by the the
dashed line.}
\label{fig:n12}
\end{center}
\end{figure}

For highly relativistic forward and reverse shocks $v_{ss_1}\approx
v_{ss_2} \approx 1/3$, $\Gamma_{s_2} \approx \Gamma_{s_1}
(n_1/n_2)^{1/2}$, $\Gr\approx \Gamma_{s_1}\Gamma_{s_2}$, 
$m_1/m_2\approx (n_1/n_2)^{1/2}$, and the ratio of the characteristic
synchrotron frequency in shell 1 to shell 2 (assuming the same magnetic
field strength) is $\sim n_2/n_1$. We note that the assumption of highly
relativistic shocks is not valid for many solutions we find for the
prompt $\gamma$-ray emissions, and that all the results reported in
paper are obtained by numerically solving the appropriate equations.
The numerically solved relationships of the ratio of the masses and
injection frequencies in the two shells are shown in
Figure~\ref{fig:n12}, and compared to the analytical estimates for
highly relativistic shocks.

In summary, for a given $\Gamma_{sh}(\gamma_i)$ we vary $n_1/n_2$ and
determine the mass ratio and the relative LF of the shell collision
($\Gr$) so that the gamma-ray production efficiency is above a
certain desired value (10\%).  All of the numerical results we show for
$\Gr$ were calculated using these steps.  Using our upper limit
on $\Gamma_2$, we also calculate the expected emission from the shell 2
in the x-ray and optical bands, and include this in our analysis.

\section{Jitter radiation process and GRBs}

A radiation mechanism, called jitter radiation, has been
suggested by \citet{medvedev} as the process for $\gamma$-ray emission
for those cases where the low energy spectrum rises more steeply than
$\nu^{1/3}$ expected of the synchrotron radiation (such spectra are said
to lie ``above the line of death" because a non-self-absorbed
synchrotron spectrum cannot have this steep of a rise). The jitter
radiation is produced when the coherence length scale for magnetic field
is short and electron trajectory is perturbed before it has traveled
a distance of a Larmor radius.  This is an attractive idea for
explaining a class of GRBs lying above the line of death, and we explore
its applicability to GRBs in this appendix. 

The peak jitter frequency in lab frame (as seen by an observer at rest
in the host galaxy) is:
\begin{equation}
\nu_j = \sqrt{16\pi q^2 \Gamma_{sh} n_e/(m_e \bar\gamma_e)} \gamma_p^2 \Gamma,
\end{equation}
where $q$ is electron charge, $\Gamma$ is the bulk Lorentz factor (LF) 
of the source, $\gamma_p = \min\{\gamma_i, \gamma_c\}$, $\gamma_c$ 
is the thermal LF of electrons
that cool on a dynamical time, $\gamma_i$ is the minimum thermal LF
of electrons behind the shock front (note that $\gamma_i = \epsilon_e
(m_p/m_e) \Gamma_{sh} (p-2)/(p-1)$; where $\Gamma_{sh}$ is the LF of the
shock front wrt the unshocked shell), $n_e$ is the comoving electron
density in the unshocked shell, and $\bar\gamma_e \sim 3$ -- 4 is the
initial effective thermal Lorentz factor of the streaming electrons. 

In order to get the gamma-ray burst spectrum below the peak to be
proportional to $\sim\nu^{1}$, we want $\nu_j \sim 10^5$ eV (the peak of 
the GRB spectrum is of order 100 keV). Therefore, from the above
equation we find the following condition on comoving electron density
in the unshocked shell:
\begin{equation}
n_e \approx {4\times10^{28} \over \Gamma^2 \gamma_p^4} (\bar\gamma_e/
    \Gamma_{sh})
\end{equation}
or the optical depth of the source to Thomson scattering is:
\begin{equation}
\tau \approx {4 n_e \Gamma_{sh} R_\gamma \sigma_T/\Gamma} \approx
        {10^{16} \bar\gamma_e t_{\gamma} \over \Gamma \gamma_p^4}
\end{equation}
where $R_\gamma$ is the distance of the source from the center of the 
explosion, and $t_{\gamma} \approx R_\gamma/(4\Gamma^2 c)$ is the GRB 
variability time scale. 

For internal shock $\Gamma_{sh}$ is of order a few, and therefore, 
$\gamma_i \approx 10^3$. In this case we find the optical depth
of the source to be
\begin{equation}
\tau \approx {10^4 \bar\gamma_e t_{\gamma} \over \Gamma}.
\end{equation}
Or $\tau \sim 1$ for $t_{\gamma} \sim 10^{-2}$ sec, and 
$\Gamma\sim 10^3$.
The next step is to estimate the LF of cooling electrons
($\gamma_c$), which can be shown to be:
\begin{equation}
\gamma_c \approx {6\pi m_e c (1+z)\over \sigma_T B^2 t_{\gamma}
                                   \Gamma (1+Y)},
\end{equation}
where B is the magnetic field in the source co-moving frame, and Y is the
Compton-Y-parameter; $Y\approx \tau \gamma_c\gamma_i$ (for $p\approx 2$).

If $\gamma_c$ is not much less than $\gamma_i$ then
Y is very large (of order a million), and most of the radiation energy
will come out as SSC photons at $> 10^2$ GeV. And moreover, electrons
cool very rapidly resulting in the synchrotron cooling frequency ($\nu_c$)
to be much less than 100 keV, and therefore the spectrum below the
peak will be $\nu^{-1/2}$ and not $\propto \nu^{1}$. Even if we take
$\gamma_c$ to be order unity, we still run into similar problems. 

We, therefore, do not include the jitter process in our analysis of 
GRB prompt radiation mechanism.

\end{document}